\documentclass[nohyper,11pt,letterpaper]{JHEP3}
\usepackage[dvips]{epsfig}
\usepackage{graphicx}
\usepackage{mathrsfs}
\usepackage{amsmath}
\usepackage{epsfig,latexsym}

\newcommand{\Res}[1]{\underset{#1}{\mathrm{Res}}}

\newcommand{\smfrac}[2]{{\textstyle \frac{#1}{#2}}}

\newcommand{\cQ}{q_\mt{7}} 

\newcommand{\be}{\begin{equation}}
\newcommand{\ee}{\end{equation}}
\newcommand{\bea}{\begin{eqnarray}}
\newcommand{\eea}{\end{eqnarray}}
\newcommand{\ba}{\begin{eqnarray}}
\newcommand{\ea}{\end{eqnarray}}
\newcommand{\nn}{\nonumber \\}

\newcommand{\beq}{\begin{equation}}
\newcommand{\eeq}{\end{equation}}
\newcommand{\beqa}{\begin{eqnarray}}
\newcommand{\eeqa}{\end{eqnarray}}
\newcommand{\beqar}{\begin{eqnarray*}}
\newcommand{\eeqar}{\end{eqnarray*}}
\newcommand{\e}{\epsilon}
\newcommand{\reef}[1]{(\ref{#1})}

\newcommand{\eg}{{\it e.g.,}\ }
\newcommand{\ie}{{\it i.e.,}\ }

\newcommand{\tlt}{{\tilde{t}}} 
\newcommand{\tx}{{\tilde{x}}}
\newcommand{\ty}{{\tilde{y}}}
\newcommand{\ts}{{\tilde{\sigma}}}
\newcommand{\tlnu}{{\tilde{\nu}}}
\newcommand{\tlgam}{{\tilde{\gamma}}}

\def\Tr{{\rm Tr}}

\newcommand{\tk}{\tilde{k}}
\newcommand{\tlq}{\tilde{q}}
\newcommand{\tom}{\tilde{\omega}}

\def\nc {N_\mt{c}}
\def\nf {N_\mt{f}}

\def\t6 {T_\mt{D6}}

\def\gym {g_\mt{YM}}

\newcommand{\mq}{M_\mt{q}}      

\newcommand{\gs}{g_\mt{s}}
\newcommand{\ls}{\ell_\mt{s}}

\newcommand{\mt}[1]{\textrm{\tiny #1}}

\renewcommand{\Re}{\mathrm{Re}\,}
\renewcommand{\Im}{\mathrm{Im}\,}

\newcommand{\labell}[1]{\label{#1}} 

\newcommand{\bi}{\begin{itemize}}
\newcommand{\ei}{\end{itemize}}
\newcommand{\ben}{\begin{enumerate}}
\newcommand{\een}{\end{enumerate}}

\newcommand{\bmtx}{\left[ \begin{array}{cc}}
\newcommand{\emtx}{\end{array} \right]}
\newcommand{\bvec}{\left[ \begin{array}{c}}
\newcommand{\evec}{\end{array} \right]}

\renewcommand{\Re}{{\mbox{Re }}}
\renewcommand{\Im}{{\mbox{Im }}}

\newcommand{\bfig}{\begin{figure}}
\newcommand{\efig}{\end{figure}}

\newcommand{\gsim}{\mathrel{\raisebox{-.6ex}{$\stackrel{\textstyle>}{\sim}$}}}

\title{Transport Properties of Holographic Defects}
\author{
Robert C. Myers and Matthias C. Wapler\\
Perimeter Institute for Theoretical Physics,
Waterloo, Ontario N2L 2Y5, Canada \\
Department of Physics \& Astronomy and Guelph-Waterloo Physics
Institute,\\
\ \  University of Waterloo,
Waterloo, Ontario N2L 3G1, Canada \\
\\E-mail: \email{
rmyers@perimeterinstitute.ca,
 mwapler@perimeterinstitute.ca }}

\preprint{arXiv:0811.0480 [hep-th]}

\date{\today}

\abstract{We study the charge transport properties of fields
confined to a (2+1)-dimensional defect coupled to (3+1)-dimensional
super-Yang-Mills at large-$\nc$ and strong coupling, using AdS/CFT
techniques applied to linear response theory. The dual system is
described by $\nf$ probe D5- or D7-branes in the gravitational
background of $\nc$ black D3-branes. Surprisingly, the transport
properties of both defect CFT's are essentially identical -- even
though the D7-brane construction breaks all supersymmetries. We find
that the system possesses a conduction threshold given by the
wave-number of the perturbation and that the charge transport arises
from a quasiparticle spectrum which is consistent with an intuitive
picture where the defect acquires a finite width. We also examine
finite-$\lambda$ modifications arising from higher derivative
interactions in the probe brane action.}


\begin{document}{\vskip 1cm}


\section{Introduction}

The AdS/CFT correspondence provides a powerful framework for the
study of strongly coupled gauge theories \cite{juan,adscft,bigRev}.
While the gauge theories that are currently amenable to such
holographic analysis are typically very different from real world
QCD, this duality may still give us some insight into the strongly
coupled quark-gluon plasma (sQGP) produced in recent experiments at
RHIC \cite{talks}. Matching AdS/CFT results with experimental data
here anticipates that the sQGP can be described by an effective CFT
which is in the same universality class as the strongly coupled
gauge theories studied holographically. Similar considerations have
recently motivated exploring the possible application of the AdS/CFT
correspondence to the study of $(2+1)$-dimensional condensed matter
systems \cite{pavel}. A variety of holographic models displaying
interesting properties, including superfluidity, superconductivity
and Hall conductivity, have now been studied
\cite{more,hallo,hall1}. Further interesting models of various types
of nonrelativistic CFT's have also been constructed \cite{nonrel}.
One advantageous aspect of the AdS/CFT correspondence is the
``uniformity'' of the calculations, \ie a single set of calculations
describes the system in different disparate regimes
($\omega\rightarrow0$ versus $T\rightarrow0$). This can be
contrasted with more conventional field theory analysis of conformal
systems in (2+1)-dimensions \cite{subir}. One might also approach
these explorations of possible ``AdS/Condensed Matter"
correspondences with some optimism because of the rich variety of
experimental systems which might exhibit conformal behaviour and
also the vast array of four-dimensional AdS vacua in the string
theory landscape \cite{land}.

In the present paper, we use the AdS/CFT correspondence to study the
charge transport properties of fields confined to a
(2+1)-dimensional defect coupled to (3+1)-dimensional
super-Yang-Mills. The defect CFT is realized by inserting $\nf$
probe D5- or D7-branes into the background of a black D3-brane. In
either construction, we are also able to introduce an additional
internal flux, in which case the defect in the CFT separates regions
where the rank of the SYM gauge group is different. The defect CFT
constructed with the D5-branes is certainly well known
\cite{lisa,hirosi,jaume}. Certain aspects of the D7-brane
construction have also been studied previously
\cite{rey,quantumhall} but we should note that the internal flux
introduced here is essential to remove an instability that would
otherwise appear in this construction. We will use linear response
theory to study the conductivity on the defect at finite frequency
and temperature and at finite wave-number, \ie the conductivity of
an anisotropic current.

An overview of the paper is as follows: In section \ref{revGeom}, we
review the holographic framework, in particular the embedding of the
probe D-branes in the AdS$_5\times S^5$ background. In section
\ref{form}, we obtain the basic results for the spectral functions,
starting with a review of the methodology and then the computation
of the transverse and longitudinal conductivities in section
\ref{goldfinger}. Here we also comment on the agreement with the
diffusion-dominated conductivity in the hydrodynamic regime, \ie in
the regime at small frequency and wave-number $\omega, q \ll T$.
This is followed by a discussion of the collisionless, ($q \gsim
T$), regime using analytical approximations in section \ref{efftf},
for both the insulating case (at small fequencies) and the optical
regime (at large frequencies). Using those results, we study the
spectrum of quasinormal modes in section \ref{quasinorm}. In section
\ref{edualmain}, we examine the effect of stringy corrections to the
gauge theory on the probe branes, which describes the behaviour of
the dual currents on the defect. In particular, electromagnetic
duality is lost when these $\alpha'$-corrections are included, which
has interesting implications for the conductivity as strong but
finite `t Hooft coupling. Finally, we consider the computation of a
topological Hall conductivity in section \ref{hall}. Section
\ref{discuss} closes with some discussion and observations about our
results. Some details of our analysis are relegated to appendices:
In appendix \ref{diffusion}, we calculate the diffusion constant for
charge transport. Appendix \ref{appcorr} presents some details of
the analysis including certain $\alpha'$ corrections in the D5-brane
worldvolume action. In appendix \ref{tanhpot}, we do an analytical
study of a slightly simplified model of the defect, which gives
further qualitative insight, and aids the numerical computation of
the quasinormal modes.

\section{Defect branes}
\label{revGeom}

The AdS/CFT correspondence is most studied and best understood as
the duality between type IIb string theory on $AdS_5\times S^5$ and
${\cal N}=4$ super-Yang-Mills theory with $U(\nc)$ gauge group. In
this context, all fields in the SYM theory transform in the adjoint
representation of the gauge group. One approach to introducing
matter fields transforming in the fundamental representation is to
insert probe D7-branes into the supergravity background
\cite{karchkatz}. However, this approach can also be used to
construct a defect field theory, where the fundamental fields are
only supported on a subspace within the four-dimensional spacetime
of the gauge theory. In particular, we will consider constructing a
$(2+1)$-dimensional defect by inserting $\nf$ D$p$-branes, with
three dimensions parallel to the SYM directions and $p-3$ directions
wrapped on the $S^5$. In the following, we work with both probe D5-
and D7-branes. If we consider the supergravity background as the
throat geometry of $\nc$ D3-branes, our defect constructions are
described by the following array:
\begin{equation}
\begin{array}{rcccccccccccl}
  & & 0 & 1 & 2 & 3 & 4& 5 & 6 & 7 & 8 & 9 &\\
\mathrm{background\,:}& D3 & \times & \times & \times & \times & & &  & & & & \\
\mathrm{probe\,:}& D5 & \times & \times & \times &  & \times  & \times & \times & &  & &  \\
& D7 & \times & \times & \times &  & \times  & \times & \times & \times & \times & &   \\
\end{array}
\labell{array}
\end{equation}

The D5-brane construction is supersymmetric and the dual field
theory is now the SYM gauge theory coupled to $\nf$ fundamental
hypermultiplets, which are confined to a (2+1)-dimensional defect.
Note that the supersymmetry has been reduced from ${\cal N}=4$ to
${\cal N}=2$ by the introduction of the defect. In the D7-brane
case, we have lost supersymmetry altogether and the defect supports
$\nf$ flavours of fermions, again in the fundamental representation
\cite{rey}. One should worry that the lack of supersymmetry in the
latter case will manifest itself with the appearance of
instabilities. However, we will explicitly show below in section
\ref{d7-geom} that this problem can be avoided. In the limit $\nf
\ll \nc$, the D5- and D7-branes may be treated as probes in the
supergravity background, \ie we may ignore the gravitational
back-reaction of the branes.

As we commented above, a similar holographic framework has been used
extensively to study the properties of the ${\cal N}=2$ gauge theory
constructed with parallel D7- and D3-branes, \ie the fundamental
fields propagate in the full four-dimensional spacetime -- \eg see
\cite{johanna,recent,long}. If a mass $\mq$ is introduced for the
hypermultiplets, it was found that the scale
$M_\mt{fun}\sim\mq/\sqrt{\lambda}$ plays a special role in this
theory. First, the ``mesons", bound states of a fundamental and an
anti-fundamental field, are deeply bound with their spectrum of
masses characterized by $M_\mt{fun}$ \cite{meson}. Next at a
temperature $T\sim M_\mt{fun}$, the system undergoes a phase
transition characterized by the dissociation of the mesonic bound
states \cite{long}. The analogous results can be verified for the
defect theories considered here. That is, the meson spectrum is
characterized by the same mass scale $M_\mt{fun}$
\cite{ramallo,holomeson} and these states are completely dissociated
in a phase transition at $T\sim M_\mt{fun}$ \cite{matt}. However,
these results are tangential to the present study, as we will only
consider the conformal regime with $\mq=0$.

Common to both of our constructions is the supergravity background
dual to ${\cal N}=4$ SYM at finite temperature. This background is a
planar black hole in $AdS_5$, corresponding to the decoupling limit
of $\nc$ black D3-branes \cite{wittt}:
\beq ds^2 = \frac{r^2}{L^2} \left( -h(r)dt^2 +dx^2+dy^2+dz^2\right)
+ \frac{L^2}{r^2} \left( \frac{dr^2}{h(r)} +r^2 d\Omega_5^2\right)\,
, \quad \quad C^{(4)}_{txyz}=-\frac{r^4}{L^4} \labell{D3geom} \eeq
where $h(r) = 1-r_0^4/r^4$. The gauge theory directions correspond
the coordinates $\{ t,x,y,z\}$. The radius of curvature $L$ is
defined in terms of the string coupling constant $\gs$ and the
string length scale $\ls$ as $L^4 = 4\pi\, \gs \nc \, \ls^4$. The
holographic dictionary relates the Yang-Mills and string coupling
constants as $\gym^2 = 4 \pi \gs$ and so we may write
$L^4=\lambda\,\ls^4$ where $\lambda=\gym^2\nc$ is the 't Hooft
coupling. As usual, we work in the supergravity approximation,
ignoring the effects of string loops or higher derivative terms
suppressed by powers of $\ls$ (except in section \ref{edualmain} and
appendix \ref{appcorr}). Hence, we are working in the limit where
both $\nc,\,\lambda\rightarrow\infty$. The background \reef{D3geom}
contains an event horizon at $r=r_0$. The temperature of the SYM
theory is then equivalent to the Hawking temperature:
\beq T = \frac{r_0}{\pi L^2}\, . \labell{Temper}\eeq

\subsection{D5-branes}\label{d5-geom}

Introducing D5-branes as in \reef{array} was the original
application of probe branes for the holographic construction of a
defect CFT -- \eg see \cite{lisa,hirosi}. The worldvolume action
which will determine the embedding of the probe D5-branes has the
usual Dirac-Born-Infeld (DBI) and Wess-Zumino (WZ) terms:
\beq I_{5}= -\nf\,T_5\int d^6\sigma\sqrt{-det\left(P[G]+2\pi\ls^2
F\right)}+\nf\,T_5\int C^{(4)}\wedge 2\pi\ls^2 F\,
.\labell{act5}\eeq
Implicitly we have assumed that the $\nf$ D5-branes are all
coincident. Hence, in principle, their worldvolume supports a
$U(\nf)$ gauge theory, however, implicitly above and in the
following, we only consider the gauge field in the diagonal $U(1)$
of this $U(\nf)$. We choose coordinates on the five-sphere in
\reef{D3geom} such that
\beq d\Omega_5 ^2 = d\psi ^2 + \cos^2 \psi
\left(d\theta^2+\sin^2\theta\,d\phi\right) +\sin^2 \psi \,
d{\Omega}_2^2 \, .
\labell{spheroid} \eeq
The D5-branes wrap the two-sphere parameterized by $\{\theta,\phi\}$
above, fill three of the gauge theory directions $\{t,x,y\}$ and
extend in the radial direction $r$. We also introduce a flux of the
worldvolume gauge field on the two-sphere:
\beq
F_{\theta\phi}=\frac{q}{2\,\nf}\,\sin\theta\,.\labell{magflux}\eeq
One may verify that this flux corresponds to dissolving $q$
D3-branes into the worldvolume of the $\nf$ D5-branes along the
$\{t,x,y,r\}$ directions, since the branes with flux sources
$C^{(4)}$ through the WZ term in \reef{act5}.

Now in general, the D5-brane embedding would be specified by giving
its profile in both the angular direction $\psi(r)$ and the D3-brane
direction $z(r)$. These embeddings all have translational symmetry
in the $\{t,x,y\}$-space, as well as invariance under $SO(3)$
rotations on the internal two-sphere. In the following, we consider
only the embeddings with $\psi=0$, \ie where the D5-brane wraps a
maximal two-sphere in the internal space. One can easily verify this
choice corresponds to a solution of the worldvolume embedding
equations. This choice also corresponds to setting the mass of the
fundamental fields to zero, \ie $\mq=0$, and so as we will describe
below, this choice also ensures that the dual field theory with the
defect remains conformal.

Hence in our analysis, we must determine the profile $z(r)$. The
induced metric on the D5-branes is now described by
\beq ds^2=\frac{r^2}{L^2} \left( -h(r)dt^2 +dx^2+dy^2\right) +
\left(\frac{L^2}{r^2h(r)} + \frac{r^2}{L^2}\partial_rz^2\right) dr^2
+L^2 \left(d\theta^2+\sin^2\theta\,d\phi\right)\,. \labell{induce}
\eeq
We can integrate over the two-sphere directions to produce a factor
of
\beq \oint_{S^2} d^2\sigma \sqrt{det_{S^2}\left(P[G]+2\pi\ls^2
F\right)} = 4\pi\,\left(\nf^2L^4+\pi^2\ls^4q^2\right)^{1/2} =
4\pi\nf L^2\,\sqrt{1+f^2} \labell{intern} \eeq
where
\beq f\equiv
\frac{\pi\ls^2}{L^2}\,\frac{q}{\nf}
=\frac{\pi}{\sqrt{\lambda}}\frac{q}{\nf}
 \labell{needy}
 \eeq
in the DBI part of the action \reef{act5}. The full D5-brane action
then becomes
\beq I_5 = -4\pi\nf\, T_5\sqrt{1+f^2}\int d^3x\,dr\,r^2
\left(1+\frac{r^4}{L^4}h(r)\partial_rz^2\right)^{1\over2} - 4\pi
\nf\, T_5 \frac{f}{L^2}\int d^3x\,dr\,r^4\,\partial_rz\,
.\labell{act6} \eeq

To simplify the analysis, we introduce the following coordinates:
\beq u=\frac{r_0}{r}\, , \qquad\qquad\chi=\frac{r_0}{L^2}z\,.
\labell{defs}\eeq
With this new notation, $h(u)=1-u^4$ and so the horizon is now at
$u=1$ while the asymptotic region is reached when $u\rightarrow0$.
The worldvolume action can now be written as:
\beq I_5 = -4\pi r_0^3\nf\,T_5\int
d^3x\,\frac{du}{u^4}\,\left[\sqrt{1+f^2}
\left(1+h(u)\chi'^2\right)^{1\over 2} +f\,\chi'\right]\,
,\labell{act5sub} \eeq
where $\chi'\equiv\partial_u\chi$. This expression is
independent of $\chi$, such that the variation with respect to $\chi'$
yields a constant of motion:
\beq \frac{1}{u^4}\,\left[\sqrt{1+f^2}\frac{h(u)\chi'}{
\left(1+h(u)\chi'^2\right)^{1/2}}+ f\right]=C\,. \labell{const1}\eeq
To avoid singular behaviour at the horizon, we need to fix the
integration constant to be $C=f$. In this case, \reef{const1} yields
\beq \chi'=-\frac{f}{\sqrt{1+f^2u^4}}\labell{deriv}\,.\eeq
Given this expression, the profile $\chi(u)$ can be expressed in
terms of an incomplete elliptic integral. However, in the following,
it will sufficient to have a closed form expression for $\chi'$. We
illustrate some typical profiles in figure \ref{chiprofile}.
\FIGURE{\includegraphics[width=0.5\textwidth]{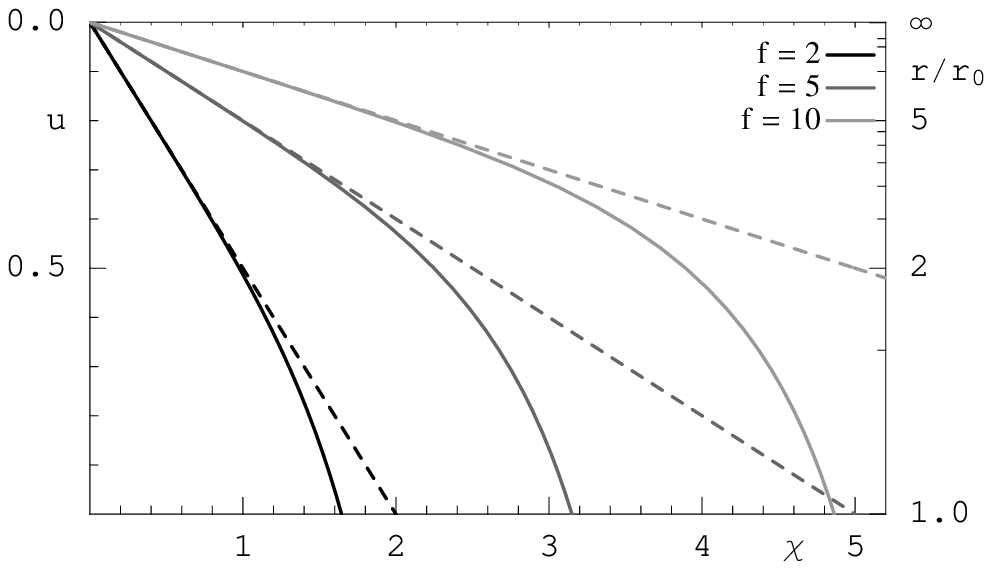}
\caption{The brane profile $\chi$ for various values of $f$ at
finite (solid) and zero (dashed) temperature. (Note the $T=0$
profiles correspond to $\chi=-f\,u$.)}\label{chiprofile}}

In terms of our original coordinates, we have
\beq \frac{\partial z}{\partial
r}=\frac{L^2\,f}{\sqrt{r^4+f^2\,r_0^4}}\,. \labell{zip}\eeq
Here we may consider the supersymmetric limit with $r_0=0$, in which
case \reef{zip} simplifies to $z=-L^2f/r$. In this case, one can
confirm that the induced metric \reef{induce} corresponds to
AdS$_4\times S^2$ with an AdS radius of curvature of $L_{{\rm
AdS}_4}=L\,\sqrt{1+f^2}$ \cite{lisa}. Hence the system inherits
SO(2,3) symmetry from the AdS$_4$ geometry, which reflects the fact
that the dual field theory remains conformal in the presence of the
defect. This conformal invariance can also be shown directly by an
analysis of the field theory \cite{joh}. Subsequently, the
construction of the fully back-reacted geometries corresponding to
the D5-branes embedded in $AdS_5 \times S^5$ demonstrate that the
preservation of the SO(2,3) symmetry is a fully nonperturbative
result \cite{jaume}.

One may note that with the supersymmetric profile, $z=-L^2f/r$,
there are an additional $q=\sqrt{\lambda}\,\nf\,f/\pi$ D3-branes
stretching from $z=0$ to $-\infty$, assuming $f>0$. Hence if one
were to include back-reaction, the asymptotic five-form flux would
be shifted from $\nc \to \nc + q$ units on this side of the space.
The same will apply at a finite temperature. Even though the brane
falls through the horizon at a finite distance in this case,
continuity at the horizon dictates that the background will carry
$\nc + q$ units of flux out to $z=-\infty$. In either case, the
natural interpretation is that the dual CFT has a $U(\nc+q)$ gauge
group in the region $z<0$, while the gauge group remains $U(\nc)$
for $z>0$.

It is interesting to pursue the interpretation of the above brane
configuration in the dual CFT further. A detailed AdS/CFT dictionary
has been developed for this defect system \cite{hirosi,joh}. In
particular, one finds that the defect lagrangian contains potential
source terms for the adjoint scalars in the SYM theory
\cite{hirosi,source}. The D5-brane carrying flux $f$ corresponds to
producing a noncommutative configuration of adjoint scalars in a
$U(q)$ subgroup of the $U(\nc+q)$ in the $z<0$ region
\cite{ramallo}. In fact, in this supersymmetric configuration, the
profile of the D5-branes can be precisely matched to the scalar
profile using noncommutative geometry \cite{neil2}:
$r^2=(2\pi\ls^2)^2\,{1\over \nf}\Tr(\Phi^2)$ where ${1\over
\nf}\Tr(\Phi^2)={q^2\over 4\nf^2}\,{1\over z^2}$.

As the D5-brane wraps a maximal two-sphere inside the $S^5$, one
might worry about the stability of this configuration. Indeed the
worldvolume field, corresponding to fluctuations in the angle
$\psi$, is found to be a tachyon with \cite{lisa}
\beq m^2_\psi=-\frac{2}{L^2(1+f^2)}=-\frac{2}{L_{{\rm
AdS}_4}^2}>-\frac{9}{4L_{{\rm AdS}_4}^2}\,. \labell{tachyon}\eeq
However, the last inequality indicates that the $\psi$-mode
satisfies the Breitenlohner-Freedman bound \cite{BF} in the
asymptotically $AdS_4$ geometry induced on the D5-brane worldvolume.
Hence this field does not in fact produce an instability.

Another concern may arise in considering the intersection of the
D5-branes with the event horizon in \reef{D3geom}. There we note
that
\beq\left.\chi'\right|_{u=1}=-\frac{f}{\sqrt{1+f^2}}\labell{note1}\eeq
or in terms of original coordinates
\beq\left.\frac{\partial z}{\partial
r}\right|_{r=r_0}=\frac{L^2\,f}{r_0^2\sqrt{1+f^2}}\,.\labell{note2}\eeq
Since the D5-brane enters the event horizon at an angle, one might
worry that the induced geometry is singular \cite{us}. However, one
can verify that this intuition is incorrect and that in fact, the
D5-brane geometry remains smooth as it crosses the horizon. Hence
the induced metric \reef{induce} describes a smooth `black hole'
geometry on the D5-brane worldvolume. A related question is: what is
the surface gravity or the temperature of the induced horizon? It is
a simple exercise to show that the relevant temperature matches that
of the bulk geometry, \ie that given in \reef{Temper}. Of course,
this reflects the fact that the defect and bulk fields will be in
thermal equilibrium, as expected.

We address one other potential concern related to the internal flux
\reef{magflux}. Throughout the paper, we will be considering finite
values of $f$, typically of $O(1)$. Hence according to \reef{needy},
we are introducing $q\sim O(\sqrt{\lambda})$ D3-branes and so one
might worry about whether it is reasonable to consider the probe
brane limit, \ie to ignore the gravitational back-reaction of the
branes. Of course, this is not a problem since the overall tension
of the D5-branes is not significantly modified by the flux, as can
be seen from \reef{act6}. The essential point is that the D3-branes
are distributed on the D5-branes over the internal two-sphere which
has an area of order $L^2\sim \sqrt{\lambda}$ and so the density of
D3-branes remains small, \ie the density is $O(f)$.

\subsection{D7 probes}\label{d7-geom}

The case of D7 probe branes is similar to the previous section with
D5-branes. The main difference lies in the internal part of the
geometry. In particular, the D7-branes wrap a(n equatorial)
four-sphere in the internal $S^5$. As before, we consider D3-branes
dissolved into the probe branes. In the present case, the D7-branes
source the three-brane charge through in the appropriate term in the
WZ action: $\frac{1}{2}(2\pi \ls^2)^2 T_7\int C^{(4)} \wedge F
\wedge F$. Hence, considering a stack of $\nf$ coincident D7-branes
with a $U(\nf)$ gauge symmetry, we introduce a nonvanishing second
Chern class on the internal four-sphere: $\cQ =
\frac{1}{8\pi^2}\oint_{S^4} Tr F\wedge F$.

The D7-branes are fixed to wrap a maximal four-sphere while the
embedding in the $AdS_5$ is described by $z = z(r)$. The induced
metric on the D7-branes becomes
 \beq ds^2=\frac{r^2}{L^2} \left( -h(r)dt^2 +dx^2+dy^2\right) +
\left(\frac{L^2}{r^2h(r)} + \frac{r^2}{L^2}z'^2\right) dr^2 +L^2
d\Omega_4^2\,. \labell{induceD7}
 \eeq
Since the present configuration contains a nontrivial nonabelian
gauge field, the worldvolume action requires a nonabelian extension
of the DBI action \cite{nonabDBI}
 \beq
I_{7}= -T_7\int d^8\sigma\,
\mathrm{STr}\sqrt{-det\left(P[G]+2\pi\ls^2 F\right)} + \frac{1}{2}
T_7 (2\pi\ls^2)^2 \int \mathrm{Tr} \left(C^{(4)}\wedge  F \wedge F
\right)\,.
 \labell{act7}
 \eeq
This action uses the proposal of a maximally symmetric gauge trace,
denoted by `STr' \cite{strace}. To be precise, the trace includes a
symmetric average over all orderings of $F_{ab}$ -- and implicitly
any appearances of the nonabelian scalars as well \cite{nreview} but
the latter will not be relevant in the present analysis. This
prescription correctly agrees with the string action to fourth order
in the field strength \cite{strace} but is known to miss certain
commutator terms which begin to appear at sixth order \cite{tilt}.
However, the contribution of such terms is typically suppressed by
factors of $1/\nf$ and so they can be safely neglected for
sufficiently large $\nf$ \cite{neil2,ConstableRobEtc}.

As before, we integrate over the internal space in the DBI action.
Here the internal $S^4$ carries an nonabelian gauge field giving the
instanton number $\cQ$. This configuration was extensively studied
in \cite{ConstableRobEtc} and hence using their results, we find
 \beqa
\oint_{S^4} d^4\Omega
&\mathrm{STr}&\sqrt{det_{S^4}\left(P[G]+2\pi\ls^2 F\right)} \nonumber \\
 & = &
\oint_{S^4} d^4\Omega\,\sqrt{g_{S^4}}\,
\sqrt{L^8+\frac{1}{2}L^4(2\pi \ls^2)^2 F_{ab} F^{ab}
+\frac{1}{64}(2\pi \ls^2)^4 (\e_{abcd}F^{ab} F^{cd})^2 }
 \nonumber\\
&=&\frac{8 \pi^2}{3}\left(\nf\, L^4 + 6 \pi^2 \ls^4 |\cQ| \right) \,
\labell{S4fact}.
 \eeqa
In the latter, we use (anti-)self-duality for the instanton
configuration: $F_{ab} = (-) {1\over2}\e_{abcd}F^{cd}$ for $\cQ
> 0$ ($\cQ < 0$). Implicitly, we are also assuming that the instanton
number is uniform on the four-sphere, which limits $\cQ \le
\nf(\nf^2-1)/6$ \cite{ConstableRobEtc}. Substituting \reef{S4fact}
and the embedding \reef{induceD7} into \reef{act7}, the action for
the background configuration becomes
\beq I_{7}= - \frac{8 \pi^2 \nf }{3} L^4 (1+ |Q|) T_7 \int d^3\sigma
dr \frac{r^2}{L^2} \sqrt{1 + \frac{r^4}{L^4} h(r) } \, - \,
\frac{8\pi^2 \nf}{3} L^4 Q\, T_7 \int d^3 \sigma dr \frac{r^4}{L^4}
z' \, , \labell{act7sub}\eeq
where we defined for convenience $Q = 6 \pi^2\frac{\ls^4}{L^4}
\frac{\cQ}{\nf}=\frac{6\pi^2}{\lambda}\frac{\cQ}{\nf}$. Now, the
computations analogous to those in section \ref{d5-geom} yield an
identical embedding $\chi (u)$ as in (\ref{deriv}) but the constant
$f$ is replaced by
 \beq f_7\equiv \frac{Q}{\sqrt{1+ 2|Q|}} \,.
 \labell{newf}
 \eeq
The microscopic interpretation of the D7-brane configuration in the
dual CFT is not as clear in the present case. However, as before,
the gauge group in the region $z<0$ will be enhanced to $U(\nc+\cQ)$
assuming $\cQ>0$. There should be source terms on the defect which
excite a noncommutative configuration of the adjoint scalars in the
transverse space. The latter can be interpreted in terms of
noncommutative geometry as giving the profile of the D7-branes, at
least to leading order in $1/\nf$ \cite{ConstableRobEtc}.

An important difference between the present case and that in the
previous section with D5-branes is that in the mass of the tachyonic
mode $\psi$ corresponding to the $S^4$ part of the D7-branes
``slipping off'' the maximal $S^4$ in the internal space. A simple
calculation reveals that
\begin{equation}
m^2_\psi \, = \, -\frac{4}{L^2 (|Q| + 1)} \, = \,
-\frac{4}{L^2_{{\rm AdS}_4}}\, \frac{|Q| + 1}{2|Q| + 1} \, .
\end{equation}
Recall that the BF bound requires $m^2 > -\frac{9}{4 L^2_{AdS_4}}$
\cite{BF} and hence is only satisfied for $|Q| > \smfrac{7}{2}$.
Hence one can trust the results in the following sections for the
D7-branes only for $f_7^2 > {49}/{32}$ and we might think of the
internal flux on the $S^4$ as creating some pressure that stabilizes
the size of the $S^4$. However, we should caution the reader that
what we have shown is that the most obvious instability is removed
for sufficiently large $Q$. While suggestive, this does not prove
the D7-brane configuration is absolutely stable.

Beyond this crucial difference, the analysis of these two systems
(\ie defects constructed with D5- or D7-branes) is completely the
same. Hence in the following, we focus on the first case of
D5-branes and only comment on differences in coefficients that may
arise for D7-branes where appropriate.


\section{Correlators}
\label{form}

In this section, we obtain examine various correlators of the
currents dual to the worldvolume gauge field $A_\mu$. First we
review the basic form of the correlators below, following
\cite{pavel}. Then we numerically compute the spectral functions in
\ref{goldfinger} and then examine the dependence of the correlators
on the temperature and the flux $f$ in \ref{efftf}.

In the following, we use holographic techniques to calculate the
retarded Green's function for a conserved current $J_\mu(x)$ on the
defect. The defect degrees of freedom form a (2+1)-dimensional CFT
which restricts the form of the correlators:
\beq C_{\mu\nu}(x-y)=-i\,\theta(x^0-y^0)\,\langle\,
[J_\mu(x),J_\nu(y)]\,\rangle\ , \labell{correl} \eeq
where translation invariance is assumed. The correlator can be
Fourier transformed to $C_{\mu\nu}(p)$ with
$p^\mu=(\omega,\vec{k})$.\footnote{We work with the mostly positive
signature so that $\eta_{\mu\nu}={\rm diag}(-1,+1,+1)$.} Now current
conservation and rotational invariance (full Lorentz invariance is
lost with $T>0$) restrict the form of the Fourier transform of this
correlator to be \cite{pavel}
\beq C_{\mu\nu}(p)=P^T_{\mu\nu}\,\Pi^T(p)+P^L_{\mu\nu}\,\Pi^L(p)\,.
\labell{form1}\eeq
where the transverse and longitudinal projectors can be written as
\beqa P^T_{ij}&=&\delta_{ij}-\frac{k_i\,k_j}{k^2}\,,\qquad
P^T_{0\mu}=0\,,\nonumber\\
P^L_{\mu\nu}&=&\eta_{\mu\nu}-\frac{p_\mu\,p_\nu}{p^2}-P^T_{\mu\nu}
\,. \labell{project}\eeqa
If we take into account that the conformal dimension of the current
$J_\mu(x)$ is 2, the components $\Pi^{T,L}$ in
\reef{form1} take the form:
\beq \Pi^{T,L}(p) = \sqrt{p^2}\,K^{T,L}(\omega/T,\vec{k}/T)\,.
\labell{form2}\eeq
In the limit of $T=0$, we have $\Pi^T(p)=\Pi^L(p)=\Pi(p)$ and
recover the Lorentz invariant correlator
\beq C_{\mu\nu}(p)=\left(\eta_{\mu\nu}-\frac{p_\mu\,p_\nu}{p^2}
\right)\,\Pi(p)\,.\eeq
In order to produce physical observables, and to interpret our
results from a condensed matter point of view, we will calculate the
conductivity from the Kubo formula
\begin{equation}
\sigma_{ij} \, = \, \frac{i}{\omega} C_{ij}\ .\labell{conduct0}
\end{equation}


\subsection{Spectral functions}
\label{goldfinger}

In this section, we compute spectral functions for excitations of
fundamental fields on the defect by studying fluctuations of the
worldvolume fields on the D5-brane probes.  In particular, we focus
on correlators of the the worldvolume vector $A_\mu$, which is dual
to the conserved current $J^\mu$ corresponding to the diagonal
$U(1)$ of the global flavour symmetry on the defect. The worldvolume
gauge field gives rise to several types of modes, one of which is a
vector with respect to the Lorentz group in the (2+1)-dimensional
defect. These modes are characterized as having only $A_{0,1,2}$
nonzero while the components on the internal two-sphere are
vanishing \cite{holomeson}. Further the radial component $A_r$ can
consistently be set to zero because we only study modes which are
constant on the internal space \cite{holomeson}.

While the full action for the gauge fields on the D5-branes receives
contributions from both the Dirac-Born-Infeld (DBI) action plus a
Wess-Zumino term, since our gauge field fluctuations have vanishing
radial and $S^2$ components, only the DBI portion of the action is
relevant in determining their dynamics. Since we only study
linearized fluctuations about the background, the gauge field action
is only needed to quadratic order, which is simply
\beq I_{gauge} \, = \, -4\pi L^2\,\sqrt{1+f^2} \nf\,T_5  \int d^3
\sigma\,dr \sqrt{- g}\, \frac{(2 \pi \ls^2)^2}{4}\, F^2 \, = \,
-\frac{1}{4g_4^2}  \int d^3 \sigma\,dr \sqrt{- g}\, F^2 \,
.\labell{gaugeAction} \eeq
Here we have integrated over the internal $S^2$ as in \reef{intern}
and use $g_{\mu\nu}$ to denote the induced metric \reef{induce} in
the AdS$_5$ directions. Above, we also defined the effective gauge
coupling for the four-dimensional Maxwell field:
\beq \frac{1}{g_4^2}\equiv 16\pi^3\ls^4 L^2\,\sqrt{1+f^2}\nf\, T_5
=\sqrt{1+f^2} \frac{2}{\pi}\,\frac{\nf\,\nc}{\sqrt{\lambda}} \,.
\labell{effect} \eeq
For the D7 case, this becomes
 \beq \frac{1}{g_4^2}\equiv
\frac{32\pi^4 \nf}{3} \ls^4 L^4(1+|Q|) T_7 =(1+|Q|)
\frac{\nf\,\nc}{3 \pi^2} \,. \labell{effectD7}
 \eeq
Note that the gauge field action \reef{gaugeAction} corresponds to
the standard Maxwell action in a four-dimensional curved spacetime.
Hence, these gauge fluctuations will exhibit electromagnetic
duality, which was shown to play and interesting role in the physics
of the conformal field theory in \cite{pavel}. We will explore this
point further in section \ref{edualmain}. Of course, Maxwell's
equations follow as
\beq
\partial_a \left({\sqrt{-g}}\, F^{ab}\right) =0. \labell{maxwell}
\eeq

Using these equations of motion, the Maxwell action
\reef{gaugeAction} becomes a total derivative and following the
standard prescription, we obtain the desired correlator from the
resulting boundary term. To proceed, let us first give the explicit
metric on the brane,
\beq ds^2 =
\frac{L^2}{u^2}\left[\frac{r_0^2}{L^4}\big(-h(u) dt^2+dx^2+dy^2 \big)+\frac{du^2}{h(u)}
\left(1+h(u)\chi'^2\right) \right]\, , \labell{newind}\eeq
in terms of the dimensionless radial coordinate $u$, as given in
\reef{defs}. Then the action \reef{gaugeAction} becomes
\beqa I_{gauge}&=& -\frac{1}{2g_4^2} \int d^3 \sigma \,du\,
\partial_a \left[\sqrt{-{g}}\, A_b \,F^{ab} \right] \, = \,
-\frac{1}{2g_4^2} \int d^3 \sigma \,  \left[ \sqrt{-{g}} g^{aa}
g^{uu} A_a \partial_u  A_a \right]^{u\rightarrow1}_{u\rightarrow0}
\nonumber \\
&=&-\frac{1}{2g_4^2} \frac{r_0}{L^2} \int d^3 \sigma \,\left[
\frac{1}{(1+h(u)\chi'^2)^{1/2}}\left( h(u) A_x
\partial_u A_x + h(u) A_y \partial_u
A_y-A_t\partial_u A_t
\right)\right]^{u\rightarrow1}_{u\rightarrow0} \ .
\labell{gaugeAct3}\eeqa
The usual AdS/CFT prescription tells us that we will only need
the contribution at the asymptotic boundary $u \to 0$
\cite{Son:2002sd}. Following \cite{Kovtun:2005ev}, we take the
Fourier transform of the gauge field,
\beq A_\mu (\sigma ) = \int \frac{d^3 k}{(2\pi)^3} e^{ i
k\cdot\sigma } A_\mu(k,u) \, , \labell{fourier}\eeq
to write the boundary action as
\beqa I_{gauge}&=&-\frac{1}{2 g_4^2 1} \frac{1}{(2 \pi)^3} \int d^3
k \,\left[\sqrt{-{g}} g^{\mu\mu}g^{uu} A_\mu(u,-k)
\partial_u A_\mu (u,k) \right]^{u\rightarrow1}_{u\rightarrow0} \nonumber\\
&=&-\frac{r_0}{2 g_4^2 L^2} \frac{1}{(2 \pi)^3} \int d^3 k
\,\left[(1+h(u)\chi'^2)^{-1/2}\left( h(u) A_x(u,-k)
\partial_u A_x (u,k) \right.\right. \nonumber\\
&&\qquad\qquad \left.\left. + h(u) A_y(u,-k) \partial_u
A_y(u,k)-A_t(u,-k)\partial_u
A_t(u,k)\right)\right]^{u\rightarrow1}_{u\rightarrow0}
\labell{gaugeAct4}\eeqa
with a single sum of $\mu$ being implicit in the first line.

Looking at the asymptotic behaviour of the fields, we write
\beq \labell{modefunc} A_{\mu}(k,u) = A_{\mu 0}(k)
\frac{A_{\mu}(k,u)}{A_{\mu}(k,u_0)} \, , \eeq
where $u_0$ is a UV regulator and it is understood that eventually
the limit $u_0 \to 0$ will be taken. We can then derive the flux
factor for, say, $A_y$ by taking variations with respect to $A_{y
0}$ \cite{Son:2002sd}:
\beq \mathcal{F}_{yy}=
-\frac{\varepsilon_0}{2}\left[\frac{h(u)}{(1+h(u)\chi'^2)^{1/2}}
\frac{A_y(u,-k)
\partial_u A_y (u,k)}{A_y(u_0,-k) A_y (u_0,k)} \right] \, ,
\labell{flux}\eeq
where $\varepsilon_0 = \frac{r_0}{g_4^2 L^2} = \frac{\pi\,T}{
g_4^2}$
--- we will show later how $\varepsilon_0$ relates to the charge permittivity.
The flux \reef{flux} should be conserved, \ie be independent of the
radius $u$. The usual AdS/CFT prescription tells us to evaluate it
at the asymptotic boundary, while applying infalling boundary
conditions at the horizon ($u = 1$), to find the retarded Green's
function \reef{correl} for the current $J_{\mu}$ in the defect CFT
\cite{Son:2002sd}:
\beqa C_{yy} = -2 \mathcal{F}_{yy} &=&
\varepsilon_0\left[\frac{h(u)}{(1+h(u)\chi'^2)^{1/2}}
\frac{A_y(u,-k)
\partial_u A_y (u,k)}{A_y(u_0,-k) A_y (u_0,k)} \right]_{u,u_0\to0}
\nonumber\\
&=& \frac{\varepsilon_0}{\sqrt{1+f^2}}\left[\frac{
\partial_u A_y (u,k)}{A_y (u,k)} \right]_{u\to0}  \,.
\labell{green}\eeqa
The other correlators $C_{\mu \nu}$ follow in general by rewriting
\reef{modefunc} as $A_\mu (k,u) = A_{\nu 0}(k) M^\nu_{\ \mu} \!
(u,k)$ \cite{pavel} and making the variation $\frac{\delta^2}{\delta
A_{\mu 0}\, \delta A_{\nu 0}}$. In our case the $t,t$ and $x,x$
correlators are given by \reef{green} with the indices appropriately
replaced:
 \beq
 C_{xx}\, =\, \frac{\varepsilon_0}{\sqrt{1+f^2}}\left[\frac{
\partial_u A_x (u,k)}{A_x (u,k)} \right]_{u\to0}  \ \mathrm{and} \ \
C_{tt}\, =\, - \frac{\varepsilon_0}{\sqrt{1+f^2}}\left[\frac{
\partial_u A_t (u,k)}{A_t (u,k)} \right]_{u\to0} \, .
 \labell{green1}
 \eeq

In order to evaluate the spectral function, we must solve the
equations of motion \reef{maxwell}. It is convenient to introduce
dimensionless coordinates by rescaling the defect coordinates as
\beq \tlt=\frac{r_0}{L^2}t\,,\qquad \tx=\frac{r_0}{L^2}x\,,\qquad
\ty=\frac{r_0}{L^2}y\,. \labell{skale}\eeq
Without loss of generality, we also assume the fluctuations only
carry momentum in the $\tx$ direction, \ie $\tk^\mu=(\tom,\tlq,0)$
--- note that, \eg $\tom= L^2/r_0\ \omega=\omega/\pi\,T$. We note that
given the Fourier transform \reef{fourier}, the vector potentials
vary as $e^{i\tk_\mu \tx^\mu}$ in the gauge theory directions.

Now the explicit equations of motion simplify to
\beqa b=u:\qquad 0 &=&
\tom A'_\tlt +\tlq\,h\, A'_{\tx}  \, , \labell{gauge1}\\
b=\tlt:\qquad 0 &=&
A''_\tlt-\frac{H'}{2H}\,A'_\tlt-\frac{H}{h}\left(\tlq^2A_\tlt+\tom \tlq
A_\tx\right)
\, , \labell{gauge2}\\
b=\tx:\qquad 0 &=& A''_\tx+
\left(\frac{h'}{h}-\frac{H'}{2H}\right)\,A'_\tx+\frac{H}{h^2}\left(\tom^2A_\tx+\tom
\tlq A_\tlt\right) \, , \labell{gauge3}\\
b=\ty:\qquad 0 &=& A''_\ty+
\left(\frac{h'}{h}-\frac{H'}{2H}\right)\,A'_\ty+\frac{H}{h^2}(\tom^2-h\,\tlq^2)A_\ty
 \, , \labell{gauge4} \eeqa
where `prime' denotes $\partial_u$ and
\beq H(u)\equiv 1+h(u)\chi'^2\,.\labell{HHH}\eeq
Before proceeding further, we make the following convenient
definition for the conductivities
\begin{equation}
\tilde{\sigma}_{ij} \, \equiv   \, \frac{i}{\tom} C_{ij} \,=\, \pi T
\sigma_{ij}  \ .\labell{conduct1}
\end{equation}
Comparing to \reef{conduct0}, here we are simply dividing by the
dimensionless frequency $\tom$, rather than $\omega$.

\subsubsection{Transverse correlator}

Let us look carefully at the $\ty$ equation \reef{gauge4}. We see
firstly, that in the limit $u \rightarrow 0$, it reduces to
\beq 0 \, =\, A''_\ty + (1+f^2)(\tom^2 - \tlq^2)A_\ty \ . \eeq
The solution of interest is then $A_\ty = A_{\ty0}\, e^{- i
\sqrt{(1+f^2)(\tom ^2 - \tlq^2)}u}$,  where the sign in the
exponential is chosen so that the solution corresponds to an
infalling wave. Given this solution, if one now calculates the
correlator with \reef{green} and applies \reef{conduct0}, the
resulting conductivity is
 \beq
\tilde{\sigma}_{yy} = \varepsilon_0 \sqrt{1-\tlq^2 / \tom^2}\,.
\labell{looow}
 \eeq
The cut in the conductivity at $\tom = \tlq$ may be surprising and
we return to this point in section \ref{colll}. We will refer to
this simple result as the low temperature approximation, reasoning
as follows: The result applies for large dimensionless
``frequencies'', \ie $|(1+f^2)(\tom ^2 - \tlq^2)|^{1/2} \gg 1$ to be
precise. However, recalling that \eg $\tom=\omega/\pi T$, if we fix
the dimensionful quantities $\{\omega, q\}$, then eq.~\reef{looow}
should apply in the limit of very low temperatures.

To solve for the full spectral functions, we must proceed with
numerical calculations. First, we impose infalling boundary
conditions at the horizon --- recall that the time-dependence of the
potentials is $e^{-i\tom\tlt}$. If we expand about $1-u\to0^+$, we
find an appropriate description of the field to be
\beq A_\ty\simeq (1-u^4)^{i\tom/4}\left(1+\beta(1-u)+\cdots\right)
\labell{nearform}\eeq
where
\beq \beta= \frac{i}{4}\tom\frac{3+5f^2}{1+f^2}+
\frac{\tlq^2}{\tom^2+4}\left(1 - i\frac{\tom}{2}\right) \,.
\labell{nearform1}\eeq
In order to implement the infalling boundary condition and to ensure
numerical stability, we choose the Ansatz
 \beq A_\ty  = (1-u^4)^{i\tom /4} e^{-\beta u} \mathcal{F}(u) \ ,
 \labell{nearform2}\eeq
and solve for $\mathcal{F}(u)$, which is nonsingular at the horizon,
with $\partial_u\mathcal{F}(u) = 0 $ at $u=1$. As the second
boundary condition, we fix the asymptotic normalization: $A_\ty |_{u
= 1} = 1$.
\FIGURE{
\includegraphics[width=1 \textwidth]{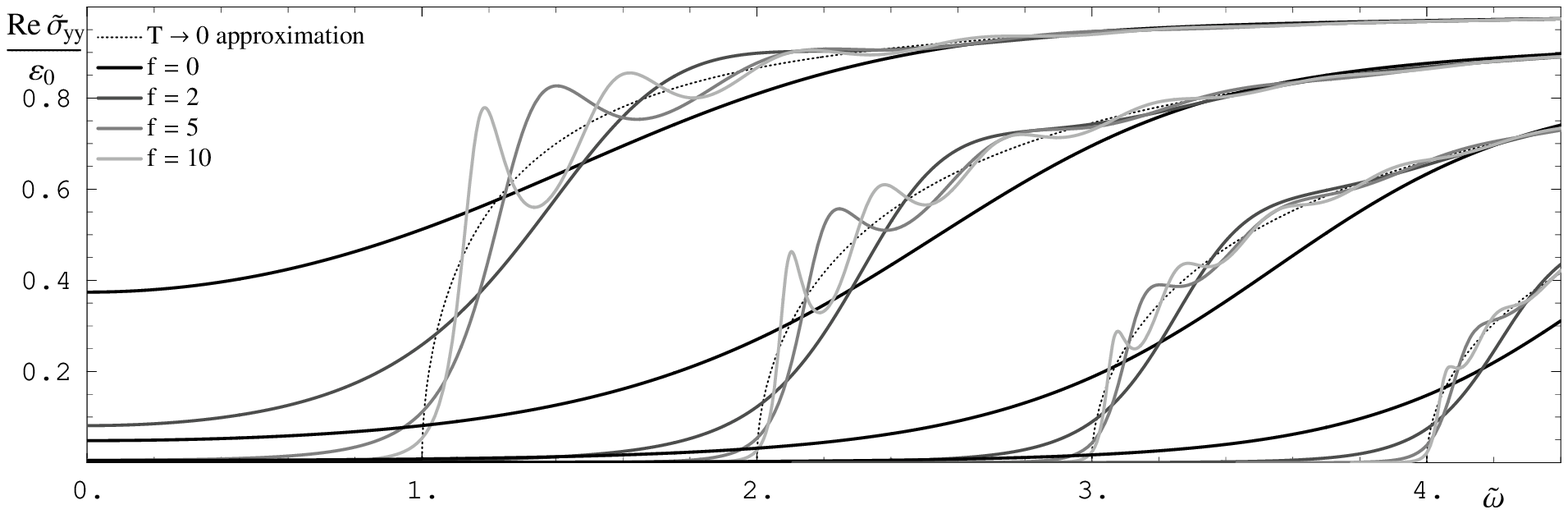}
\caption{The normalized transverse conductivity $\Re
\tilde{\sigma}_{yy}(\tom)/\varepsilon_0$ at $\tlq \in \{1,2,3,4\}$
for various values of the flux $f$. The low temperature
approximation \reef{looow} is shown as the dotted line for each
$\tlq$.} \label{condfig}}

Figure \ref{condfig} shows $\Re \tilde{\sigma}_{yy}( \tom
)/\varepsilon_0$ for various values of $\tlq$ and $f^2$. We see that
at $f=0$ the spectral functions are similar to those in
\cite{pavel}. However, as $f$ increases, they approach the low
temperature limit \reef{looow} more closely, and show some
oscillatory behaviour at $\tom > \tlq$. We will discuss aspects of
this behaviour in section \ref{efftf}.

\subsubsection{Longitudinal correlator}
Now, let us consider the $tt$ equation \reef{gauge2}. It is easy to
see that (\ref{gauge1}-\ref{gauge3}) are not independent, and we
cannot produce a second order equation involving $A_{\tilde{t}}$
only. However, we can produce one for $A'_{\tilde{t}}$:
\begin{equation}
0 \, = \, A'''_{\tilde{t}} \, - \, \Big(\frac{h}{H} \big(\frac{H}{h}
\big)' + \frac{H'}{2 H}\Big)A''_{\tilde{t}} \, + \,
\Big(\frac{h}{H} \frac{H'}{2 H} \big(\frac{H}{h} \big)' -
\big(\frac{H'}{2 H} \big)' \Big)A'_{\tilde{t}} \, + \,
\frac{H}{h^2}\big(\tom^2 - \tlq^2 h \big)A'_{\tilde{t}} \, , \labell{at_prime}
\end{equation}
which simplifies to
\begin{equation}
0=\left(\frac{A'_\tlt}{\sqrt{H}}\right)''+
\left(\frac{h'}{h}-\frac{H'}{2H}\right)\,
\left(\frac{A'_\tlt}{\sqrt{H}}\right)'
+\frac{H}{h^2}(\tom^2-h\,\tlq^2)\frac{A'_\tlt}{\sqrt{H}} \ ,
\end{equation}
and hence is the same as \reef{gauge4} for $A_\ty$ replaced by
$A'_\tlt/\sqrt{H}$. Let us set $A'_\tlt=c\,\sqrt{H}\,A_\ty$ with
some constant $c$ to be determined. Now, to find ${
\partial_u A_\tlt (u,k)}/{A_\tlt (u,k)}$
at $u=0$, we employ \reef{gauge2} and $A''_\tlt = c
\sqrt{H}\big(A'_\ty +\frac{H'}{2H} A_\ty\big)$ as in \cite{pavel}.
It follows then from $h'|_{u=0} = 0=H'|_{u=0}$ that
\begin{equation}
c= \frac{\sqrt{1+f^2}}{A'_\ty |_{u=0}} \left(\tlq^2A_{\tlt 0}+\tom
\tlq A_{\tx 0}\right) \, .
\end{equation}
Hence, we can read off $\left[\frac{\partial_u A_\tlt (u,k)}{A_\tlt
(u,k)} \right]_{u\to0}$ and $\left[\frac{
\partial_u A_\tlt (u,k)}{A_\tx (u,k)} \right]_{u\to0}$ from
\begin{equation}
A'_\tlt =   \frac{\sqrt{1+f^2} \sqrt{H} A_\ty}{A'_\ty |_{u=0}}
\left(\tlq^2A_{\tlt 0}+\tom \tlq A_{\tx 0}\right) \ .
\end{equation}
Finally as in \cite{pavel}, we find
\beq C_{tt} = - \varepsilon_0^2\, \tlq^2/C_{yy} \qquad \mathrm{and}
\qquad C_{xx} = - \varepsilon_0^2\, \tom^2/C_{yy} \, .
\labell{translongreln} \eeq

Applying \reef{conduct1}, these results yield interesting relations
for the corresponding conductivities. In particular,
\reef{translongreln} yields
\beq \tilde{\sigma}_{xx}=\varepsilon_0^2/\tilde{\sigma}_{yy}\ .
\labell{conduce3} \eeq
We can also consider a low temperature limit as above. However, this
is most easily derived by combining \reef{looow} and \reef{conduce3}
to find
 \beq
\tilde{\sigma}_{xx} = \frac{\varepsilon_0}{\sqrt{1-\tlq^2 /
\tom^2}}\,. \labell{low22}
 \eeq
Again, we will return to discuss the cut appearing in the
conductivity at $\tom = \tlq$ in section \ref{colll}; and the
conductivity can only be found in general from numerical
calculations. Some typical results for (the real part of)
$\tilde{\sigma}_{xx}$ are shown in figure \ref{plotlong}. We note
that again that as the flux $f$ increases, our results approach the
low temperature approximation \reef{low22}, together with some
``oscillatory'' behaviour similar to that found in the transverse
case. In contrast to the results in the previous section, the
conductivity here diverges as $\tom \to \tlq$, as can be anticipated
from \reef{low22}.
\FIGURE{
\includegraphics[width=1 \textwidth]{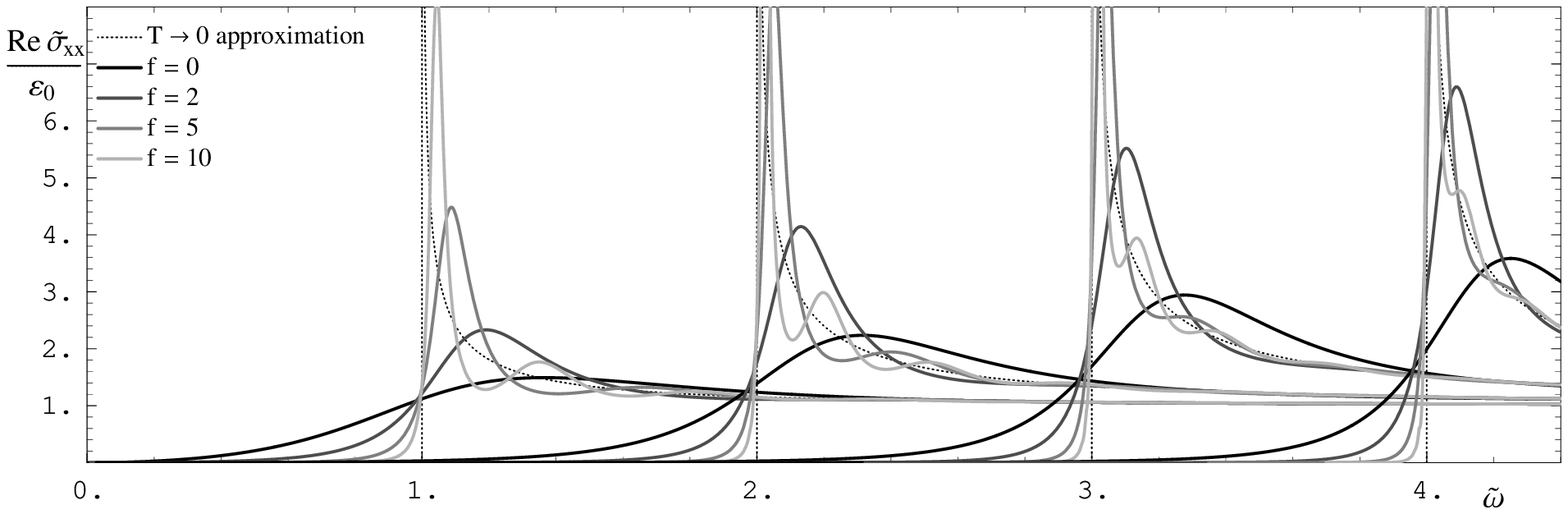}
\caption{The normalized longitudinal conductivity $\Re
\tilde{\sigma}_{xx}(\tom)/\varepsilon_0$ at $\tlq \in \{1,2,3,4\}$
for various values of the flux $f$. The low temperature
approximation \reef{low22} is shown as the dotted line for each
$\tlq$.} \label{plotlong}}

We can find the permittivity $\varepsilon$ from the hydrodynamic
limit $T \gg \omega,q$ \cite{hydropaper},
\beq -\, \Im C_{tt} \, = \, \frac{\varepsilon D\, \omega q^2
}{\omega^2 + (D q^2)^2}  \, = \, \frac{\varepsilon \,\pi DT  \, \tom
\tlq^2 }{\tom^2 + (\pi\,DT\, \tlq^2)^2} \, . \labell{diffpol} \eeq
In this regime, the spectral function is dominated by the diffusion
pole $\tom=-i \pi\,D T\,\tlq^2$, as dictated by Fick's law
\reef{fick}. The diffusion constant is $D = \frac{\sqrt{1+f^2}}{\pi
T} I(f)$,  as we calculate in appendix \ref{diffusion}, where we
also define the function $I(f)$. Comparing to our numerical results
for the spectral functions as shown in figure \ref{diffplot} for
various values of $f$ and $\tlq \ll 1$, we find $\varepsilon  =
\frac{\varepsilon_0}{ I(f)}$. We can verify the latter from the
definition of the permittivity \cite{Chaikin},
 \beq
 \varepsilon \, = \, \frac{1}{T} \lim_{\tom,\tlq \to 0} C_{tt} \, ,
 \eeq
which is in perfect agreement with the numerical result.
\FIGURE{
\includegraphics[width=0.49 \textwidth]{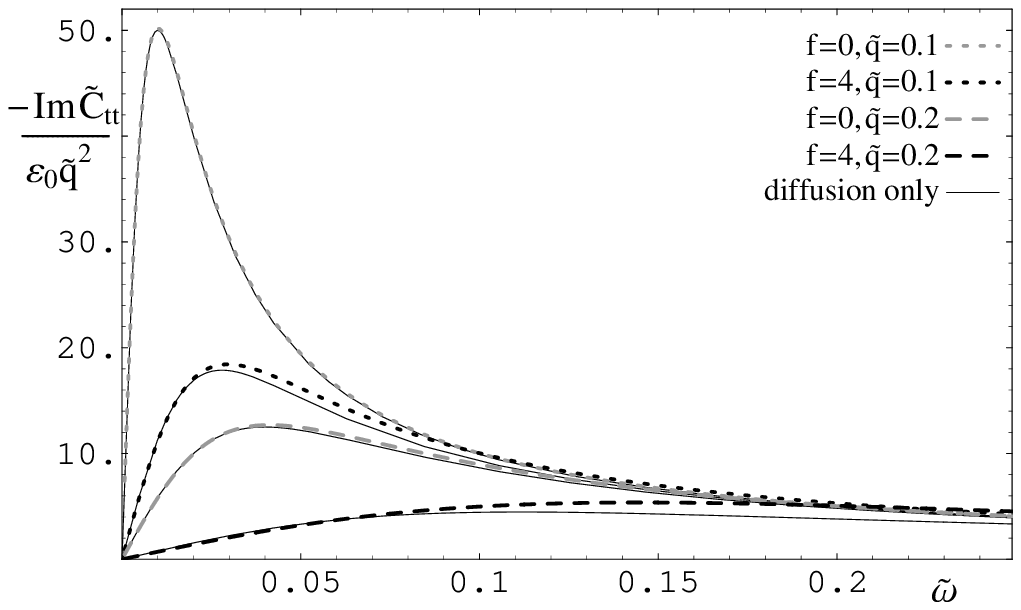}
\caption{$\frac{-\Im \tilde{C}_{tt}(\tom)}{\varepsilon_0 \tlq^2}$
for $f\in \{ 0,4\}$ and $\tlq \in \{0.1,0.2\}$. The solid lines
correspond the approximate result \reef{diffpol} with only the
diffusion pole appearing in the hydrodynamic
limit.}\label{diffplot}}

\subsection{Temperature and $f$ dependence}\label{efftf}

In the previous sections, we found an interesting dependence of the
conductivity on the temperature and the flux $f$. These properties
characterize the nature of the defect, as shown more in detail in
figure \ref{condapprox}. There we see that at low $T$ or large $f$,
there is a conduction threshold at $\omega=q$. We can interpret this
as the energy required to excite a collective excitation of the
conducting mode. In the regime $\omega<q$, the conductivity appears
exponentially suppressed as one might expect with a chemical
potential $q$. That is, this exponential suppression in the
low-temperature ``DC limit'' might be interpreted as the Boltzmann
tail of some thermal distribution function. Examining this behaviour
in more detail in the next section suggests the introduction of an
effective temperature, which seems to play an interesting role in
the subsequent analysis. Examining the conductivity at low $T$ or
large $f$ also reveals ``oscillations'' in the spectral curves. The
frequency of this oscillations has a non-trivial dependence on $f$
and seems to depend inversely on the temperature, as one might
expect from the general scaling properties. Their amplitude is
roughly independent of $f$, but depends on some positive power of
the temperature and decreases with increasing $\frac{\tom}{\tlq}$.
In the following, we will also extract some quantitative
approximation to this pattern.
\FIGURE{
\includegraphics[width=0.48\textwidth]{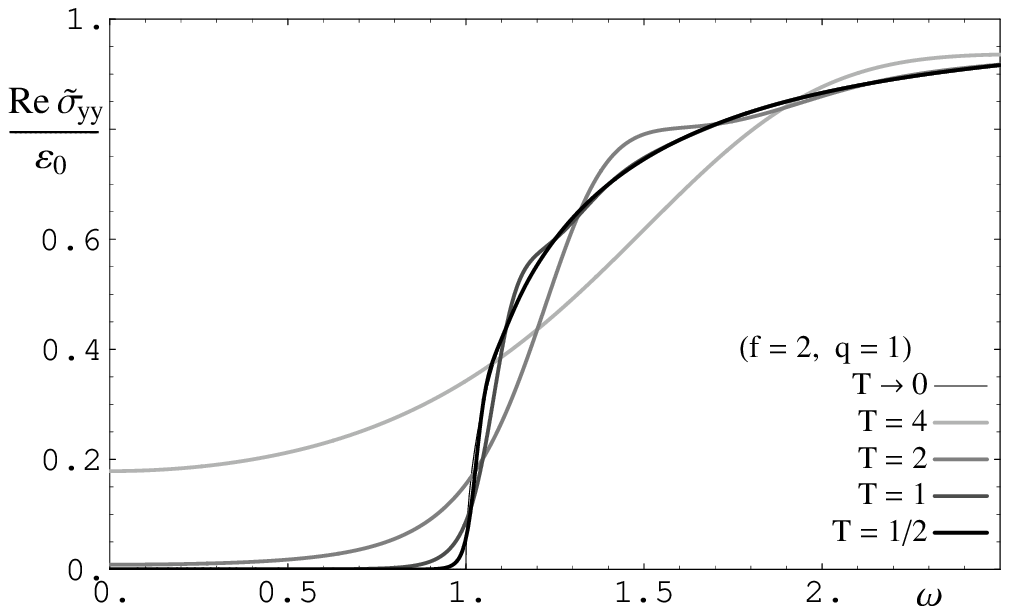}
\includegraphics[width=0.48\textwidth]{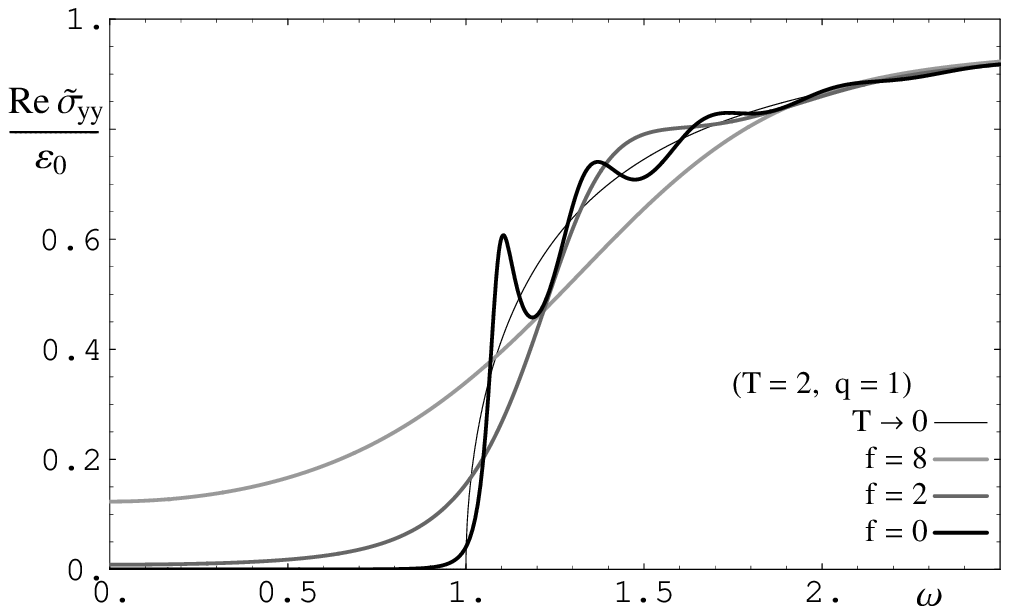}
\caption{On the left, we plot the normalized transverse conductivity
$\Re \tilde{\sigma}_{yy}(\omega)/\varepsilon_0$ for $q=1$ and $f=2$
at various temperatures, in terms of the frequency $\omega$. On the
right: $\Re \tilde{\sigma}_{yy}(\omega)/\varepsilon_0$ for $q=1$ and
$T=2$ for various values of $f$. In both plots, the $T\rightarrow 0$
limit \reef{looow} is shown with the narrow black curve.}
\label{condapprox}}

First, we study these two effects analytically as a perturbation
around the zero temperature limit. Next, we will show how they arise
from poles in the spectral functions that can be interpreted in the
field theory as the quasiparticle states of the resonances on the
defect and arise on the gravity side through the quasinormal modes
of the vector field. Finally, we will demonstrate the latter by
reconstructing the location of the poles in the complex frequency
plane from the data on the real axis and also by analytically
solving a toy model that is very similar to our present problem.
\subsubsection{Effective temperature}\label{efftf2}
First let us study the temperature dependence of the DC limit. This
can be easily done by finding an approximate solution in the $q \gg
T$ limit. For simplicity, we will also take $\omega \ll T$. We will
concentrate on the transverse correlators, but we will see that the
conductivity is obtained from a small perturbation around a large
background, such that by \reef{translongreln}, a similar behaviour
applies to the longitudinal correlator in the limit that we will
consider. Before proceeding, it is useful to recall here that
$h(u)\equiv1-u^4$ while $H(u)$ is given by \reef{HHH}, so that
asymptotically as $u\rightarrow0$, $H\rightarrow \sqrt{1+f^2}$ while
near the horizon where $u\rightarrow1$, $H\rightarrow 1$.

Let us first re-express \reef{gauge4} in terms of the Ansatz $A_\ty
= A_{\ty0}\, e^{\int^u \zeta}$, such that $C_{yy} =
\frac{\varepsilon_0}{\sqrt{1+f^2}} \lim_{u \rightarrow 0}\zeta$:
\begin{equation}\labell{nu_eq}
\zeta'  + \zeta^2 \, + \, \left(\frac{h'}{h} - \frac{H'}{2H} \right)
\zeta + \frac{H}{h}\left(\frac{\tom ^2}{h} - \tlq^2 \right) \, = \,
0 \ .
\end{equation}
We see that for large $\tlq$ (\ie $q\gg T$), this equation is
dominated by the second and last terms, such that an approximate
solution is $\zeta = \pm \zeta_0 $,  $ \zeta_0 \equiv  \tlq
\sqrt{\frac{H}{h}}$. Implicitly, here we have chosen the branch
corresponding to $A_\ty$ decaying near the horizon. Further, as we
will see below, this also corresponds to an infalling boundary
condition at the horizon. The terms that we ignored are then of the
order $\tlq h^{-3/2}$, such that the approximation is valid in the
region $1 - u \gg \tlq^{-2}$. The subleading terms in $\zeta$ are of
the order $\tlq^0 u^3 h^{-1}$.

Next, we study the linearized equation for a small perturbation
$\zeta \rightarrow \zeta_0 + \epsilon$:
\begin{equation}\labell{epseqn}
\epsilon' \, + \, \epsilon \left(\frac{h'}{h} - \frac{H'}{2H}- 2
\zeta_0 \right) \, + \, \frac{\tom^2 H}{h^2} - \zeta_0' +
\left(\frac{h'}{h} - \frac{H'}{2H} \right) \zeta_0 \, \equiv \,
\epsilon' \, - \, \epsilon \alpha(u) \, - \, \beta (u)
\end{equation}
with the general solution
\begin{equation}\labell{eps_approx}
\epsilon \, = \,  e^{\int_0^u \! d \bar{u} \, \alpha(\bar{u})}
\left(\epsilon_0  \, + \, \int_0^u \! d\acute{u} \, e^{-
\int_0^{\acute{ u}} \! d \bar u \, \alpha(\bar u)} \beta (\acute u)
\right) \ .
\end{equation}
In the limit that we considered for $\zeta_0$ this reduces simply to
$ \epsilon \, = \,  \epsilon_0 e^{2 \int_0^u\! d \bar u \,
\zeta_0} $ because of the exponential suppression in the last term
in \reef{eps_approx}. As one would have physically expected, this
perturbation grows as one approaches the horizon, and decays away
near infinity. The subleading terms from the part of $\alpha(u)$
that we ignored in the integral in the exponent is again of order
$\tlq^0 u^3 h^{-1}$.

To find $\epsilon_0$, fix the $\omega$ dependence and further
constrain the subleading terms, we proceed by considering an
approximate solution in the region $h = 1 - u^4 \ll 1$, which has
overlap with $h \gg \tlq^{-2}$. The equation we need to solve is now
\begin{equation}
4 \partial_h \zeta \, - \, \zeta^2 \, + \, \frac{4}{h} \zeta \, + \,
\frac{\tlq^2}{h} - \frac{\tom^2}{h^2} \, = \, 0 \, ,
\end{equation}
which has a general analytic but not very illuminating solution in
terms of Bessel functions, allowing for a combination of infalling
and outgoing waves at the horizon $u=1$. Choosing an infalling
boundary condition leaves us with
\begin{equation}\labell{NHGapp}
\zeta \, = \, - \frac{ \tlq^2 h\, {}_{0}\! F_1 \big(2 + i
\frac{\tom}{2} ; \tlq^2 \frac{h}{16}\big) \, - \, 16 i
\frac{\tom}{2} \big(2 + i \frac{\tom}{2} \big) {}_{0}\! F_1 (2 + i
\frac{\tom}{2} ; \tlq^2 \frac{h}{16})}{8  h \big(2 + i
\frac{\tom}{2} \big) {}_{0}\! F_1 \big(1 + i \frac{\tom}{2} ; \tlq^2
\frac{h}{16}\big)} \ ,
\end{equation}
where ${}_{0}\! F_1 (a ; x)$ is the confluent hypergeometric limit
function \cite{fluepi}. To match with the $h \ll 1$ regime of the
asymptotic solution, we begin by expanding to first order in $\tom$
and then do an expansion around $\tlq^2 h \gg 1$, which gives us
\begin{equation}
\zeta \, \sim  \, - \frac{\tlq}{\sqrt{h}} + \frac{1}{h} + \cdots \,
- \, i \tom \tlq \frac{16 \pi}{\sqrt{h}} e^{- \tlq \sqrt{h}} \, + \,
\cdots \ .
\end{equation}
Hence the full solution for $h \gg \tlq^{-2}$ is:
\begin{equation}
\zeta \, = \, - \tlq \sqrt{\frac{H}{h}} + \frac{1}{h} + A(u,\tlq, f)
\, - \, i \tom \tlq \frac{16 \pi}{\sqrt{h}} e^{- 2 \tlq \int_u^1
\sqrt{\frac{H}{h}}} \big(1 + B(u, \tlq, f) \big)\ , \labell{tempnu}
\end{equation}
where $A(u,\tlq, f)$ is some function that behaves away from the
horizon as $\le \mathcal{O}(u^3, \tlq^0)$ and $B(u, \tlq, f)$
behaves as $\le \mathcal{O}(u^0, \tlq^0)$. Near the horizon, \ie for
$h \ll \tlq^{-2}$, the solution behaves as $\zeta \sim -i
\frac{\tom}{h} - \frac{2 - i \tom}{8 + 2\tom^2} + \cdots$. As a
consistency check in the region $1 \gg h \gg \tlq^{-2}$, it is easy
to verify that the (small) imaginary and (dominating) real parts do
indeed satisfy \reef{eps_approx} when taking into account the
next-to-leading terms.

From \reef{tempnu}, we find that the leading term in the
conductivity is
 \beq
\sigma_{yy}\sim 16 \pi \varepsilon_0 \tlq\, e^{-2 \tlq \int_u^1
\sqrt{\frac{H}{h}}}\ .
 \labell{magic99}
 \eeq
Inspired by a Boltzmann factor, and by the zero-temperature
conduction threshold $\omega_0 = q$, we can interprete the
exponential factor as $\exp[-{q}/{T_{eff}}]$ where
\begin{equation}\labell{efftempint}
T_{eff} \, = \, \frac{\pi}{2}\,T \,  \left(\int_0^1 d u
\sqrt{\frac{H( u)}{h( u )}}\right)^{-1} \ .
\end{equation}
We note that the integral is finite since the integrand converges as
$h^{1/2}$ at $u\to 1$. There are two limits in which we can evaluate
this integral analytically: $f = 0$ and $f\gg1$. In these limits one
finds
\begin{eqnarray}
f=0:&&T_{eff} = T \frac{\sqrt{\pi}}{2}\,\frac{\Gamma\left(
\frac{3}{4}\right)}{\Gamma\left( \frac{5}{4}\right)} \sim  1.198\,T
\labell{teff0}\\
f\gg1:&&T_{eff}  \simeq
\frac{\Gamma\left({3}/{4}\right)^2}{\sqrt{\pi}}
\frac{T}{\sqrt{f}}\left(1 +
\left(\frac{4\Gamma(3/4)^2}{\pi^{3/2}}-\frac{1}{\sqrt{2}}\right)
\frac{1}{\sqrt{f}} \right)
\labell{teffbig}\\
&&\qquad\
\sim0.847\,\frac{T}{\sqrt{f}}\left(1+0.372/\sqrt{f}\right)\ .
\nonumber
\end{eqnarray}

\FIGURE{
\includegraphics[width=0.45\textwidth]{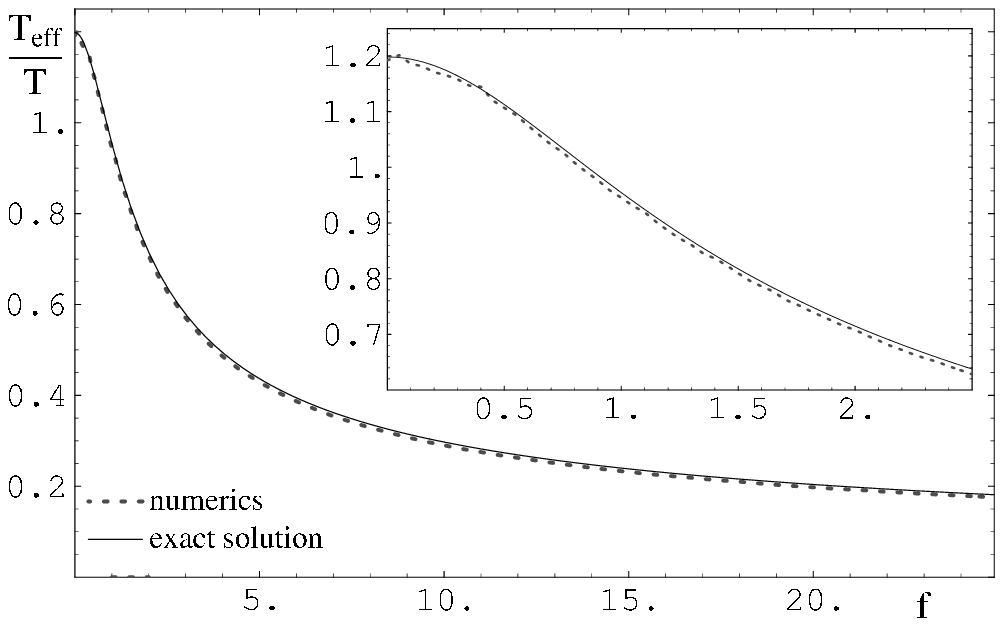}
\caption{Ratio of the ``effective temperature'' derived from the
temperature and $\tlq$ dependence at $\tlq \gg 1 \gg \tom$ to the
blackhole temperature as a function of the flux $f$. We show the
exact expression derived from the $\tlq \rightarrow \infty$ limit
and the numerical estimate at finite $\tlq$.}\label{efftplot}}
Comparing these results with the numerics, we find good convergence
in a consistent manner of both the profile of $A_y(u)$ and the
effective temperature with increasing $\tlq$. Since the
approximation that gave us the integrand in \reef{efftempint} is
valid up to roughly $h \gtrsim \tlq^{-2}$, we expect that $T_{eff}
/T$ measured at finite $\tlq$ has a relative accuracy of roughly
$\tlq^{-1}$. A simple way to estimate the effective temperature from
the conductivity is to compute $\partial_{\tlq} \log
\frac{\tilde{\sigma}_{yy}}{\tlq} \sim \frac{\pi}{T_{eff}}$ at large
values of $\tlq$. The factor of $\tlq^{-1}$ that we have included
here ensures that the convergence to the actual value of $T_{eff}$
is faster than logarithmic in $\tlq$. In figure \ref{efftplot}, we
show the comparison to the numerical estimate computed at
$q/T_{eff}\sim 47$, that is the best numerically stable estimate,
and demonstrate how the estimates converge to the exact results.

As illustrated in figure \ref{efftplot}, our new effective
temperature does not match the actual temperature of the system,
except for $f\simeq0.85$. At this point, we emphasize that, as
discussed in section \ref{revGeom}, the degrees of freedom on the
defect are in equilibrium with the thermal bath of adjoint fields
with temperature $T$. Of course, $T_{eff}$ is still a scale that
seems to play an interesting role in the defect conformal field
theory, as we will see in the following. Again, the reason that we
assign this scale the appellation of ``effective temperature'' is
that it appears to play the role of a temperature when the
conductivity \reef{magic99} is interpreted as a Boltzmann
distribution. It would be interesting if one could also give a
physical interpretation to the pre-factor $16 \pi \tlq$ in front of
the exponential in \reef{magic99}.

\subsubsection{Resonances on the defect}\label{resdef}
Next, we study the oscillatory behaviour of the spectral functions
at $\tom > \tlq$, using two different methods. Our results for the
transverse correlator $C_{yy}$ obviously can also be translated to
give us the longitudinal correlators $C_{xx}, C_{tt}$ using
\reef{translongreln}. Hence we will only discuss the former case.

We begin with the WKB-like expansion that gave \reef{epseqn}, which
was the starting point for the effective temperature above. Now
however, we do not have the scale $h \sim \tlq^2$ where we can match
the near-horizon approximation to the asymptotic approximation.
Furthermore, the dominant solution for $\epsilon$ is now oscillatory
since $\zeta \sim \zeta_0 = i\frac{\sqrt{H}}{h} \omega \sqrt{1 - h
\frac{\tlq^2}{ \tom^2}}$, rather than exhibiting the exponential
decay found above. The latter also reduces the validity of the
approximation that led to \reef{epseqn} and further we have to worry
about the logarithmically diverging integral $\int \zeta_0\propto
\ln h$ as $u \to 1$. So let us take the solution \reef{eps_approx},
but now with
 \beq
\zeta_0 = i\frac{\sqrt{H}}{h} \omega \sqrt{1 - h \frac{\tlq^2}{
\tom^2}} \, , \ \ \alpha =  2 \zeta_0 - \left(\ln \frac{h}{\sqrt{H}}
\right)' \, , \ \ \beta = \zeta_0' + \left(\ln \frac{h}{\sqrt{H}}
\right)' \zeta_0 \,
 \eeq
and match this in the limit $u \rightarrow 1$ to the appropriate
expansion of \reef{NHGapp}:
 \beq
\zeta \, \sim \, - \frac{i \tom}{h} \, + \, \frac{i \tlq^2}{2 \tom -
4 i} \, +\mathscr{O}(h) \, = \, -\zeta_0 \, + \,
\frac{\tlq^2}{\tom^2 + 2 i \tom} \, + \, \mathscr{O}(h) \, .
 \eeq
Now, we see that the divergent oscillations from the $e^{\int
\alpha}$ terms in \reef{eps_approx} must cancel, and the
approximation $\epsilon \ll \zeta_0$ should be valid near the
horizon. Taking the limit $\lim_{u\to 1} \epsilon =
\frac{\tlq^2}{\tom^2 + 2 i \tom} \equiv \epsilon_H$ of
\reef{eps_approx}, and solving for $\epsilon_0$ gives us then
 \bea
\epsilon|_{u = 0} \, = \,  \epsilon_0 & = &
\epsilon_H e^{-\int_0^1 du \alpha(u)} \, +\, \int_0^1
du \epsilon_H e^{-\int_0^u d\tilde{u} \alpha(\tilde{u})} \beta(u)
 \nonumber \\
& = & \epsilon_H \, - \, \int_0^1 du \, h \sqrt{\frac{1+f^2}{H}}
e^{-2\int_0^u d\tilde{u} \zeta_0} \left(\epsilon_H \alpha -
\frac{1}{2} \zeta_0 \frac{h' \tlq^2/\tom^2}{1-h\tlq^2 /\tom^2}
\right) \, .
 \labell{epsosc}\eea
It turns out that the $\mathscr{O}(h^{-1})$ divergent terms in
$\epsilon_H \alpha + \beta$ in \reef{epsosc} do indeed cancel, such
that the integral converges with the integrand $\propto h^2
h^{i\tom/2}$ as $u\to 1$. Unfortunately, we were not able to
evaluate this integral analytically, even in the limits where
various quantities involved getting large or small. We show this
approximate result \reef{epsosc} compared to the full numerical
result in figure \ref{compplot}.

Because of the rapid convergence as $u\to 1$, we see however that
most of the contribution to the integral comes from regions where
$h\sim 1$, in particular for large $f$. Hence as a very crude
approximation, we can set $h = 1$ and hence $\beta = 0$, which
allows us to compute the integral analytically:
 \beq\labell{epsintapp}
\epsilon_0 \, \sim \, \epsilon_H \sqrt{1 + f^2} e^{-2 i \tom
\sqrt{1+f^2} I(f) \sqrt{1 - \tlq^2/\tom^2}} \ .
 \eeq
While we do not expect this latter expression to give us the correct
phase and amplitude information, we still anticipate that this
result gives a good approximation for the frequency of the
oscillations, $2 \sqrt{1+f^2} I(f) \sqrt{1 - \tlq^2/\tom^2}$.

There is an alternative way of seeing more physically from the bulk
point of view, how the finite temperature effects arise by casting
the equation of motion for $A_y$ \reef{gauge4} in the form of the
Schr\"odinger equation, as suggested in \cite{RobSpec}:
 \beq\labell{schroeq}
\left( - \partial_\rho ^{\, 2}  +  h\, \tlq^2\right) A_y \, = \,
\tom^2 A_y 
\qquad {\rm where}\ \ \rho \, = \, \int_0^u d\tilde{u}
\frac{\sqrt{H}}{h} \ .
 \eeq
\FIGURE{\includegraphics[width=0.48\textwidth]{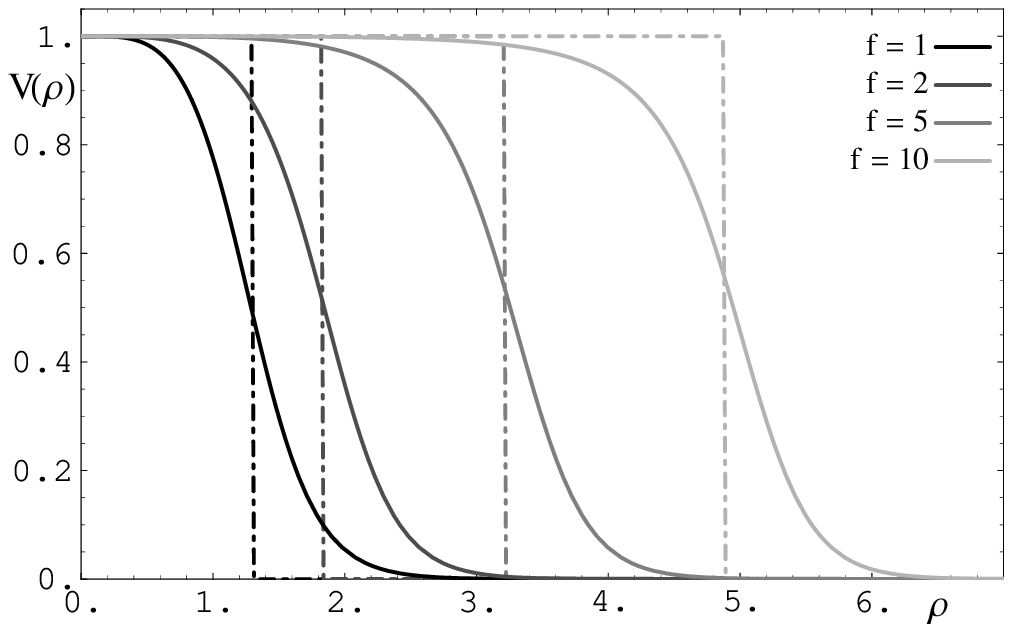}
\caption{The effective Schr\"odinger potential $V(\rho) = \tlq^2 h$
for the gauge field on the brane. We set $\tlq=1$ in the plot.}
\label{potplot}}
\noindent In terms of this new radial coordinate, the horizon gets
mapped to $\rho \to \infty$, and we see that $\rho$ is rapidly
varying only for $u \lesssim f^{-1/2}$ and for $h \ll 1$. This
suggests that for large $f$ we can approximately split the problem
in two regions: An asymptotic one where $h\sim 1$ and $\rho \sim
\rho_\infty(u) = \int_0^u d\tilde{u} \sqrt{H} $ and the near-horizon
region, where $h \ll 1$ and $\rho \sim \rho_{H}(u) = \rho_0 -
\frac{1}{4} \ln (h)$ for some $\rho_0$. Going even further in our
approximation, we assume a square potential $V = \tlq^2$ for $\rho <
\rho_\infty (1) = \sqrt{1+f^2} I(f)$ and $V = 0$ for $\rho >
\rho_{\infty}(1)$, which is displayed in figure \ref{potplot}, where
we see that this approximation is indeed justified. At this point,
we might also observe that the effective Schr\"odinger potential
appearing here is very similar in structure to that found for
supergravity modes \cite{sugra} and for mesonic modes, as discussed
in \cite{fast}.

With the square potential, it is trivial to find the solution for
infalling boundary conditions at the horizon:
\begin{displaymath}
A_y \, \sim \, \left\{ \begin{array}{lll} {\textstyle A_{0}\, 2
\sqrt{1 - \tlq^2 / \tom^2}
e^{- i \tom (\rho - \sqrt{1+f^2} I(f))} } &:& \rho < \sqrt{1+f^2} I(f) \\
{\textstyle  A_0 \Big(\big(1+\sqrt{1 - \tlq^2 / \tom^2}\big)e^{- i \tom \sqrt{1 - \tlq^2 /
\tom^2} (\rho - \sqrt{1+f^2} I(f)) }} && \\ \ \ \ \ \ {\textstyle - \big(1-\sqrt{1 - \tlq^2 /
 \tom^2}\big) e^{ i \tom \sqrt{1 - \tlq^2 / \tom^2} (\rho - \sqrt{1+f^2} I(f))} \Big) } &:&
 \rho > \sqrt{1+f^2} I(f) \end{array}\right. \ .
\end{displaymath}
Keeping in mind the change of coordinates, this gives us in terms of
the Ansatz that we used for the perturbative treatment
\begin{displaymath}
\zeta|_{u=0} \, = \, - i \tom \sqrt{1 + f^2} \times \left\{ \begin{array}{lll}
\frac{1  - \tlq^2/\tom^2}{1 - \frac{\tlq^2}{\tom^2} \cos^2 \big(\tom \sqrt{1 + f^2} I(f)
 \sqrt{1 - \tlq^2/\tom^2}\big)}&:& \tom > \tlq\\
\frac{1 - \tlq^2/\tom^2}{\cosh^2\big(\tom  \sqrt{1 + f^2} I(f) \sqrt{ \tlq^2/\tom^2 - 1}
 \big) - \tlq^2/\tom^2} &:& \tom < \tlq  \end{array}\right. \ . \labell{squarecon}
\end{displaymath}
The solution for $\tom > \tlq$ has the same location of the maxima
as \reef{epsintapp}, up to a small shift because of the overall
slope of the curve, but it is missing an exponential suppression
factor (for increasing frequencies) in the amplitude because we
approximated the smooth potential by a discontinuous one. For $\tom
\ll \tlq$, we also find the exponential suppression that leads to
the effective temperature computed at $\tom \rightarrow 0$,
\reef{tempnu}. Hence, we can clearly see how both effects arise from
a resonant mode on the width of the defect, and from tunnelling
through the defect region, respectively. In appendix \ref{tanhpot},
we approximate the potential by a hyperbolic tangent, for which we
can find an analytic solution, and find that is very closely
reproduces the exact result with a significant deviation only at
frequencies $|1- \tlq^2/\tom^2| \ll 1$ where the spectral function
is most sensitive to the details of the potential.

We show a comparison of the conductivity obtained from the different
approximations in figure \ref{compplot}. As expected, the
perturbative approximation in \reef{epsosc} gives a very close
approximation for small perturbations around the $T\to 0$ result,
$\left|\frac{C_{yy} - \lim_{T\to 0} C_{yy}}{\lim_{T\to 0} C_{yy}}
\right| \ll 1$, but deviates significantly wherever the finite
temperature effects become important. The analytical result
\reef{tanhpotcon} from the approximate $\tanh$ potential
\reef{tanhpoteq} however, provides a good fit for small $\tlq$ and
all values of $\tom$. For larger $\tlq \gtrsim \pi$ and $\tom >
\tlq$, there is a significant phase shift proportional to the
separation of the resonances but their amplitudes, separation and
the tailoff for $\tom < \tlq$ fit very closely. This is because the
phase $\phi$ is sensitive to absolute changes in the integral of the
potential, $\delta \phi \propto \int\delta  V \propto \tlq^2$, such
that already small deviations in $V/\tlq^2$ may have a big effect.

\FIGURE{
\includegraphics[width=0.48\textwidth]{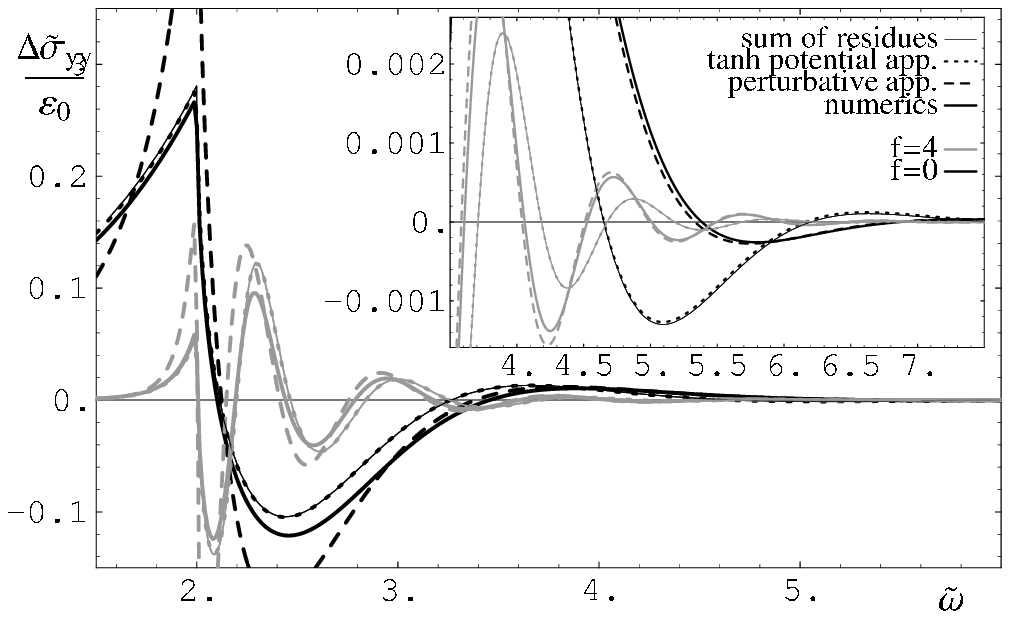}
\includegraphics[width=0.48\textwidth]{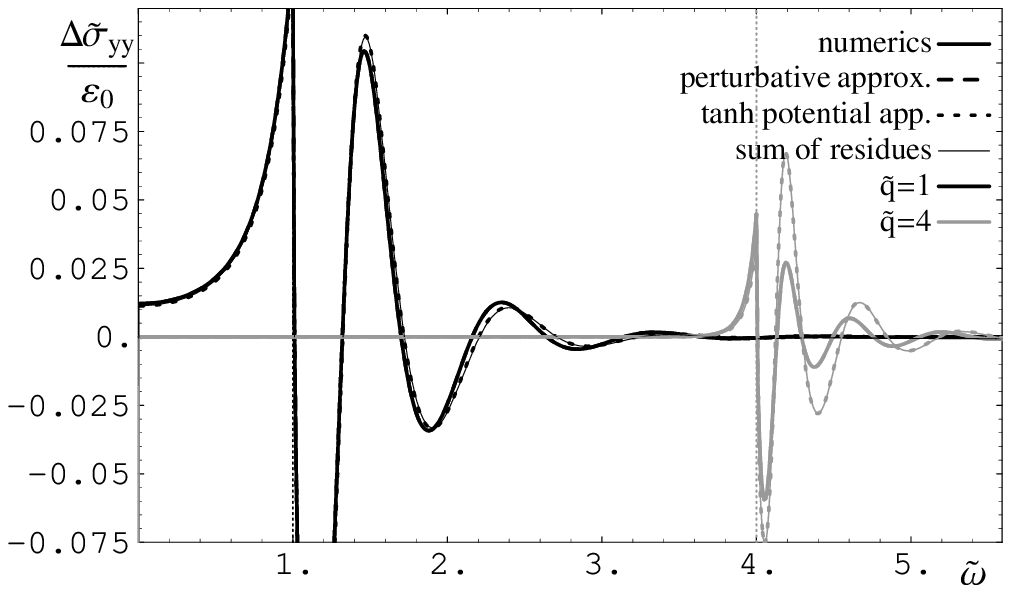}
\caption{Comparing the conductivity obtained from the numerics, from
the approximation \reef{epsintapp}, from the approximation using a
$\tanh$ potential \reef{tanhpotcon} and from the sum of residues
obtained from its poles \reef{expolesumlong}. Here we focus on the
oscillatory behaviour by plotting $\Delta \tilde{\sigma}_{yy}
\equiv\Re\left(\tilde{\sigma}_{yy}(\tom) -
\varepsilon_0\sqrt{1-\tlq^2/\tom^2}\right)$, \ie we subtract off the
low temperature limit \reef{looow}. Left: $f\in \{ 0,4\}$ and $\tlq
=2$. Right: $\tlq \in \{2,4\}$ and $f=4$. }\label{compplot}}

\section{Quasinormal modes and quasiparticles}\label{quasinorm}

In general, the thermal correlators will have poles in the lower
half of the complex frequency plane --- \eg see discussion in
\cite{RobSpec} or \cite{fast}. The positions of these poles
characterize the energy and lifetime of various excitations in the
system. When one of these poles is close to the real axis, the
spectral function will show a distinct peak and the corresponding
excitation can be interpreted as a quasiparticle. That is, the
excitation satisfies Landau's criterion for a quasiparticle that the
lifetime is much greater than the inverse energy. As illustrated in
figures \ref{condfig} and \ref{plotlong}, which essentially plot the
spectral function, the defect theory is developing metastable
quasiparticles in the large $f$ regime. Hence it is of interest to
examine the pole structure of the correlators and the spectrum of
quasiparticles in the defect conformal field theory. This gives us
not only more information on the defect field theory, but also
allows us to speculate more on the nature of the defect.

In principle, we could always find the poles in the thermal
correlators by simply numerically computing them over the entire
complex frequency plane. Of course, such a brute force approach
would present an enormously challenging problem at a technical
level. However, since the correlators should be meromorphic, we can
alternatively try to extract this information by fitting along the
real axis, the spectral function derived from an approximate
analytical solution of poles and positive powers  -- an approach
similar in spirit to that followed in \cite{fast}. To do so, we use
the complex ``rest frame'' frequency $\tlnu = \sqrt{\tom^2 -
\tlq^2}$ which maps $[0, \tlq ] \rightarrow [i \tlq , 0]$ and
$[\tlq,\infty[ \rightarrow [0, \infty [$. The motivation to do so is
the fact that the resonance pattern is most suitably characterized
by $\frac{\tom}{\tlnu}\tilde{\sigma}_{yy} -1$ as a function of
$\tlnu$, as shown in figure \ref{resfig}. There we see that even at
finite temperature this quantity varies only slowly with varying
$\tlq$. Certainly, in the low temperature limit, we expect Lorentz
invariance to be restored and then correlators will naturally depend
on the combination $\tom^2-\tlq^2$, as is implicit in \reef{looow}
and \reef{low22}.
\FIGURE{
\includegraphics[width=0.48\textwidth]{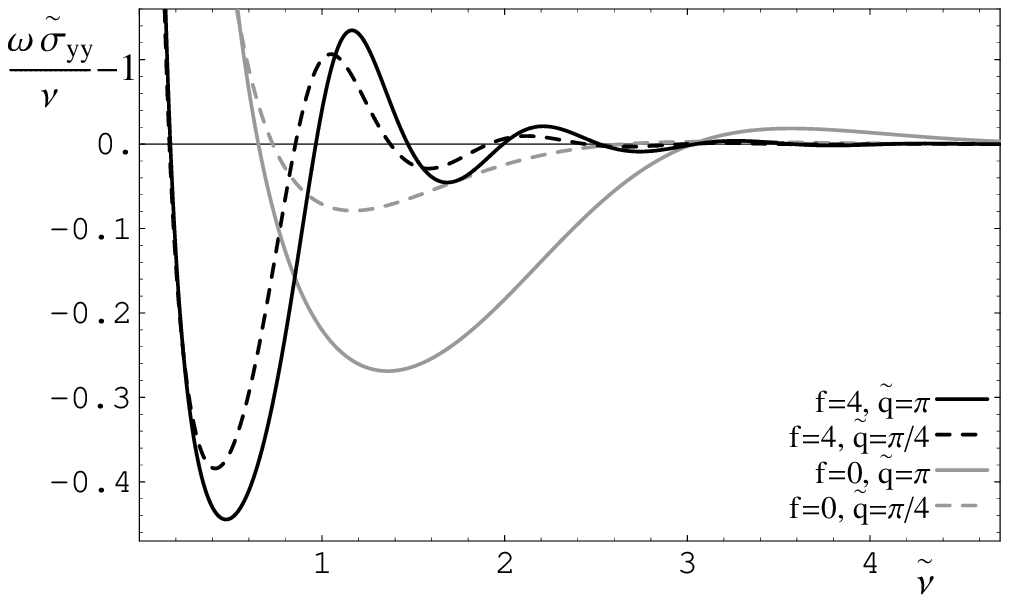}
\caption{$\frac{\tom}{\tlnu}\tilde{\sigma}_{yy}(\tlnu)$ for $\tlq
\in \{\pi/4,\pi\}$ and $f\in\{0,4\}$. This form of the spectral
function is very similar for different values of $\tlq$.}
\label{resfig}}

\subsection{Finding the Ansatz}

The strategy that we will take to find the poles is to take a
suitable Ansatz for the location of the $n^{th}$ pole in the complex
frequency plane, $\tlnu_n = \pm \tlnu_0(n) - i \tlgam_0(n)$, and for
the corresponding residue, and allow for the parameters to vary
slowly. If the Ansatz is good enough, and the parameters vary slowly
enough, then we can fit the conductivity resulting from a sequence
with constant parameters $(\tlnu_0, \tlgam_0)$ to the numerical
result using only the data in the region around the $n^{th}$
``resonance''. This data can be parametrized by the amplitude of the
resonance around the background and by the gap between the
resonances. The resulting parameters $(\tlnu_0(n), \tlgam_0(n))$
then give the location of the pole $\tlnu_n$, and it's residue.

A suitable guess for the full Ansatz is
 \beq
C_{y y} \, = \, - \varepsilon_0 \, \sum_{n \ge 1} \frac{1}{\pi}\left(
\frac{n (\tlnu_0 + i \tlgam_0)^2}{\tlnu + n (\tlnu_0 + i \tlgam_0)}
- (\tlnu_0 + i \tlgam_0)   \, + \,      \frac{n (\tlnu_0 -  i
\tlgam_0)^2}{\tlnu - n ( \tlnu_0 -i \tlgam_0)} + (\tlnu_0 - i
\tlgam_0)  \right) \ , \labell{gen_ansatz}
 \eeq
where the constant terms were introduced to cancel the otherwise
divergent behaviour of the series, since the pole terms do not decay
for large $n$. The condition that allows us to locally treat the sum
as an infinite series with constant $\{\tlnu_0,\tlgam_0\}$ is now
$\partial_n \log \{\tlnu_0(n),\tlgam_0(n)\} \ll 1$ and $\partial_n
\log \{\tlnu_0(n),\tlgam_0(n)\} \ll \frac{\tlnu_0^2}{n^2
\tlgam_0^2}$. Rewriting \reef{gen_ansatz} in a more suggestive form,
we find for $\tlnu \in \mathbb{R}$, \ie $\tom > \tlq$
 \bea
\Im C_{y y} \, = \, \varepsilon_0\, \Im \tlnu \cot\left(\pi
\frac{\tlnu}{\tlnu_0 + i \tlgam_0}\right) \,
=  \, \varepsilon_0\, \Im \tlnu \frac{\sin \frac{2 \pi \tlnu
\tlnu_0}{\tlnu_0^2 + \tlgam_0^2} \, -\, i \sinh \frac{2 \pi \tlnu
\tlgam_0}{\tlnu_0^2 + \tlgam_0^2}}{\cosh \frac{2 \pi \tlnu
\tlgam_0}{\tlnu_0^2 + \tlgam_0^2} \, -\, \cos \frac{2 \pi \tlnu
\tlnu_0}{\tlnu_0^2 + \tlgam_0^2}} \ ,
 \labell{lin_ansatz}
 \eea
such that we get the conductivity
 \beq
\tilde{\sigma}_{yy} \, = \,\varepsilon_0\, \frac{\tlnu}{\tom} \frac{
\sinh \frac{2 \pi \tlnu \tlgam_0}{\tlnu_0^2 + \tlgam_0^2}}{\cosh
\frac{2 \pi \tlnu \tlgam_0}{\tlnu_0^2 + \tlgam_0^2} \, -\, \cos
\frac{2 \pi \tlnu \tlnu_0}{\tlnu_0^2 + \tlgam_0^2}} \,,
 \eeq
which turns out to be finite at $\tlnu \rightarrow 0$. These
exponentially suppressed resonances are characteristically what we
expect and we can, in principle, fit the parameters $\tlnu_0$ and
$\tlgam_0$ to the resonance pattern.

To be more precise however, we need to go back to the original
``physical'' frequency $\tom$. Keeping the location of the poles and
the residue fixed, the sum becomes now
 \bea\labell{omansatztrans}
C_{y y} = \varepsilon_0\, \frac{\tom}{\pi} \log \left(\frac{\tlnu_0 - i \tlgam_0}{\tlnu_0 + i
\tlgam_0}\right)
 \ + ~~~~~~~~~~~~~~~~~~~~~~~~~~~~~~~~~~~~~~~~~~~~~~~~~~~~~~~~~~~~~~~~ \\
\varepsilon_0\, \sum_{n \ge 1}\frac{1}{\pi}\left(  \frac{n (\tlnu_0
-  i \tlgam_0)}{\big(\tlq^2 + n^2 (- \tlnu_0 +  i \tlgam_0)^2
\big)^{1/2}}\left( \frac{n (\tlnu_0 -  i \tlgam_0)^2}{\tom - \big(
\tlq^2 +  n^2 ( \tlnu_0 - i \tlgam_0)^2\big)^{1/2}} + \frac{n
(\tlnu_0 -  i \tlgam_0)^2}{ \big( \tlq^2 +  n^2 ( \tlnu_0 - i
\tlgam_0)^2\big)^{1/2}}  \right)
  \right. \nonumber \\
  \left.  -\,  \frac{  n (\tlnu_0 + i \tlgam_0)}{\big(
\tlq^2 + n^2 (\tlnu_0 + i \tlgam_0)^2 \big)^{1/2}}\left( \frac{n
(\tlnu_0 + i \tlgam_0)^2}{\tom + \big( \tlq^2 + n^2 (\tlnu_0 + i
\tlgam_0)^2 \big)^{1/2}} - \frac{n (\tlnu_0 + i \tlgam_0)^2}{
 \big( \tlq^2 + n^2 (\tlnu_0 + i \tlgam_0)^2 \big)^{1/2}}\right) \right)  , \nonumber
 \eea
where the term $\frac{\tom}{\pi} \log \frac{\tlnu_0 - i
\tlgam_0}{\tlnu_0 + i \tlgam_0}$ cancels the unphysical negative DC
conductivity in the $\tlq \gg \tlnu_0, \tlgam_0$ limit that would
arise otherwise. Note there is still a logarithmic divergence in the
real part, that we are not interested in. This sequence does not sum
to any known analytic expression, but the integral approximation can
be computed straightforwardly analytically, such that in order to
eventually study the sequence numerically, we will only sum the
first few hundred poles and add a small ``background'' contribution
from the rest of the poles using the integral.

Following the same considerations, we also find an Ansatz for the
longitudinal correlator,
 \bea\labell{omansatzlong}
C_{x x} & = & \varepsilon_0\,\sum_{n \ge
1}\frac{1}{(n\!-\!\frac{1}{2})\pi}\left( \frac{(n\!-\!\frac{1}{2})
(\tlnu_0 -  i \tlgam_0)}{\big(\tlq^2 + \big(n\!-\!
\frac{1}{2}\big)^2 ( \tlnu_0 -  i \tlgam_0)^2 \big)^{1/2}} \
\frac{1}{\tom - \big( \tlq^2 +  \big(n\!-\!\frac{1}{2}\big)^2 (
\tlnu_0 - i \tlgam_0)^2\big)^{1/2}} \right.
\nonumber \\
&&\quad - \left. \frac{(n\!-\!\frac{1}{2}) (\tlnu_0 + i
\tlgam_0)}{\big( \tlq^2 + \big(n\!-\! \frac{1}{2}\big)^2 (\tlnu_0 +
i \tlgam_0)^2 \big)^{1/2}} \ \frac{1}{\tom + \big( \tlq^2 +
\big(n\!-\! \frac{1}{2}\big)^2 (\tlnu_0 + i \tlgam_0)^2 \big)^{1/2}}
\right) \ ,
 \eea
which converges and needs no regularization or terms with positive
powers of $\tom$. In terms of $\tlnu$, the poles are located at
$(n-1/2)(\pm \tlnu_0 - i \tlgam_0)$, with residues
$\frac{1}{n-1/2}$, as we expect by \reef{translongreln} from the
ansatz for $C_{yy}$.

In order to finally obtain the location of the poles and their
residue, we split the spectral function at the minima into segments
around each maximum and simply fit them to our Ansatz giving us a
set of parameters that we attribute to the local properties of the
sequence at the most nearby pole, as described in the beginning of
the section. We obtain both $\tlnu_0$ and $\tlgam_0$ and an overall
factor $(1+\epsilon_\mathscr{R})$ (or
$(1+\epsilon_\mathscr{R})^\star$ on the negative branch) for the
residues. The latter is needed because the resonance pattern is
exponentially suppressed already for reasonably small $\tlnu \gtrsim
\pi$, such that the background of the fitted sequence needs to be
adjusted to in precise agreement with the background of the data, in
order to extract the relevant information which is contained in the
resonances. As it turns out that $|\epsilon_\mathscr{R}| \ll 1$, we
will not comment about its value for the rest of the paper, because
it is irrelevant for both the quantitative and the qualitative
discussions. In principle, one can introduce more parameters, such
as an overall shift in the frequency, but this would not improve the
results, since in practice, it simply introduces extra degeneracy in
parametrizing the fit.

\subsection{Quasiparticles from the collisionless
regime}\label{colll}

Given the Ansatz for the transverse and longitudinal correlators
above, we will now discuss the results for determining the positions
of the poles. The results are only displayed for Re~$\tlnu>0$ but as
shown in \reef{gen_ansatz}, there is a corresponding set of poles
with Re~$\tlnu<0$. In this section, we focus on the collisionless or
short-wavelength regime with $\tom\gg1$ and $\tlq\gsim1$.

As a first test, we compare the fitted location of the poles to
their exact location for the $\tanh$ potential in appendix
\ref{ex_pol_str}. We expect that this gives us a good estimate for
the quality of the fit for the actual spectral functions. Some
typical results are shown in the first plot of figure
\ref{poleplot}. We find that for $f=0$, the fit is very poor, with
the $\tlq=\pi/4$ result being worse than the $\tlq=\pi$ case, and
there is a small deviation for $f=4$ at large $\tom$, again with a
slightly better fit for larger $\tlq$. Apart from that, \ie for
large $f$ or large $\tlq$ and small $\tom$, the fit is very good.
This is just what we would have expected from our conditions for the
validity of the Ansatz as smaller $\tlq$ imply more rapidly varying
$\tlnu_0,\tlgam_0$ at least for the first few poles and both small
$f$ and $\tlq$ and large $\tom$ move the poles further away from the
real axis. Furthermore, for large $\tom$, the amplitude of the
resonance pattern becomes quickly suppressed and so it is subject to
systematic deviations and noise.

Now, let us look at the qualitative behaviour. We see that, as
anticipated with the Ansatz, the poles lie roughly equally spaced on
a straight line, \ie they are resonances in a region of fixed width
with fixed ``mass'' to inverse lifetime ratio. With increasing $f$,
both the separation of the poles and the slope of the line of poles
decreases, \ie the poles are moving closer to the real axis. Of
course, these changes are reflected by the appearance of distinct
peaks in the previous plots of conductivity at large $f$. This
behaviour is roughly independent of $\tlq$, and there is an overall
shift depending on $\tlq$, that is larger for smaller values of $f$.
One might expect both the decreasing energy gap and the increasing
mass to width ratio since the length scale due to the width of the
defect increases and the shape of the step in the potential is
approximately fixed. The deviation of the poles from a straight line
is stronger for large $\tlq$ and reflects the fact that the shallow
potential at small $\tlq$ is fully probed at small $\tlnu$, whereas
at large $\tlq$, the resonances at small $\tlnu$ are only sensitive
to the details of the top of the potential, and only probe the
steeper regions at higher $n$. This effect is obviously more visible
at large $f$ because of the closer spacing of the resonances. We
also see that the poles of the transverse correlator lie roughly
half-way between the poles of the longitudinal correlator, as
anticipated in \reef{translongreln}. Comparing the various
approximations to the location of the poles of the actual spectral
function, we see the behaviour that we saw in figure \ref{compplot}
encoded in a different way. Here, we see the shift of the poles of
the $\tanh$ potential that is more significant for larger $\tlq$.

As an aside, let us briefly return to the cuts that appeared in the
the transverse and longitudinal conductivities, in \reef{looow} and
\reef{low22}, with the limit $T\to 0$. Given the present analysis,
it is natural to conclude that this cut arises through an
accumulation of poles near $\tom \sim \tlq$. Assuming the locations
of the poles in the $\tlnu$ plane, $\tlnu_n$, to be roughly
independent of $\tlq$, we find that for $\tlq \gg 1$ we get $\tom_n
= \sqrt{\tlnu_n^2+\tlq^2} \sim \tlq + \tlnu_n^2/\tlq$. This then
leads to an infinite number of poles accumulating near $\tom = \tlq$
as $T\to 0$ and resulting in a cut.
\FIGURE{
\includegraphics[width=0.48\textwidth]{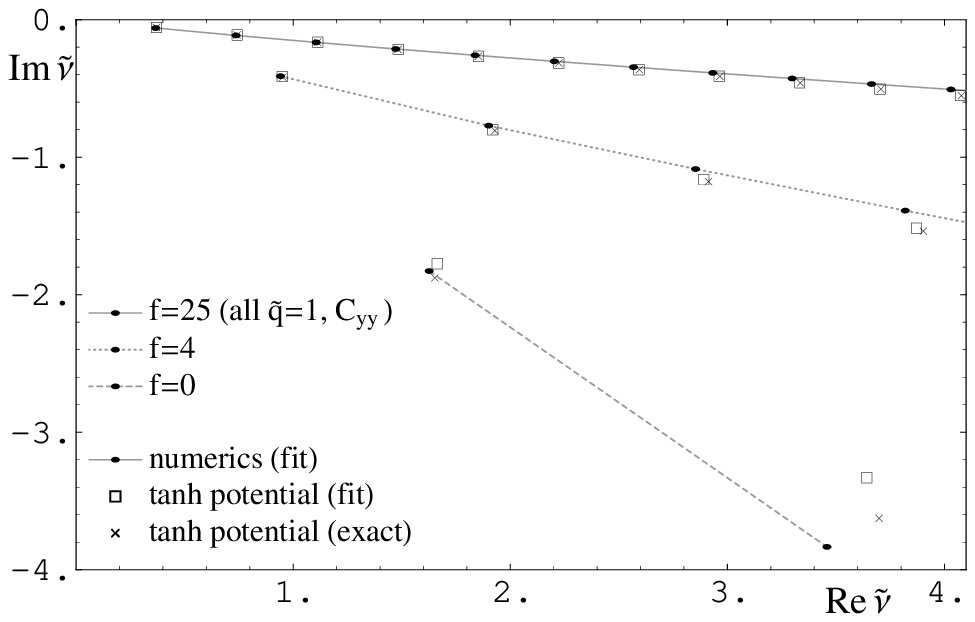}
\includegraphics[width=0.48\textwidth]{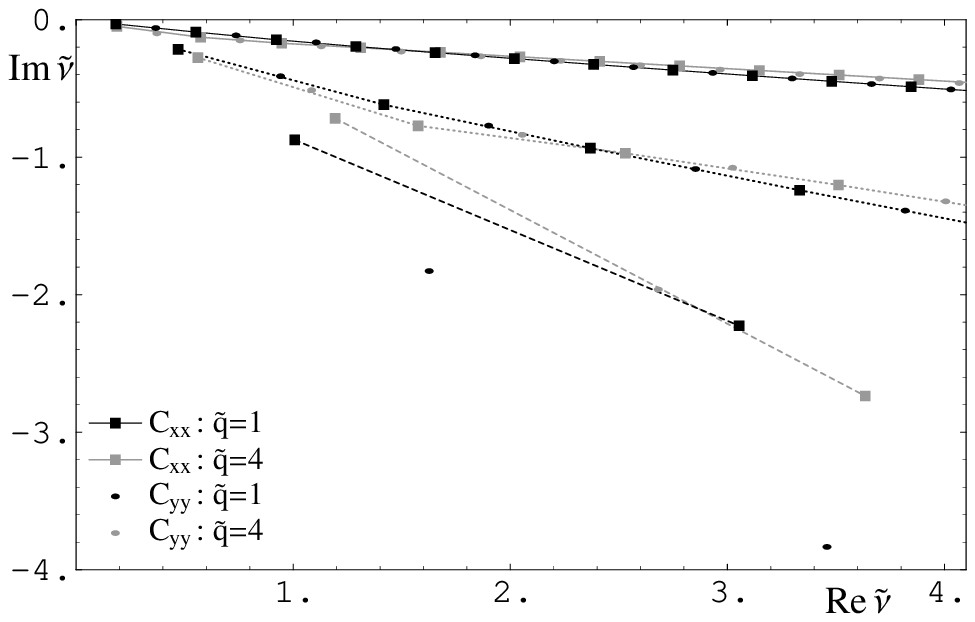}
\caption{Left: Comparing the exact location of the poles of the
transverse correlator to the fit, using the approximate
$\tanh$-potential of appendix \ref{tanhpot}, at values
$f\in\{0,4,25\}$ and $\tlq = 1$. Right: Comparing the poles of the
transverse correlator with the ones of the longitudinal correlator
at $f\in\{0,4,25\}$ and $\tlq \in \{1,4\}$. The lines are only shown
to guide the eye as to which poles correspond to the same values of
$f,\tlq$.} \label{poleplot}}

While the Ansatz \reef{gen_ansatz} fixes the poles along a straight
line with a fixed spacing, \ie $\tlnu = n (\tlnu_0 - i \tlgam_0)$ in
the Re~$\tlnu>0$ region, we fit the parameters locally to each peak
of the spectral function and so the fitted poles deviate slightly
from this simple Ansatz. Keeping in mind the limitations, let us try
to extract some quantitative information on these deviations. In
particular, at large $n$, the poles approach a straight line of the
form $\tlnu_n = \delta \tlnu + i \delta \gamma + n(\tlnu_0 + i
\tlgam_0)$ ($\tlnu_n = \delta \tlnu + i \delta \gamma +
(n-1/2)(\tlnu_0 + i \tlgam_0)$ in the longitudinal case). To extract
this information, we use different techniques in different regimes,
which we outline to forewarn the reader about the validity of the
results. For large $f$, where we have at least the first 5
well-fitted poles, we ignore the first 0-3 poles, leaving us at
least 5 poles, such that we can fit the asymptotic lines plus a
decaying exponential to $\Re \tlnu_n$ and $\Im \tlnu_n$ and still
get information about the accuracy. For some cases, the exponential
fit fails, and we resort to fitting a straight line and estimate the
accuracy from the second derivative in the location of the poles.

The assumption of an exponential deviation from a straight line may
seem somewhat arbitrary, but it turns out to be the right choice, as
it is the only natural candidate whose results are independent
within errors from the particular choice of the number of poles used
for the regression. The location of the poles from the $\tanh$
potential, for example, contains by this criterion an
$\mathscr{o}(\ln n)$ term as expected from \reef{tanhexpoleapp}.

In the borderline case, where there are 4 poles, we extract the
uncertainties by fitting the last 3 poles to the asymptotic line
with the deviation estimated by the fit with 4 points. For 3 poles
only, we still get a rough estimate for the asymptotic limit (from
the last 2 points) and for the accuracy by including the first
point. For $f=0$, we always find only the first two poles, so we can
give only an order of magnitude guess for the rest of the sequence.
Finally, we estimate the uncertainties from the errors in the fit of
the sequence of poles and from the deviation of the estimated
location of poles to their exact location in the case of the $\tanh$
potential. We use the latter also to add a shift to try to correct
for systematic errors in the fit of the Ansatz (\ref{omansatztrans},
\ref{omansatzlong}) to the numerical data. We are somewhat sloppy
with the uncertainties in the sense that we do not distinguish
between random and systematic errors. So we assume that the accuracy
of the fits is limited by systematic uncertainties in the
convergence towards the asymptotic straight line, which may result
in a slight overestimating of uncertainties in the averaged data
that we present below. As expected, the results from the transverse
and longitudinal poles are identical within the errors and so we
average over them.

Let us now examine some of the results of our fitting in figure
\ref{fpoleplot}. In the first two plots, we show results for the
energy gap between the quasinormal modes, $\tlnu_0$. In particular
for large $f$, we see that the asymptotic behaviour of $\pi/\tlnu_0$
matches a simple straight-line fit: $\pi/\tlnu_0=c_1 \sqrt{f}+c_0$
with $c_1\simeq 1.821$ and $c_0\simeq-0.539$. In the first plot,
this behaviour seems to match well with the asymptotic behaviour of
$\pi T/(2 T_{eff})$ and $\pi D(f) T$. Note, however, that the second
plot shows that upon closer examination, the deviation between
$\pi/\tlnu_0$ and the curves set by these scales in the large $f$
regime seems to be beyond the errors expected for our numerical fit
to $\tlnu_0$. Note that large $f$ behaviour in \reef{teffbig} gives
$\pi T/(2 T_{eff})\simeq 1.854\sqrt{f}-0.690$, while
\reef{diffusion3} yields $\pi D T\simeq 1.854\sqrt{f}-1$. The
asymptotic behaviours for these quantities have precisely the same
slope and the difference is in the constant term (and the subleading
$1/\sqrt{f}$ terms), as can be seen in figure \ref{fpoleplot}. This
slope is only a fair match for that found in our straight-line fit.
We expect that this is because of subleading $1/\sqrt{f}$ terms and
that we would see better convergence at larger values of $f$. In any
event, it seems then that $\tlnu_0$ is closely related to other
characteristic scales in the defect theory.  Note that here since
$\tlnu_0$ appears to be independent of $\tlq$ within the errors (see
figure \ref{kpoleplot}), the data in figure \ref{fpoleplot} is
averaged over $\tlq \in {\pi/2,\pi,2\pi}$.

We also show the overall shift $\delta \tlgam$ and the ratio
$\tlgam_0/\tlnu_0$ separately in figure \ref{fpoleplot} for the
cases $\tlq \in {\pi/2,\pi,2\pi}$. In each case, these parameters
show a $1/\sqrt{f}$ falloff for large $f$. In particular, this means
that the width $\tlgam_0$ is falling as $1/f$ and so we see the
origin of the quasiparticle peaks in the spectral curves. In each
plot, we also show $\pi T/(2 T_{eff})$ for each case and see there
is good agreement within the estimated errors of the numerical
results. This is just what we expect, since the detailed shape of
the step in the effective Schr\"odinger potential and hence the
ratio between the two modes in the asymptotic region of the
potential, is to a good approximation independent of $f$. The slowly
varying part of this ratio gives rise to the finite shift and the
exponential suppression factor gives rise to $\tlgam_0/\tlnu_0$,
which are in the limit of large $\tom$ proportional to the inverse
of the width of the asymptotic region of the potential. This can be
more easily seen from the expressions in appendix \ref{ex_pol_str}.
From the boundary point of view, it comes at no surprise that the
overall shift of the poles is proportional to the overall energy
scale, and that the quasiparticle excitations become more stable
with increasing $f$, which is proportional to the width of the
potential step. One could make a similar plot of $\delta\tlnu$ but
we do not show the results here. While on the whole the trends
appear similar to those for $\delta\tlgam$, the values are typically
smaller by a factor of roughly 2 while the relative errors are
larger by a similar factor. Hence at least for the smaller values of
$\tlq$, the results are consistent with zero shift.
\FIGURE{
\includegraphics[width=0.48\textwidth]{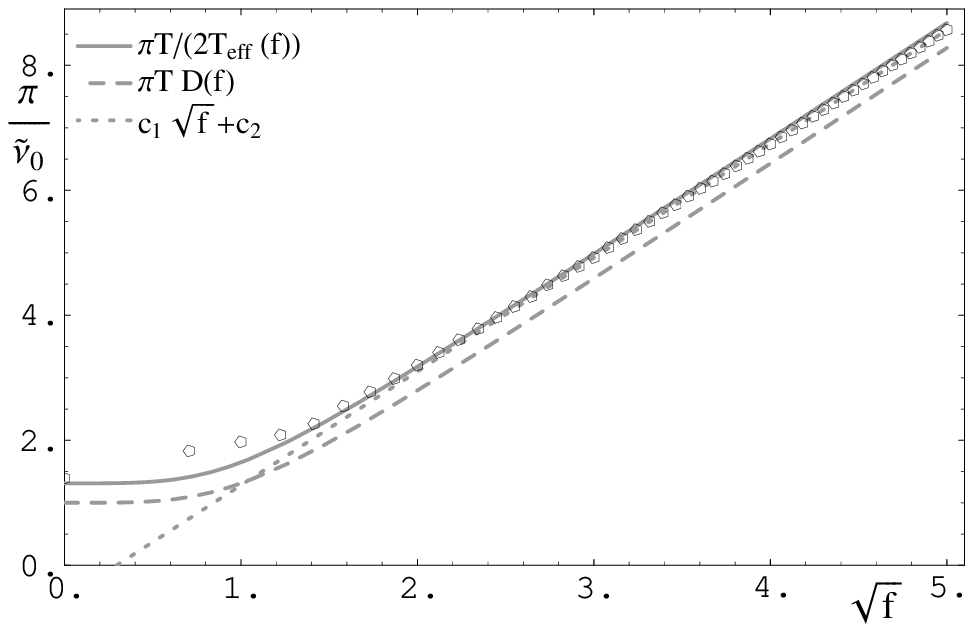}
\includegraphics[width=0.48\textwidth]{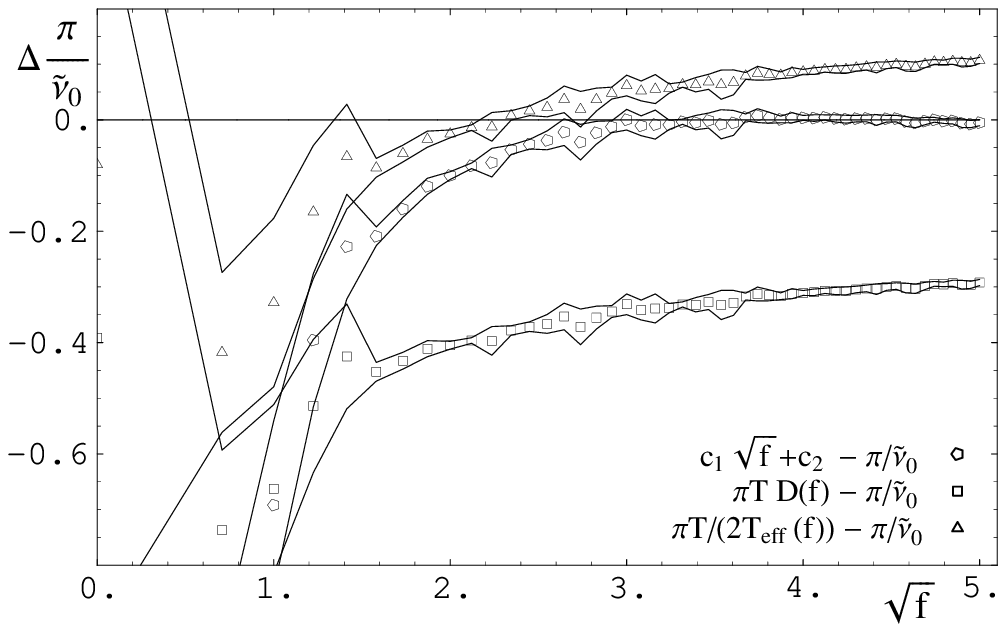}\\
\includegraphics[width=0.48\textwidth]{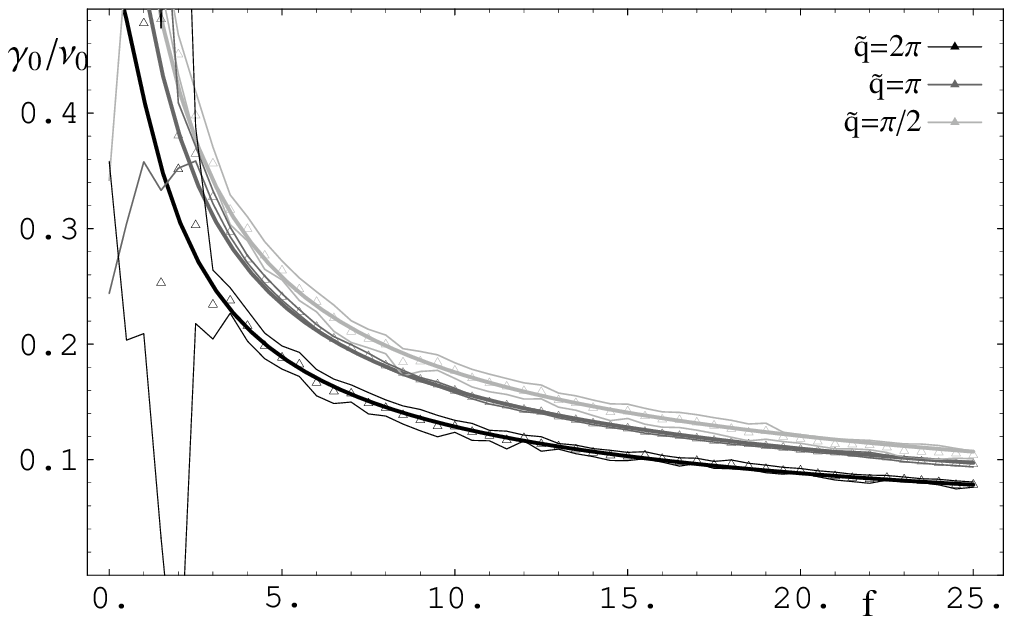}
\includegraphics[width=0.48\textwidth]{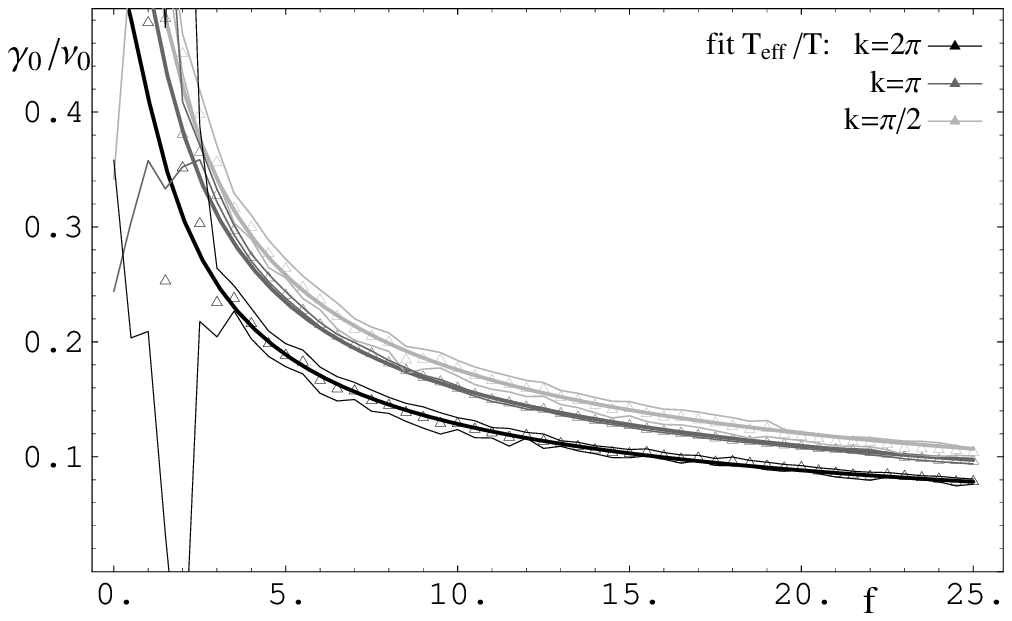}
\caption{Top left: The energy gap between resonances $\tlnu_0$
presented as $\pi/\tlnu_0$. The asymptotic $\sqrt{f}$ behaviour is
fit is several ways. Top right: The difference between the previous
fits and the numerical results for $\pi/\tlnu_0$. The data in the
top two plots is averaged over $\tlq \in {\pi/2,\pi,2\pi}$. Bottom
left: The ``mass to width'' ratio $\tlgam_0/\tlnu_0$ for the
sequence of poles for several values of $\tlq$. Bottom right: The
$f$-dependence of the overall shift of the poles $\delta\tlgam$. The
points are the average numerical data and the narrow lines indicate
the $1\sigma$ uncertainties, which includes both random and
systematic errors.} \label{fpoleplot}}

Now let us turn to the $\tlq$ dependence of the quasinormal modes.
Because of the good agreement of the $f$ dependence with
$T_{eff}/T$, we improve the accuracy of our results by taking a
(weighted) average over the suitably scaled values of the
characteristic quantities for $f\in \{9,16,25,36 \}$. In figure
\ref{kpoleplot}, we show the $\tlq$ dependence of the same
quantities as in figure \ref{fpoleplot}. First, we see that
$\tlnu_0$, which is supposed to depend only on the width of the
potential, is within the uncertainties independent of $\tlq$. Any
change in $\tlq$ however scales only the height of the potential
step.

From the results in appendix \ref{ex_pol_str}, we would expect that
varying $\tlq$ changes only the overall shift of the poles, but we
know already that the full result has fundamentally different
characteristics coming from the shape of the potential step because
of the absence of a significant $\mathscr{o}(\ln n)$ term. In
general, however, $\tlnu_0$ should not change significantly, since
we consider here only $f\ge 9$, so the potential is already so wide
that small details of probing the potential step should not change
the the quasinormal modes too much. Both the shift, and the
deviation from the linear Ansatz conspire to give us both the right
``low temperature background'' with approximately symmetric
oscillations around it as in figure \ref{resfig}. From the fact that
this behaviour resembles that in the resulting conductivity from
\reef{lin_ansatz}, one should assume that there are small shifts and
deviations for small $\tlq$. One also expects the shift to grow not
faster than $\propto \ln \tlq$, provided that the ratio of the two
modes in the asymptotic region depends at most on some power of the
height of the step in the potential.

In figure \ref{kpoleplot}, we find roughly this behaviour of the
shift, with small $\delta \tlnu,\tlgam$ at small $\tlq$ and an
indication of some converging or slowly growing behaviour at large
$\tlq$. We also find a small drop in $\tlgam_0/\tlnu_0$ with some
converging behaviour at large $\tlq$. In principle, we could try to
use this information to try to reverse engineer the calculations in
appendix \ref{ex_pol_str}, \ie to reconstruct the ratio of the
incoming and outgoing modes. For example, the absence of a
significant $\mathscr{o}(\ln n)$ term tells us that there is no
significant $\tlnu$ dependence, the approximately constant (in $f$)
ratio $\delta\tlnu/\delta\tlgam$ shows us that the ratio of the
modes has a complex phase (and also its value) but there is nothing
really interesting to learn from this. A somewhat interesting point
though is that the change in $\tlgam_0/\tlnu_0$ tells us that at
small $\tlq$, the potential ``appears smoother'' than at large
$\tlq$.
\FIGURE{
\includegraphics[width=0.48\textwidth]{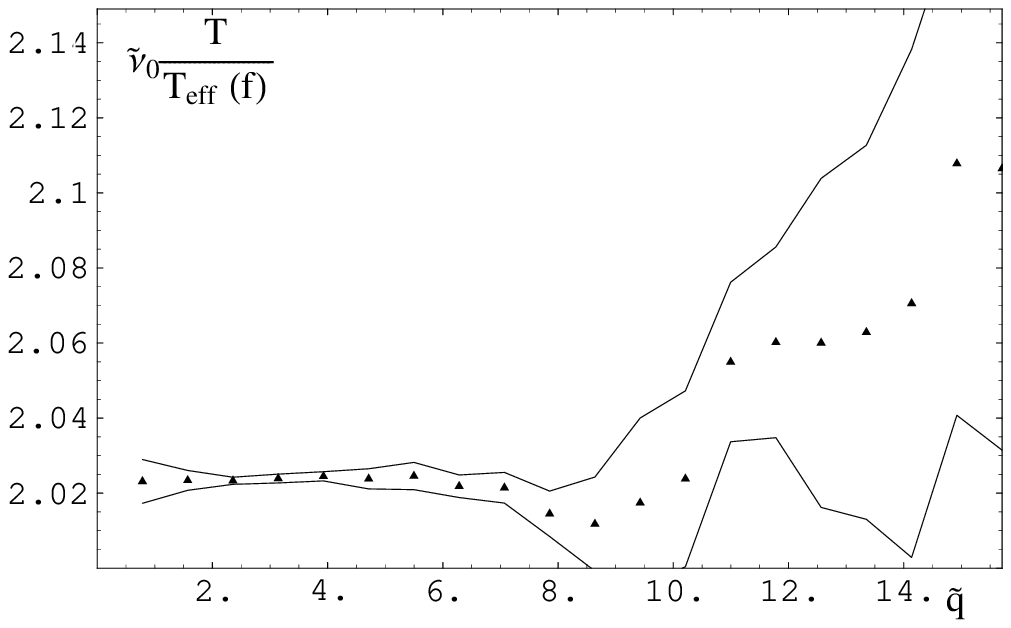}
\includegraphics[width=0.48\textwidth]{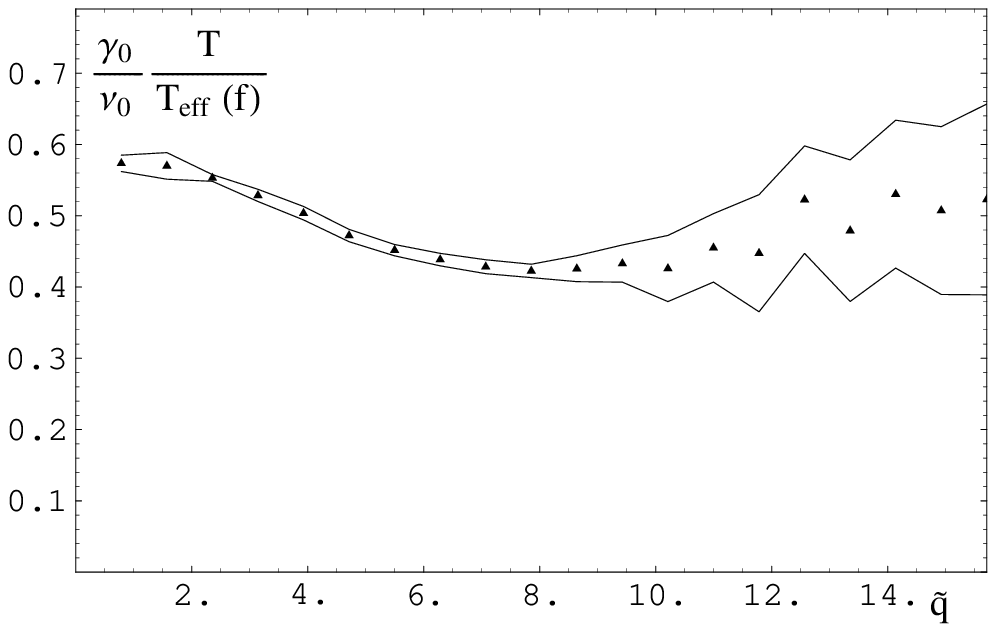}\\
\includegraphics[width=0.48\textwidth]{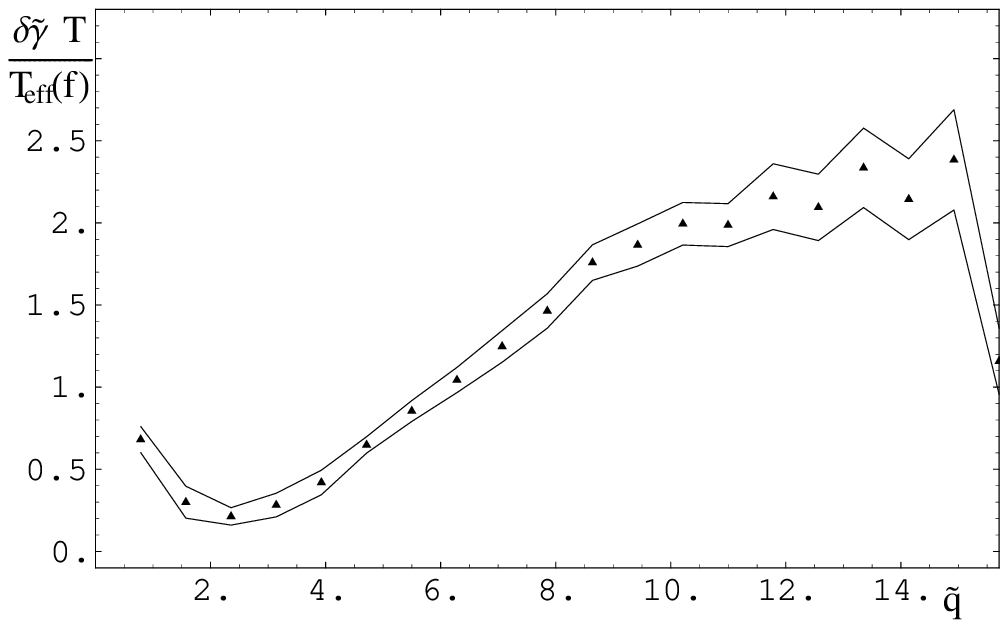}
\includegraphics[width=0.48\textwidth]{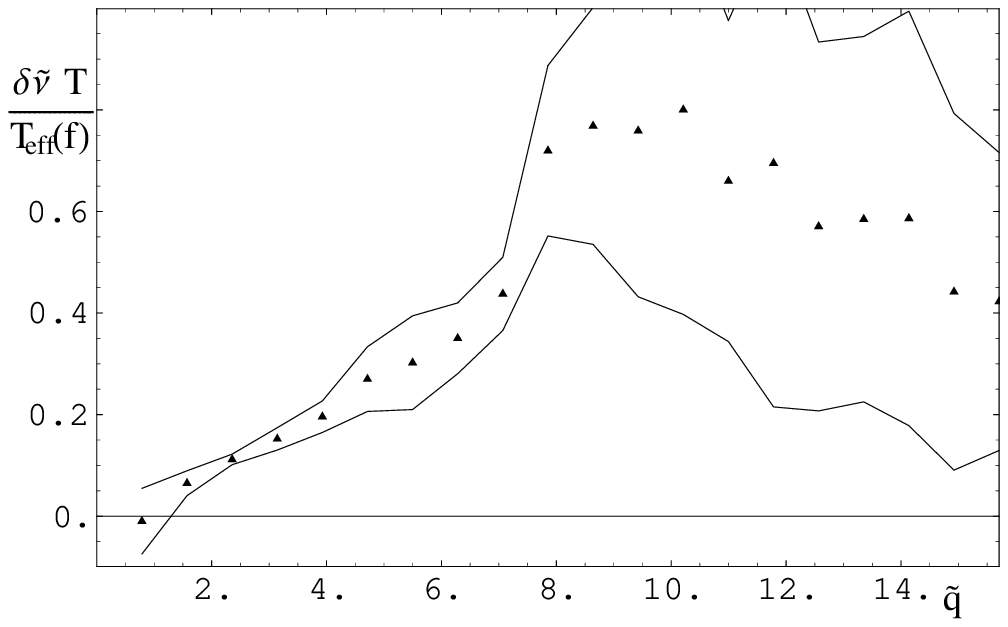}
\caption{Average over quantities appropriately scaled with
$T_{eff}(f)/T$ for $f\in \{9,16,25,36 \}$: Top left: (In)dependence
of $\tlnu_0$ on $\tlq$. Top right: $\tlgam_0/\tlnu_0$. Bottom: The
complex shift $\delta \tlnu$ and $\delta \tlgam$.}
\label{kpoleplot}}

\subsection{Poles in the hydrodynamic regime} \label{diffuse}
In this section, we focus on the hydrodynamic regime where
$\tom,\tlq\ll1$. In this regime, the diffusion pole \reef{diffpol}
dominates the structure of the correlators. One might wonder, why we
have not included the diffusion pole into the sum for $C_{yy}$, as
in \reef{gen_ansatz}. As we will show below, this is because the
diffusion pole disappears at a critical value of the wave-number,
$\tlq_c$, which is below the values of $\tlq$ that we have
considered to this point. Below $\tlq_c$ there are two poles on the
imaginary axis in the $\tom$ plane, one of them being the diffusion
pole, and the other one at larger absolute imaginary values of $\tom$, which
decreases slowly as $\tlq$ grows, as shown in figure \ref{impoles}.
While the diffusion pole is in perfect agreement with what we
expected, the second pole, corresponding to rapid (\ie on thermal
scales) decay of long-range modes, is somewhat puzzling. In
particular, it has a non-trivial $f$ dependence at small values of
$\tlq$. It seems that for large $f$, the lifetime of those modes is
not anymore proportional to the length scale of the defect, but
increases less rapidly.

At $\tlq_c$, there is a branch cut, and the poles move away from the
imaginary axis out  into the complex plane to turn into the first
quasiparticle poles, \ie the $n=1$ poles in \reef{gen_ansatz}. Hence
at this point, the transport changes from the collision dominated
phase to the collisionless phase. This is a good example of the
interplay between various length scales. We can interpret this on
the one hand as the height of the effective potential being smaller
or larger than the inverse length scale of the defect (and hence the
effective temperature) and on the other hand as separating between
between modes smaller and larger than the size of the defect. From a
hydrodynamic viewpoint, however this branch cut gives us
approximately the mean free path, which is in strongly coupled
systems proportional to, and of the same order as, the temperature
scale, and we see an approximate scaling of $\tlq_c$ with the
effective temperature.
\FIGURE{
\includegraphics[width=0.48\textwidth]{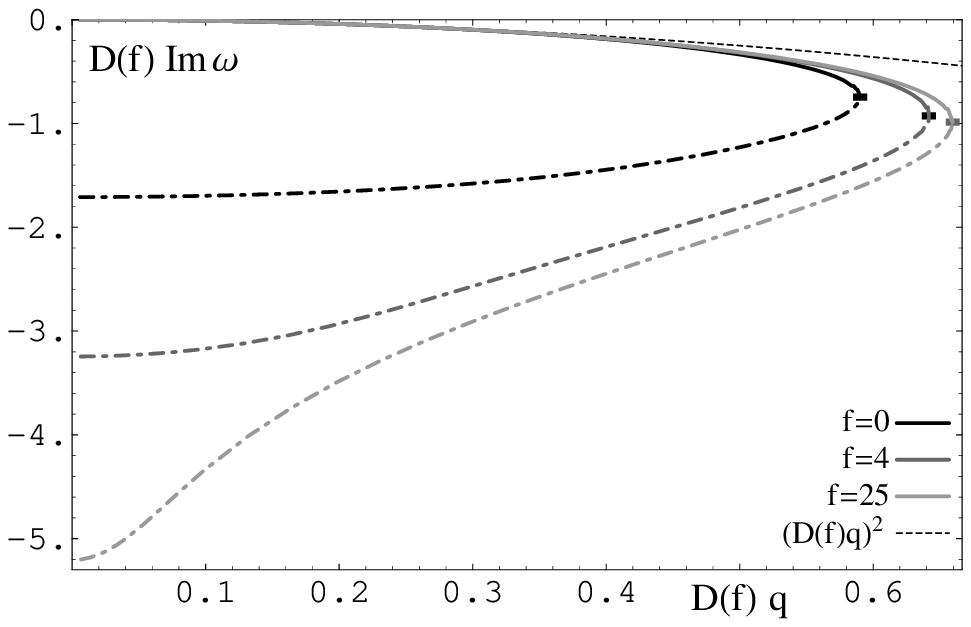}
\includegraphics[width=0.48\textwidth]{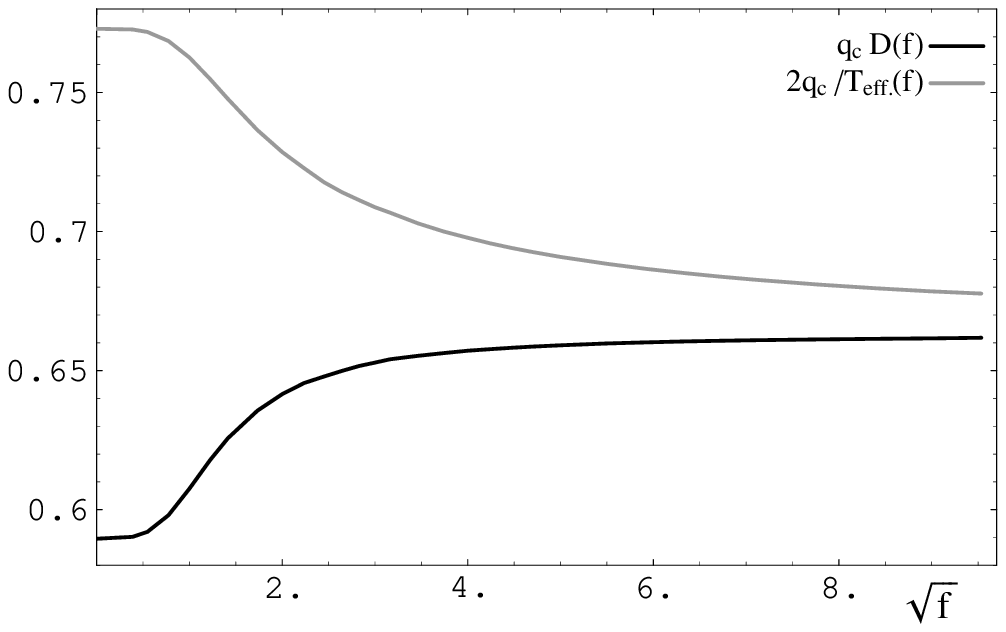}
\caption{Left: The location of the poles on the imaginary axis below
$q_c$, rescaled by the diffusion constant. The dashed line is
what one expects for the diffusion pole. Right: The critical values $q_c$ at
which the purely dissipative poles disappear, multiplied with
several length scales. } \label{impoles}}

On the right in figure \ref{impoles}, we compare $\tlq_c$ with the
various length scales in the problem, as we did before for the
spacing of the quasiparticle masses in figure \ref{fpoleplot}. Since
we are in the completely opposite regime in terms of length scales
of the perturbations, it is no surprise that there is significant
disagreement between the scaling of $\tlq_c$ and $\tlnu_0$, but the
disagreement is surprisingly small. In addition to the opposite
limit of the size of the perturbations, the data in figure
\ref{fpoleplot} contains only frequencies, which one can interpret
as being related directly to the width of the defect, whereas here
we consider the $f$-dependence of relevant values of $\tlq$, which
measure scales along the defect. Overall, it seems that in the limit
of large $f$, $\tlq_c\simeq2/(3\,D)$ or $\tlq_c\simeq4T_{eff}/3$.
The relative factor between these two expressions is not surprising
given that, in the previous section, we noted that $D = 2/T_{eff}$
as $f\to \infty$. Further, given our previous expressions for $D$
and $T_{eff}$, we note that $\tlq_c\propto T/\sqrt{f}$ for large
$f$, \ie $\tlq_c$ decreases as $f$ grows. Then as the plot shows, up
to an overall numerical factor, most features of the $f$-dependence
of $\tlq_c$ can be related to either of these other physical scales.

In principle, the decreasing residue of the poles with increasing
$n$ allows us to track the location of the first few poles of the
longitudinal correlator even further, directly by fitting a sequence
of Lorentzians, but we will not bother about such a detailed
discussion of the hydrodynamic regime in this paper. It is
interesting to see however, how the small-$\tlq$ limit of the shift
$\delta \tlnu +  i \delta \tlgam$ shown in figure \ref{kpoleplot}
qualitatively agrees with a shift towards the bifurcation point.

It is interesting to note that this pairing of the diffusion pole
with a fast dissipative mode was also recently found in the
quasinormal mode spectrum of black holes in AdS$_4$ \cite{ads4}.
However, an infinite number of pairs of poles were identified there,
appearing along the imaginary axis. In that case, the critical
wave-number at which the higher pairs meet at smaller and move off
into the complex plane decreases for pairs higher up along the
imaginary axis. We looked for similar higher dissipative modes in
the present framework but it seems that the diffusion mode and its
partner are the only modes appearing on the imaginary frequency
axis.

\section{Electromagnetic duality and perturbative corrections}\label{edualmain}

At the outset of our analysis, we set $\mq=0$ to maintain conformal
invariance in the defect system. In the brane construction, this
means the internal geometry is fixed and the low energy effective
action on the effective four-dimensional brane reduces to Maxwell
theory \reef{gaugeAction} with a fixed coupling (independent of the
radius).\footnote{In the D5-brane embeddings for $\mq\ne0$, the size
of the internal $S^2$ varies and so the effective coupling of the
Maxwell theory \reef{gaugeAction} depends on the radius. As
explained in \cite{pavel}, the gauge field equations are no longer
duality invariant and as a result the correlators discussed here are
independent.} Hence resulting equations of motion are invariant
under electromagnetic duality, which has interesting implications
for the transport coefficients, as emphasized in \cite{pavel}.

Given the Maxwell action \reef{gaugeAction}, the gauge field
equations can be expressed as
 \beq
\nabla^\mu F_{\mu\nu}=0\,,\qquad
\nabla^\mu\widetilde{F}_{\mu\nu}=0\qquad {\rm with}\
\widetilde{F}_{\mu\nu}=\frac{1}{2}\,\varepsilon_{\mu\nu\rho\sigma}
F^{\rho\sigma}\ . \labell{eomA}
 \eeq
Hence we have electromagnetic duality with $F_{\mu\nu}$ and
$\widetilde{F}_{\mu\nu}$ satisfying the same equations of motion.
Implicitly, we used this duality in deriving the relation between
the transverse and longitudinal correlators \reef{translongreln},
\ie the key step was demonstrating the $A_\tlt$ and $A_\ty$
equations, \reef{gauge2} and \reef{gauge4}, could be put in the same
form. As in \cite{pavel}, this result \reef{translongreln}
subsequently restricts the transport coefficients to satisfy
\beq \Pi^T(\tom,\tlq)\,\Pi^L(\tom,\tlq)=-\varepsilon_0^2 (\tom^2-\tlq^2)
\ .\labell{constraint1}\eeq
Since with $\tlq=0$, we have $\Pi^T(\tom,0) = \Pi^L(\tom,0)$, it
follows that:
 \beq
\ts(\tom)=i\frac{\Pi(\tom,0)}{\tom}=  \varepsilon_0 = \pi \,D\,T \,
\varepsilon\,   ~~~ \mathrm{or} ~~~ \sigma(\omega) = D\,\varepsilon
\,. \labell{conduct} \eeq
That is, $\sigma(\omega)$ is independent of frequency and
temperature. One can show that this remarkable result is consistent
with the Einstein relation,\footnote{See e.g. \cite{Chaikin},
section 7.4 for a suitable discussion of the Einstein relation.} as
noted already in \cite{pavel}.

However, as for any low energy action in string theory, we must
expect that there are higher derivative interactions correcting the
Maxwell action \reef{gaugeAction}. In fact, the action \reef{act5}
implicitly captures an infinite set of these stringy corrections, as
would be illustrated if we expanded the DBI term in powers of $F$.
This expansion would also demonstrate that these higher order terms
are suppressed by factors of $\alpha'=\ls^2$. In terms of the dual
CFT, the contributions of these $\alpha'$ interactions will provide
corrections to the leading supergravity results for a finite 't
Hooft coupling. However, none of the higher order terms coming from
the DBI action will modify the two-point correlators in the planar
limit, \ie in the large $\nc$ limit, because these interactions all
involve higher powers of the field strength. One must keep in mind
though that, as already alluded to in section \ref{d7-geom}, the DBI
action does not capture all of the higher dimension stringy
interactions. The full low-energy action includes additional terms
involving derivatives of the gauge field strength
\cite{Kitazawa,koerber}, as well as higher derivative couplings to
the bulk fields, \eg curvature terms \cite{rsquared,bad}. In
principle, any such interaction, which is quadratic in $F$, has the
potential to make finite $\lambda$ corrections to the correlators
which we have studied above.

In appendix \ref{appcorr}, we identified a particular higher
derivative term which makes a quadratic correction
\reef{gaugeActionmod1} to the four-dimensional low energy action.
This term makes the leading correction to the correlators, at least
when the internal flux is nonvanishing. Including this term, the
vector equations of motion become
 \beq
\nabla^\mu F_{\mu\nu}= \xi \, L^2\, \nabla^\mu\Box
F_{\mu\nu}\,,\qquad \nabla^\mu\widetilde{F}_{\mu\nu}=0\qquad {\rm
with}\ \xi= 
\frac{\zeta(3)}{2\pi \sqrt{\lambda}}\frac{ f^2}{
\sqrt{1+f^2}}\ . \labell{eomA1}
 \eeq
We can recognize the higher derivative term as a string correction
by recalling that $L^2/\sqrt{\lambda}=\ls^2$. Clearly these
equations are no longer invariant under the replacement: $F_{\mu\nu}
\rightarrow \widetilde{F}_{\mu\nu}$. One could attempt an
$\alpha'$-corrected electromagnetic duality by defining
$\widetilde{X}_{\mu\nu}=\left(1-\xi\, L^2 \Box\right)F_{\mu\nu}$.
Formally treating $\xi$ as a small expansion parameter, one can
rewrite \reef{eomA1} as
 \beq
\nabla^\mu X_{\mu\nu}=-\xi\,L^2\,\nabla^\mu\Box X_{\mu\nu}\,,\qquad
\nabla^\mu\widetilde{X}_{\mu\nu}=0\ . \labell{eomA2}
 \eeq
Hence an exchange $F_{\mu\nu}\rightarrow X_{\mu\nu}$ does not quite
leave the equations of motion invariant either, \ie the sign of the
$\xi$ term changes between \reef{eomA1} and \reef{eomA2}. This
then confirms the initial intuition that the $\alpha'$-corrected low
energy theory describing the four-dimensional dynamics of the vector
field is no longer invariant under electromagnetic duality. Hence we
can longer expect \reef{constraint1} and \reef{conduct} to apply
when finite $\lambda$ corrections are taken into account for the
defect CFT.

In the following, we examine in more detail the effect of this
leading finite $\lambda$ correction. For
simplicity,\footnote{Similar considerations apply to the
longitudinal conductivity but the calculations are somewhat more
involved.} we focus on the modifications to the transverse
conductivity $\ts_{yy}$. Here we simply present the results of our
numerical calculations. The preliminary analysis determining the
analytic form of the transverse correlator is given in appendix
\ref{appcorr}. There are two distinct contributions to the
modification of the correlator. First, since the bulk action
contains an additional term, there are new surface terms
\reef{newer2} which must be evaluated in the holographic
calculation. Remarkably, as described in the appendix, the net
effect of this contribution is to shift the permittivity
 \beq\labell{permmodm}
\varepsilon_0 \to \varepsilon_0 \left(1 -
\frac{1}{\sqrt{\lambda}}\frac{ f^2\, \zeta(3)}{\pi(1+f^2)^{3/2}}
\right) = \varepsilon_0 \left(1 - \frac{2\,\xi}{1+f^2} \right) \, .
 \eeq
\FIGURE{\includegraphics[width=0.49\textwidth]{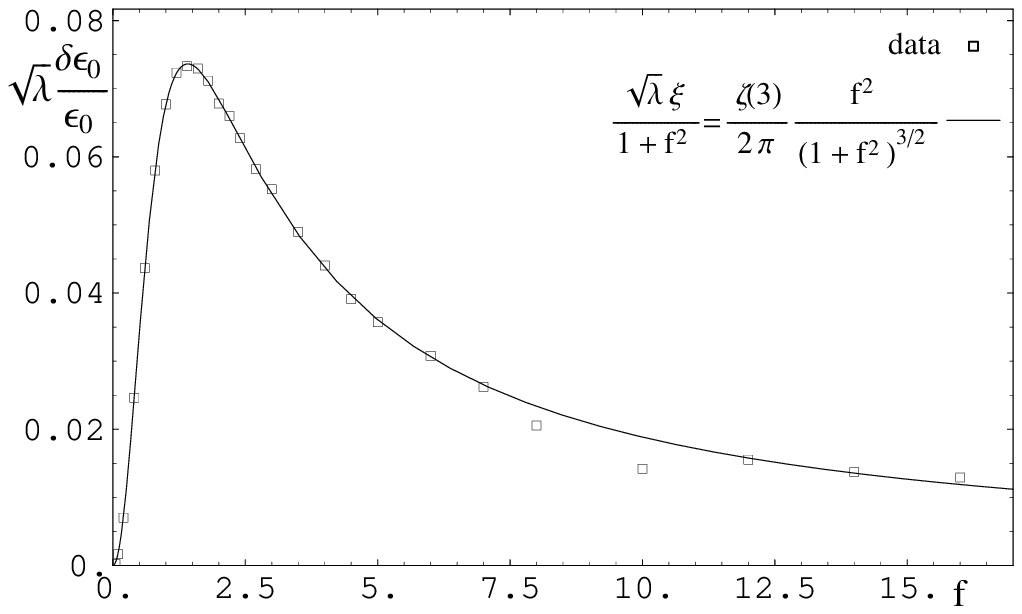}
\caption{The modification of $\varepsilon_0\equiv
\ts(\tom\rightarrow\infty)$ as a function of $f$ from the
$1/\sqrt{\lambda}$ corrections.}\label{epsmodf}}
\noindent The second modification of the correlator arises because
the bulk equations of motion have been corrected, as in
\reef{eomA1}. Hence the solutions for the vector are modified and
this change of the solution alone leads to changes in the correlator
coming from the leading supergravity expression \reef{green}. Now,
we have some ambiguity in how we might define $\varepsilon_0$ in the
theory with finite-$\lambda$ corrections. Recall that this quantity
originally appeared in \reef{flux} but above was simply related to
the conductivity \reef{conduct} found in the infinite $\lambda$
limit. Hence a convenient choice, which we adopt at finite
$\lambda$, is: $\varepsilon_0\equiv\ \ts(\tom\rightarrow\infty)$.
Then our numerical results indicate that this second
finite-$\lambda$ correction also shifts $\varepsilon_0$ precisely as
in \reef{permmodm} except for a factor of $-3/2$. The total shift is
shown in figure \ref{epsmodf} and the result seems to match
precisely $-1/2$ times the shift given in \reef{permmodm}.

Given that the invariance under electromagnetic duality is lost at
finite $\lambda$, the frequency independence of the conductivity
$\ts(\tom)=\ts_{yy}(\tom,\tlq=0)$ found in \reef{conduct} is also
lost as shown in figure \ref{corrk0}. Note that here we are plotting
the change arising from the inclusion of the finite $\lambda$
corrections, \ie $\delta\ts(\tom) = \ts(\tom) - \varepsilon_0$ where
our subtraction includes the finite-$\lambda$ correction to
$\varepsilon_0$, as described above. Note that in the figure, the
factor $1/\xi \propto \frac{\sqrt{1+f^2}}{f^2}$ is included to
cancel the $f$ dependence coming from the factor in front of the
higher order term in \reef{maxwellmod}. While the resulting
conductivity shows an oscillatory behaviour, we note that the DC
conductivity, \ie at $\tom=0$, is generally smaller than at high
frequencies, \ie for $\tom\rightarrow\infty$. The net difference is
plotted in figure \ref{corrdropf}, as a function of $f$. As shown,
the numerical results are very well fit with a simple analytic form
proportional to $f^2/(1+f^2)$. Note that the first few points in
this plot (including where the difference becomes positive) are not
reliable, because of the high sensitivity to errors in $\delta
\varepsilon_0$, which was only computed approximately in a numerical
calculation.
\DOUBLEFIGURE{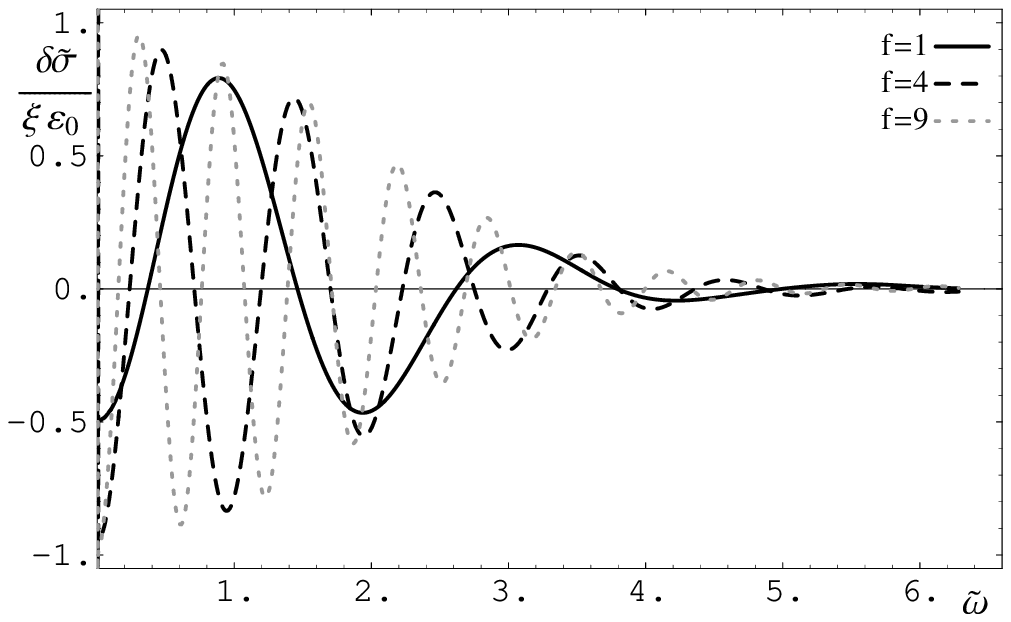,width=0.49\textwidth}{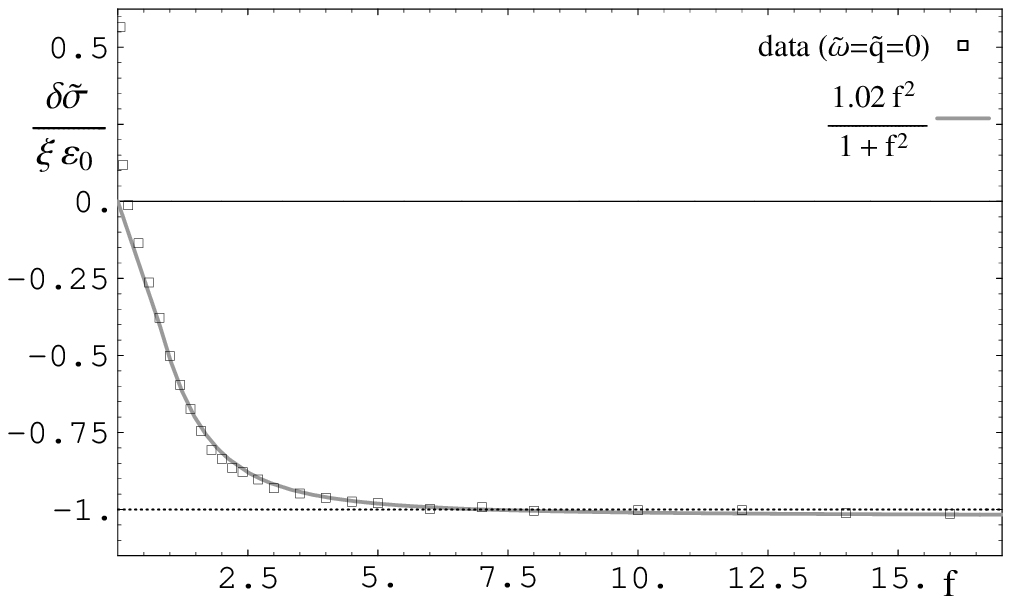,
width=0.49\textwidth}{The finite-$\lambda$ correction to the
conductivity $\delta\ts(\tom)$ for various values of $f$.
\label{corrk0}}{$\ts(\tom=0) - \ts(\tom\rightarrow\infty)$, the
change in conductivity from $\tom=0$ to $\tom\rightarrow\infty$ at
finite $\lambda$. \label{corrdropf}}

One of the interesting features that figure \ref{corrk0} seems to
exhibit is that the oscillations of $\ts(\tom)/(\xi\varepsilon_0)$
for various values of $f$ are all contained within some universal
envelope, that is decaying with $\tom$. In fact, this same envelope
also applies for the conductivity at finite values of $\tlq$, as
illustrated in figure \ref{correnvf}. In this figure, we are showing
${\delta \ts_{yy}}/(\varepsilon_0 \xi)$ for $f\in\{1,4,9,16\}$ and
$\tlq \in \{0,\pi/4,\pi/2,\pi,2\pi\}$, where $\delta
\ts_{yy}\equiv\ts_{yy}(\tom,\tlq)-
\ts_{yy}(\tom\rightarrow\infty,\tlq)$. Again, the factor $1/\xi$ is
included in the figure to cancel the $f$ dependence explicitly
appearing in the higher order term in \reef{maxwellmod}. In
particular, we see here that the envelope appears to be independent
of $\tlq$. However, as shown in figure \ref{smallfdel}, for
sufficiently small $f$ ($f=1/4$ in the figure), there exists a
critical value of $\tlq$, above which the conductivity is no longer
bounded by this universal envelope. Note that for the same values of
$f$, the curves below the critical value of $\tlq$ are still bounded
by the envelope. However, note that both $\xi,\varepsilon_0\,\propto
f$ for large $f$ and so the amplitude of oscillations in $\ts(\tom)$
alone is actually growing with $f$.
\DOUBLEFIGURE{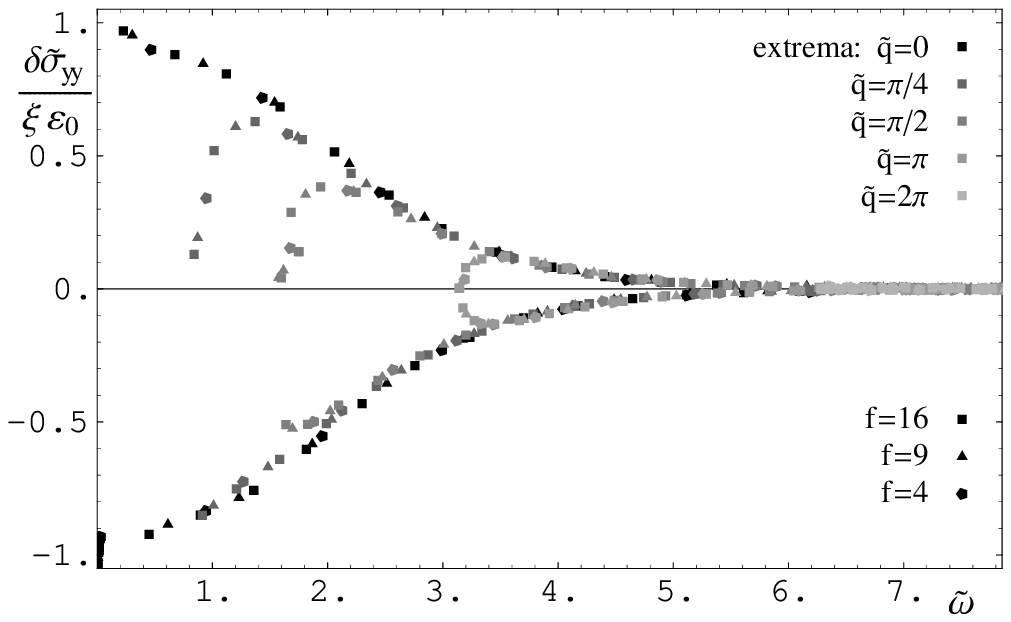,width=0.49\textwidth}
{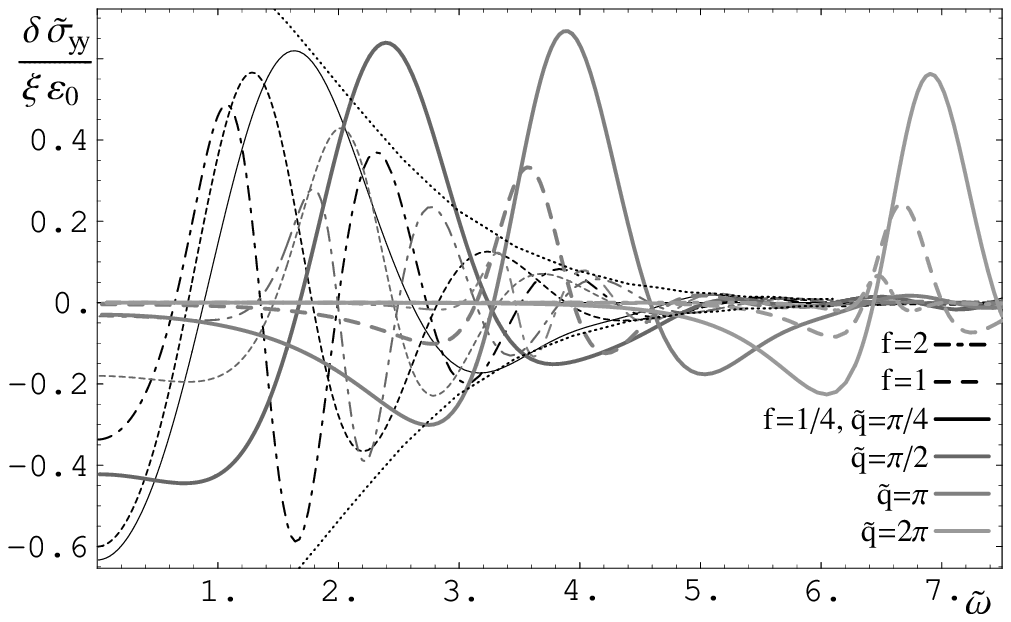,width=0.49\textwidth} {The extrema of
$\frac{1}{\varepsilon_0 \xi}\delta\ts_{yy}$ for $\tlq \in
\{0,\pi/4,\pi/2,\pi,2\pi\}$ and ``large'' $f\in \{4,9,16\}$. We see
that all the curves are approximately bound by some universal
envelope function, that decays exponentially in $\tom$.
\label{correnvf}} {$\frac{1}{ \varepsilon_0 \xi}\delta\ts_{yy}$ for
$\tlq \in \{\pi/4,\pi/2,\pi\}$ and $f\in \{1/4,1\}$. For $f=1/4$ and
sufficiently large $\tlq$, the conductivity is not bounded by the
universal envelope (shown in thin dots). \label{smallfdel}}

At finite $\tlq$, the leading order result (for infinite $\lambda$
and $\nc$) for $\ts_{yy}$ also exhibited similar damped oscillations
which were confined within a certain envelope, as discussed in
\ref{resdef}. Comparing this previous envelope with that for
$\delta\ts_{yy}$ (for large $f \gtrsim 2$), we see that the previous
one does not depend only on $\tom$, in contrast to the behaviour
found above. The exponential decay of the amplitude at large $\tom$
is also slower than here than with the envelope for the leading
infinite-$\lambda$ result. This would imply that the
finite-$\lambda$ corrections become more and more significant at
large $\tom$, while they become less significant with increasing
$\tlq$.

The ``frequency'' of the oscillations is approximately the same for
the leading term and the finite-$\lambda$ correction. Figure
\ref{corrresf} shows this in more detail by plotting
$\frac{\sqrt{\lambda} }{ \varepsilon_0 }\delta \ts_{yy}$ in terms of
$\sqrt{\tom^2 - \tlq^2}$ and also for comparison the oscillations of
the infinite-$\lambda$ or ``zero'th order'' result.

\FIGURE{\includegraphics[width=0.49\textwidth]{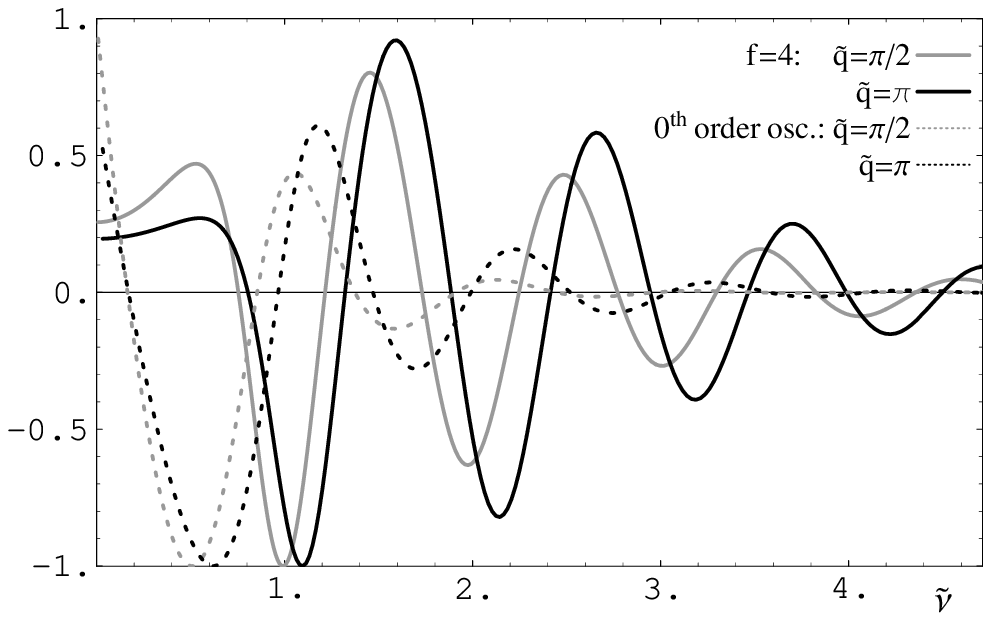}
\includegraphics[width=0.49\textwidth]{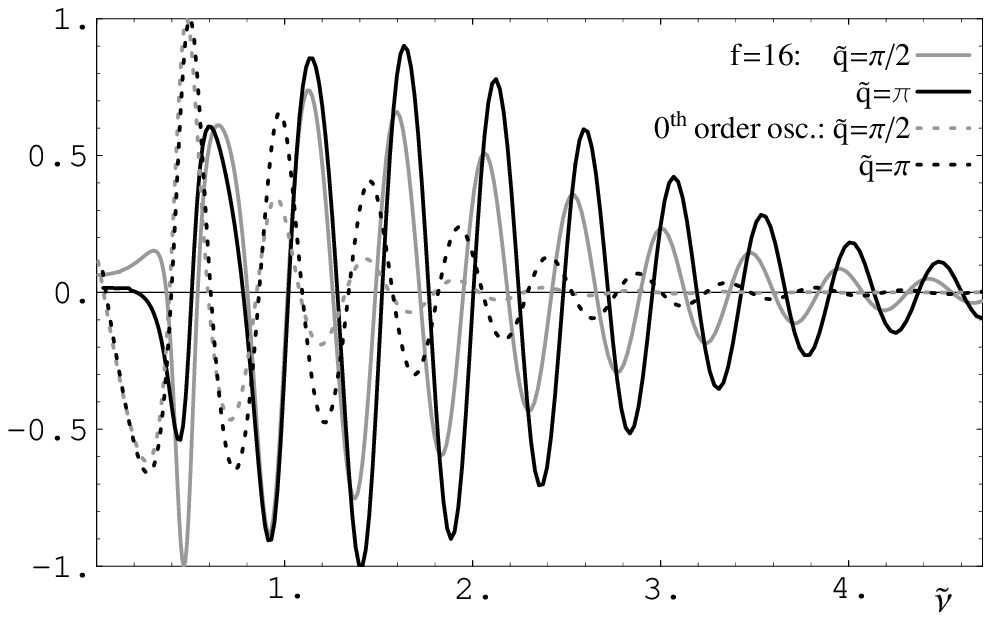}
\caption{Left:$\sqrt{\lambda}\delta\sigma_{yy}$
in the rest-frame frequency $\tlnu$ for $\tlq \in \{\pi/2,\pi\}$
and $f=4$, scaled to $1$. The dotted lines show
$\sigma_{yy}-\sqrt{1-\tom^2/\tlq^2}$, also scaled to 1.
Right: $f = 16$}\label{corrresf}}

We find that there is a phase shift of between $\pi/2$ and $\pi$ in
the oscillations, implying that they will shift towards larger
$\tom$ and decrease in amplitude. We can also see that there is some
tendency for a smaller phase shift (i.e. less/no decrease in
amplitude, less shift) as $f$ and $\tom$ increase. In terms of the
location of the poles, this implies a shift towards larger real and
imaginary frequencies and an increased spacing between the
quasinormal modes, again more significant for large $f$, small
$\tlq$ and large $\tom$. Another point to view this is that there is
a finite-$\lambda$ behaviour, that becomes important for the higher
resonances. In principle one could quantify this more precisely by
doing a perturbative treatment of the methods used in section
\ref{quasinorm}, but we will not discuss this here. The shift
$\varepsilon_0$ can be absorbed into the residue. In terms of the
potential, this implies that the potential becomes narrower,
especially at small $f$ and small $\tlq$ (or large $T$), which
simply means that the length scale that we attributed to the strong
coupling decreases and disappears. The $\tom$ dependence also
implies that the potential becomes smoother at finite coupling.

\FIGURE{\includegraphics[width=0.49\textwidth]{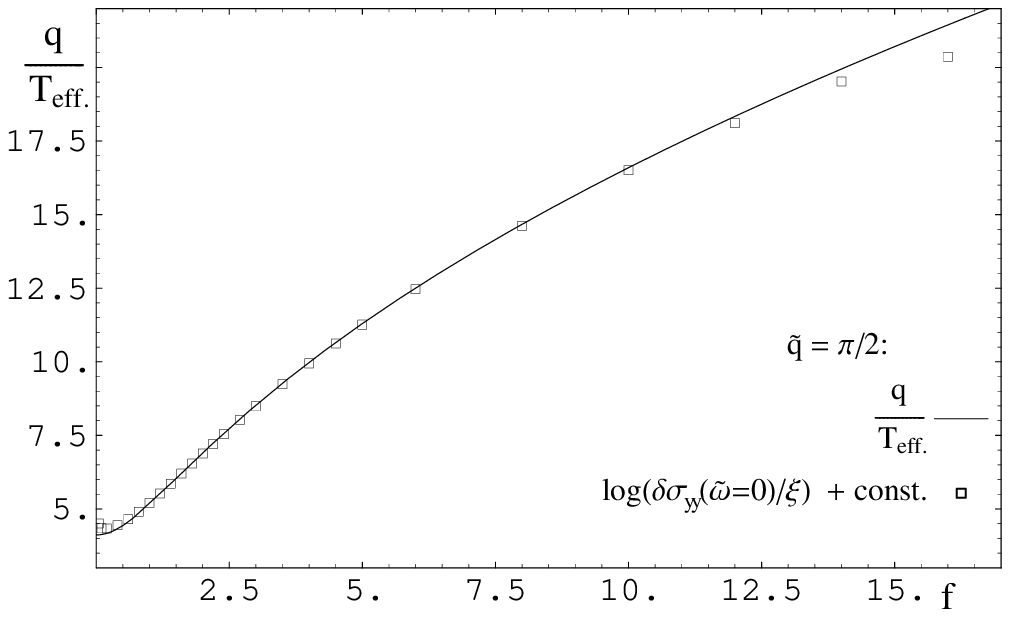}
\includegraphics[width=0.49\textwidth]{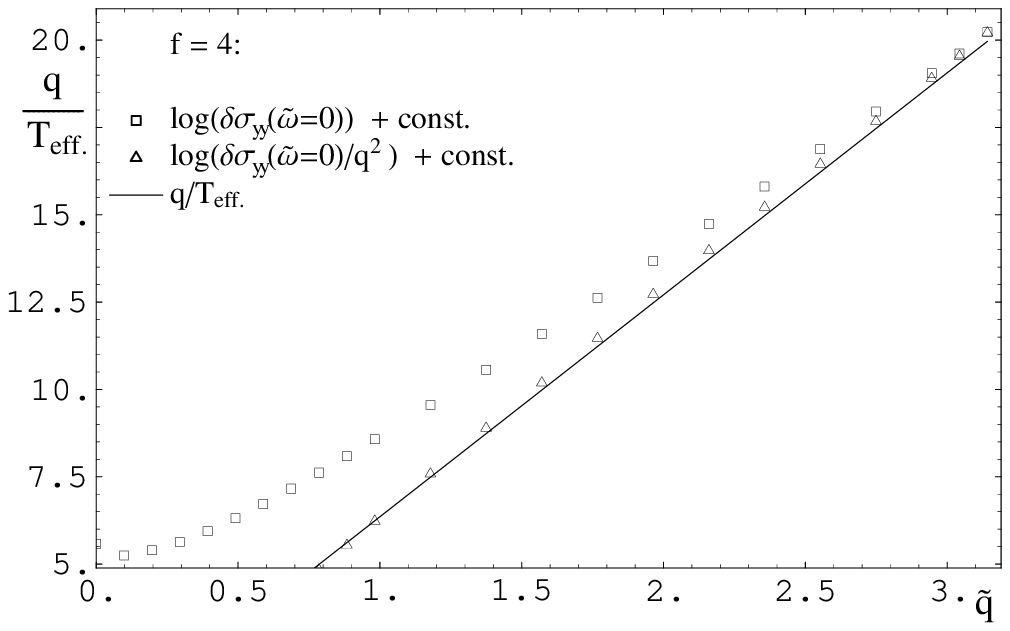}
\caption{Left: $\log\, \delta
\tilde{\sigma}_{yy}(\tom = 0) /\xi$ (modulo a constant shift) and
$\frac{\tlq}{T_{eff}(f)}$ for $\tlq = \pi/2$. The accuracy of the
last two points are very sensitive to possible errors in $\delta
\varepsilon_0$. Right: $\log\, \delta \tilde{\sigma}_{yy}(\tom =0)$
and $\frac{\tlq}{T_{eff}(f)}$ for $f=4$. $\log\, \delta
\tilde{\sigma}_{yy}(\tom= 0) /\tlq^2$ is shown to demonstrate the
slow convergence due to polynomial factors of
$\tlq$.}\label{corrdropfsupp}}

For $\tom - \tlq < 0$, the finite-$\lambda$ correction becomes
quickly negative and exponentially suppressed with increasing $f$
and $\tlq$, roughly as described by the ``effective temperature'',
such that the exponential suppression does not get broken but is
possibly modified. We show the shift
$\frac{\delta
\tilde{\sigma}_{yy}(\tom = 0)}{\varepsilon_0 \xi} $
as a function of $f$ for $\tlq = 0$
and a function of $\tlq$ for $f = 1$ in figure \ref{corrdropfsupp}.
Recall that figure \ref{corrdropf} shows the same results for $\tlq
= 0$.

The form of $\delta \ts_{yy}$ for $f\lesssim 2$ is more similar to
the resonances associated with the infinite-$\lambda$ result for
$\ts_{yy}$, as the amplitude seems to decay exponentially with
$\tlnu$ and depends only polynomially on $\tlq$, as we show in
figure \ref{smallfdel}. Just as for the large $f$, the decay is
slower than the one in the $0^{th}$ order term. This demonstrates
that the effects of finite $\lambda$ are more significant for small
$f$, where the length scales are still dominated by the $f=0$ length
scale, that we can attribute to the strong coupling. Further it
shows that the length scale due to the ``width'' of the defect has a
tendency to persist.

\section{Hall conductivity} \label{hall}

The conductivity in section \ref{form} is diagonal reflecting the
parity invariance of the defect theory. Recently AdS/CFT techniques
were applied to study Hall conductivity in the three-dimensional
conformal field theories dual to an AdS$_4$ background
\cite{hallo,hall1}. The construction in \cite{hallo} involved
breaking the parity invariance by introducing a background magnetic
field. Of course, these calculations with a dyonic black hole could
be easily emulated here by introducing additional background gauge
fields on the AdS part of the  probe branes --- this would be a
relatively straightforward extension of the analysis in, \eg
\cite{findens}. In \cite{hall1}, parity invariance is broken by the
introduction of an auxiliary gauge field with a nonzero
$\theta$-term. This construction is closely related to the following
where we produce an off-diagonal conductivity by the addition of a
topological $\theta$-term to the four-dimensional SYM action
\cite{sl2z}. A related model of the quantum hall effect based on a
probe brane construction appears in \cite{quantumhall}.

To introduce an $xy$ component to the conductivity, we begin by
considering the Chern-Simons part of the D5-brane action. In
particular, the latter includes the following term:
\beq \Delta I= \frac{(2\pi\ell_s^2)^3}{3!}\nf\,T_5\int
C^\mt{[0]}\,F\wedge F\wedge F \labell{wzterm}\eeq
where $C^\mt{[0]}$ is the RR scalar. Now the background
\reef{D3geom} remains a consistent solution of the type IIB
supergravity equations if we set this scalar to some arbitrary
constant, \ie $C^\mt{[0]}=a$. Of course, this choice corresponds to
adding a topological $\theta$-term to the action of the dual SYM
theory \cite{sl2z}. Now if we recall the magnetic flux
\reef{magflux} on the internal $S^2$, the above contribution
\reef{wzterm} reduces to the following four-dimensional action
\beq \Delta I=(2\pi)^4\ell_s^6\, T_5\,2\pi\,a\,q\int F\wedge F\,,
\labell{chern2}\eeq
where $q$ is the magnetic flux quantum number \reef{magflux},
indicating the number of D3-branes dissolved into the D5-brane. Thus
upon integrating out the $S^2$ part of the probe brane geometry,
this term \reef{wzterm} has become a topological $theta$-term for
the four-dimensional worldvolume gauge fields. Since it is a
topological term, it does not modify the equations of motion
(\ref{gauge1}--\ref{gauge4}) for the gauge field. However, it does
produce an additional boundary term,
\beq \Delta I=2(2\pi)^4\ell_s^6\, T_5\,a\,q\int d^3\sigma\,\left[
A_y\left(\partial_t A_x-\partial_x A_t\right)\right]_{u\to0^+}\,,
\labell{chern3}\eeq
which will modify the correlators. Note that we have simplified the
above expression by assuming that in the cases of interest (as in
previous sections) the gauge fields are independent of $y$.
Introducing the Fourier transform \reef{fourier}, this boundary term
becomes
\beq \Delta I=-i4\pi\ell_s^6\, T_5\,a\,q \int d^3 k \,\left[
A_y(-k,u)\left(\omega\, A_x(k,u)+k_x\,
A_t(k,u)\right)\right]_{u\to0^+}\,. \labell{chern4}\eeq
Now following the same steps as in section \ref{goldfinger}, we
arrive at the following off-diagonal contributions to the retarded
Green's function:
\beq C_{xy} = i 8 \pi (2\pi\ell_s^2)^3 T_5\, a\,q \,\omega \,,
\qquad
C_{ty} = -i 8 \pi (2\pi\ell_s^2)^3 T_5\, a\,q \, k_x\,.
\labell{offd} \eeq
Note that in the $T=0$ limit, we expect this contribution to the
Green's function can be assembled in the Lorentz invariant
expression:
\beq \Delta C_{\mu\nu}=i\, \alpha\,\varepsilon_{\mu\nu\rho}\,k^\rho
\,,\labell{dG2}\eeq
where $\alpha$ is the dimensionless constant:
\beq \alpha=8 \pi (2\pi\ell_s^2)^3 T_5\, a\,q  = \frac{2\,a\, q}{\pi
\gs} = \frac{8\, a\, q}{\gym^2} \,. \labell{const98}\eeq
The corresponding analysis in the D7 framework gives
\beq \alpha_\mt{7}=16 \pi^2 (2\pi\ell_s^2)^4 T_7\, a\,\cQ = \frac{2
a\, \cQ}{\pi \gs} = \frac{8\, a\, \cQ}{\gym^2} \,.
\labell{const98D7}\eeq
While in principle, this form \reef{dG2} need not be preserved at
finite temperature, our results \reef{offd} calculated at finite $T$
indicates that in fact the form is preserved. Of course, this
independence of the temperature is undoubtedly related to the
topological nature of the $\theta$-term which is responsible for
this off-diagonal contribution. It is amusing to note that since $q$
(and $\cQ$) is an integer,  \reef{const98} and \reef{const98D7} take
just the form of the integer quantum Hall effect, \ie $\sigma_{xy} =
\frac{n \, e^2}{2 \pi}$ with $n \in \mathbb{Z}$. By this analogy, we
would associate $e^2 = {4 a}/{\gs}$.

\section{Discussion} \label{discuss}

In this paper, we used holographic techniques to investigate the
transport properties of certain defect CFT's. In particular, we
studied matter on a $(2+1)$-dimensional defect emersed in a heat
bath of $(3+1)$-dimensional $N=4$ super-Yang-Mills plasma. Our
analysis covers two distinct defect CFTs. The first was realized by
embedding $\nf$ probe D5-branes in AdS$_5\times S^5$,  as described
in section \ref{d5-geom} and in this case, the system (at $T=0$)
preserves eight supersymmetries. The second system involves
embedding $\nf$ probe D7-branes in the AdS$_5\times S^5$ background
and the resulting defect CFT preserves no supersymmetries. In both
cases, the theory could be deformed by introducing an additional
internal flux on the probe branes. In the dual CFT, the defect then
separated regions where the rank of the SYM gauge group was
different. As described in section \ref{d7-geom}, this flux was
crucial to remove an instability which would otherwise appear with
the D7-brane construction. Perhaps surprisingly, the transport
properties of both defect CFT's were essentially identical.

Overall, our analysis revealed the expected diffusion-dominated
hydrodynamic limit at small wave-numbers and we found a smooth
crossover to a collisionless regime at the large wave-numbers. In
the latter regime, the defect theory exhibits a conduction
threshold, given by the wave-number $q$ of the current and the
system is approximately described only in terms of the ``rest-frame
frequency'' $\nu = \sqrt{\omega^2 - q^2}$.

In many respects, our results coincided with those in \cite{pavel},
where holographic techniques were used to study a purely
$(2+1)$-dimensional system with sixteen supersymmetries. Hence
maximal supersymmetry (or supersymmetry, in general) does not seem
to be a key feature for producing the interesting behaviour of these
holographic models. Instead many properties seem to emerge from the
infinite $\nc$ and infinite $\lambda$ limits, that are implicit in
making a supergravity analysis of the AdS dual. In section
\ref{edualmain}, we elucidated one such effect that arises purely
from the large-$\nc$ and large-$\lambda$ limits, namely the
frequency independence of the conductivity, $\ts(\tom)$. In
\cite{pavel}, this effect was described as a consequence of the
electromagnetic duality of the gauge theory giving the dual
description of the CFT currents. We were able to explicitly show
that this duality is lost when stringy corrections are included in
the worldvolume gauge field action and explicitly calculated the
frequency dependence in $\ts(\tom)$ arising from the corresponding
finite-$\lambda$ corrections to the conductivity. As described in
appendix \ref{appcorr}, one can well imagine that there will be
other interactions which, although they appear to be of higher order
in the $\alpha'$ expansion, provide further corrections to the
conductivity $f^n/\sqrt{\lambda}$, where $n>2$. Hence, our results
in section \ref{edualmain} are only the leading corrections when $f$
is small but finite. There will also be curvature interactions to
the worldvolume action of the probe branes \cite{rsquared,bad}.
These will also produce finite-$\lambda$ corrections but in contrast
to the previous discussion, the latter will not be enhanced by
factors of $f$ and first appear only at order $1/\lambda$. While a
completely consistent set of curvature terms is not known at this
time, the effect of certain representative terms was considered in
\cite{matt}.

While certain aspects of charge transport were similar for the
present defect CFT's and the maximally supersymmetric CFT studied in
\cite{pavel}, we also found some profound differences. The most
prominent is the dependence of our results on the internal flux $f$,
certainly a difference since no such parameter exists in the
maximally supersymmetric case. For example then, with a large $f$,
strong quasiparticle peaks appeared in the spectral functions and
conductivities. Similarly, certain phenomena in the defect CFT
seemed to be controlled by a new dynamically generated scale in this
regime, \ie a scale distinct from the temperature $T$. We denoted
this scale as the effective temperature $T_{eff}$ in section
\ref{efftf2}. For small $f$, $T_{eff}\sim T$ to within factors of
order one. However, for large $f$, one finds that $\pi T/(2
T_{eff})\simeq k\,\sqrt{f}$ where $k=4 \Gamma(5/4)^2 / \sqrt{\pi}$,
as shown in \reef{teffbig}. While this seems a curious way to
present the ratio of $T$ and $T_{eff}$, it was found in section
\ref{colll} that precisely this combination played a role in fixing
the spacing of the quasiparticle poles. Further, as also noted
there, precisely the same behaviour was found at large $f$ for the
diffusion constant: $\pi\, D\,T\simeq k\,\sqrt{f}$.

A more intuitive picture as to the origin of this dynamical scale
comes from considering the probe brane geometry, as in section
\ref{revGeom}. Recall that when $f$ is nonvanishing, the defect
separates a region where the SYM has gauge group $U(\nc+q)$ from one
where the gauge group is $U(\nc)$. However, on the side where the
rank of the gauge group is enhanced, the defect also excites a
noncommutative configuration of adjoint scalars in a $U(q)$ subgroup
of the full $U(\nc+q)$. At $T=0$, this configuration extends to
infinity with $\Tr(\Phi^2)={q^2\over 4\nf}\,{1\over z^2}$. In
particular, this configuration preserves the conformal symmetry, \ie
does not introduce a new scale. The scalar profile is reflected in
the radial profile of the probe D-branes which also extends out to
$z\rightarrow\infty$ when $T=0$ and $f$ is nonvanishing. However,
when the temperature is nonvanishing, the probe branes fall into the
horizon at a finite value of $z$. For large $f$, one finds
 \beq
 z_{max}\,T = k\,\sqrt{f}\,,\label{zmax}
 \eeq
where $k$ is precisely the same constant found above. The natural
interpretation of this profile is that at finite temperature,
thermal fluctuations decohere the scalar fields at some distance
away from the defect. That is, at finite $T$, the defect can only
excite a coherent configuration of scalars out to $z_{max}$.
However, at small $f$, $z_{max}$ vanishes whilst the dynamical length scale doesn't vanish. We can interpret that by some extra contribution to the width that arises from excitations in the bulk fields that are induced at strong coupling from the presence of the charged matter degrees of freedom on the defect.

One interpretation then is that the defect effectively acquires a
finite width when $T$ is nonvanishing. This intuitive picture may
seem more reasonable, if we recall that the system is at (extremely)
strong coupling and so any probe exciting of the defect fields will
actually excite a complicated mixture of defect fields and ``bulk"
SYM fields. This picture of finite-width for the defect seems to
match well with the results for the quasiparticle spectrum on the
defect. In particular, we found that both the conduction threshold
and the resonance peaks are well-described by a quasiparticle
``tower'' with equally spaced ``rest frame'' energy and constant
``mass to inverse lifetime ratio''. The length scale that is suggested from this spectrum is very similar to $\pi D T$ and $\pi T/(2 T_{eff})$ (and at large $f$ also similar to $z_{max}$) plus a small constant.

In terms of the effective
Schr\"odinger analysis, \eg see appendix \ref{tanhpot}, the
quasinormal spectrum arises in the gravity side from interference on
a potential barrier in the asymptotic region. From the profile of
the brane, this translates into an interference or resonance in the
region around the defect. These two dual pictures of the origin of
the spectrum seem orthogonal. The intuitive CFT picture involves a
width and implicitly, a potential, in the space transverse to the
defect, while the effective Schr\"odinger analysis constructs an
effective potential in the radial or ``energy scale'' direction. It
would certainly be interesting to have a clearer connection between
these two descriptions.

The hydrodynamic and collisionless regimes are cleanly separated at
a critical wave-number where the diffusion pole disappears, as
observed in section \ref{diffuse}. There, we found that the
diffusion mode is partnered with another dissipative mode on the
imaginary axis. These two poles coalesce at the critical wave-number
and move out into the complex plane for larger $q$. Hence
precisely at the critical wave-number $q_c$, the corresponding thermal
correlator will exhibit a curious double pole on the imaginary axis. Interestingly, $q_c$ has a similar qualitative $f$ dependence as the other (inverse) length scales and is also quantitatively close, as $q_c /(\pi T) \sim 0.67/( k \sqrt{f})$. This supports again the concept that the properties of the defect are controlled by one fundamental length scale, that can be related to the effective width of the defect.

In section \ref{hall}, we outlined a topological Hall conductivity,
which is induced when the defect is coupled to the SYM gauge theory
with a topological $\theta$-term. Of course, it would also be
interesting to study the Hall conductivity induced by a background
magnetic field on the defect. Other interesting directions would be
to study the effect of a finite mass for the defect fields or of a
finite chemical potential. With these additional parameters, there
are many non-trivial physical properties of the defect related to
condensed matter physics that one can study through the transport
properties. We will address these questions in an upcoming paper
\cite{matt2}.

\acknowledgments We thank Cliff Burgess, Jaume Gomis, Sean Hartnoll,
Matt Headrick, Gary Horowitz, Pavel Kovtun, Volodya Miransky, Markus
Muller, Andrei Parnachev, Subir Sachdev and Aninda Sinha for helpful
discussions and useful comments. Research at the Perimeter Institute
is supported in part by the Government of Canada through NSERC and
by the Province of Ontario through MRI. We also acknowledge support
from an NSERC Discovery grant and from the Canadian Institute for
Advanced Research. We would also thank the Kavli Institute for
Theoretical Physics for hospitality during the initial stages of
this project. Research at the KITP was supported in part by the NSF
under Grant No. PHY05-51164.

\appendix

\section{Diffusion constant on the defect}\label{diffusion}

The worldvolume gauge field corresponds to a conserved current on
the defect in the dual CFT. In the hydrodynamic regime, one then
expects to see the diffusion of the conserved charge according to
Fick's law:
\beq \partial_t\, j^0 = D\ \vec\nabla^2 j^0\,.\labell{fick}\eeq
This expectation can be confirmed in a holographic context
\cite{Kovtun:2003wp,Policastro:2002se,andrei2} and, in fact, the
computation of the diffusion constant $D$ can be performed in a
number of different ways. In the following, we use the membrane
paradigm approach.

The computation of the diffusion constant via the membrane paradigm
was discussed in \cite{Kovtun:2003wp} where explicit formulae for
various transport coefficients in terms of metric components for a
wide class of backgrounds were derived.  There, the authors
considered perturbations of a black brane background and a formula
for the diffusion constant (eq.~(2.27) in \cite{Kovtun:2003wp})
resulted from a derivation of Fick's law. An analogous computation
can be performed for the D5-branes' vector field for black hole
embeddings considered here, with the result
\beqa D &=& \left.
\frac{\sqrt{-g}}{\sqrt{h}}\frac{1}{g_{xx}\sqrt{-g_{tt}g_{\rho\rho}}}
\right|_{\rho=1}\ \int d\rho \left( -g_{tt}\, g_{\rho
\rho}\,\frac{\sqrt{g_{\rm int}}}{\sqrt{-g}}\right)
\nonumber \\
&=& \frac{(1+f^2)^{1/2}}{\pi T}  \int_0^1 \frac{du}{\sqrt{1+f^2u^4}}
\equiv \frac{(1+f^2)^{1/2}}{\pi T}\,I(f)\,. \labell{membraneD} \eeqa
In the first expression above, the metric $g$ is the induced metric
on the D5-branes \reef{induce} and $g_{\rm int}$ is the determinant
of the metric on the internal two-sphere (with unit radius). The
integral can be evaluated analytically yielding a hypergeometric
function:
\begin{equation}
I(f) \, = \, {}_2F_1 (\frac{1}{4},\frac{1}{2};\frac{5}{4}; - f^2)
\labell{integral} \end{equation}
\FIGURE{
\includegraphics[width=0.48\textwidth]{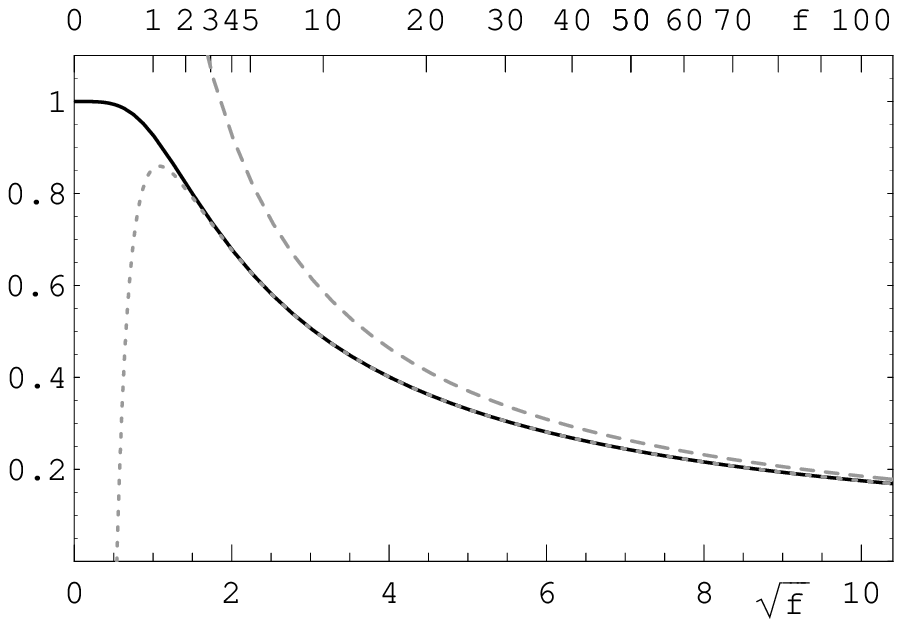}
\caption{The integral $I(f)$ plotted as a function of $\sqrt{f}$: The solid line is the exact result as
given in eq.~\reef{integral}. The upper dashed (lower dotted) line
corresponds to the first term (first two terms) in the large $f$
expansion in eq.~\reef{bigff}.} \label{diffusionii}}
\noindent Figure \ref{diffusionii} shows a plot of $I(f)$.

This same integral in eq.~\reef{membraneD} reappears at various
points in our analysis and so it is useful to gain some better
intuition for this expression. First, let us rewrite the integral as
 \beq
I(f) \, = \, f^{-1/2}\int_0^{f^{1/2}} \frac{ds}{\sqrt{1+s^4}} \
.\labell{muck}
 \eeq
Now, we find that we can expand the integrand around $s = 0$ as
\begin{equation}
\frac{1}{\sqrt{1+s^4}} \, \sim  \, 1 \, - \, \frac{1}{2} s^4 \, + \,
\frac{3}{8} s^8 \, + \, \ldots \,
\end{equation}
and around $s = \infty$
\begin{equation}
\frac{1}{\sqrt{1+s^4}} \, \sim  \, \frac{1}{s^2}\big( 1 \, - \,
\frac{1}{2} s^{- 4} \, + \, \frac{3}{8} s^{-8} \, + \ldots\big) \, .
\end{equation}
At $s = 1$, the convergence of both sequences goes as $\frac{1}{n!}$
and a more precise approximation is
\begin{equation}
\frac{1}{\sqrt{1+s^4}} \, \sim \,\frac{1}{\sqrt{2}}(1- (s-1)
+\ldots) \ .
\end{equation}
Combining \reef{muck} with these expansions, we can find the
integral in various approximations
 \bea
  f \ll 1\, : ~~~~ I(f)
&\sim& 1  \, - \, \frac{1}{10} f^2 \, + \, \frac{3}{72} f^4 \, +
\ldots  \\
f \gg 1\, : ~~~~ I(f) &\sim& c_\infty f^{-1/2} \, - \,
 f^{-1} \, + \, \frac{1}{10}f^{-3} - \, \frac{3}{72} f^{-5}\, + \ldots  \labell{bigff}\\
f \sim 1\, : ~~~~ I(f) &\sim& f^{-1/2} \Big(c_1  \, + \,
\frac{1}{\sqrt{2}} \big((f^{1/2} - 1) \, - \, \frac{1}{2} (f^{1/2} -
1)^2 \, + \ldots\big) \Big)
 \eea
 where $c_1 = \int_0^{1}
\frac{ds}{\sqrt{1+s^4}} = 2 \pi^{-1/2}\Gamma^2
\big(\frac{5}{4}\big)$ and $c_\infty = \int_0^{\infty}
\frac{ds}{\sqrt{1+s^4}} = 2c_1 \simeq 1.854$. Further, we note that
the expansion about $s=0$ is just the expansion of the
hypergeometric function as a hypergeometric series.

Hence, at small and very large $f$, we have for the diffusion
constant
\beq D\,T\rightarrow\left\lbrace
\begin{matrix}
\frac{1}{\pi}&\qquad{\rm with}\ f\rightarrow 0\,,\cr
\frac{c_\infty}{\pi}\sqrt{f}&\qquad{\rm with}\ f\rightarrow\infty\,.
\cr
\end{matrix}
\right. \labell{diffusion3}\eeq
We note that the $f=0$ result is different from but close to the
value found for M2-brane hydrodynamics \cite{M2hydro}:
$D\,T=3/4\pi$. Also note that in the limit of large $f$, the
diffusion constant grows as $\sqrt{f}$.

\section{Corrections to D5-brane action}\label{appcorr}

The worldvolume action \reef{act5} is a low-energy effective action
which captures the interactions of the massless open string modes
supported on the D5-branes. The ``stringy'' nature of the underlying
theory will in principle produce an infinite series of higher
dimension terms that are suppressed at low energies by the inverse
string tension, \ie $\alpha'\equiv\ls^2$. However, in practice, one
typically only includes a specific set of terms to a certain order
in the $\alpha'$ expansion. In fact, the action \reef{act5}
implicitly captures an infinite set of these stringy corrections, as
can be seen by the explicit factor of $\ls^2$ accompanying the gauge
field strength in the DBI action. Further, we might add that this
entire series of terms plays a role in our analysis, as the full
square-root form of the action is used in \reef{intern} to normalize
the DBI contributions. However, as already alluded to in section
\ref{d7-geom}, the DBI action does not capture all of the higher
dimension stringy interactions. Beyond the nonabelian commutator
terms referred to there, the full low-energy action includes
additional terms involving derivatives of the gauge field strength
\cite{Kitazawa,koerber}, as well as higher derivative couplings to
the bulk fields \cite{rsquared,bad}. In terms of the dual CFT, the
contributions of these $\alpha'$ interactions will provide finite
$\lambda$ corrections to the leading supergravity results.

In the following and in section \ref{edualmain}, we focus our
attention on a particular new term involving derivatives of the
field strength, which modifies the vector correlators on the
D5-branes. Our results below give the leading $1/\lambda$
corrections when $f$ is finite. Our calculations consider explicitly
those leading contributions for the transverse correlator. As we
discuss in section \ref{edualmain}, the higher derivative
interaction also upsets the electromagnetic duality on the AdS$_4$
part of the worldvolume.

An effective action to describe open string gauge fields at higher
order in $\alpha'$ has been extensively studied in the literature,
\eg see \cite{Kitazawa,koerber}. However, since we consider only the
linearized equations of motion (or the quadratic action) to compute
the correlators in the $AdS_4$ directions, the leading term at
$\mathscr{o} (\alpha'^3)$ is \cite{Kitazawa}
 \beq\labell{gaugeActionmod}
-\frac{1}{4 g_4^2}\, \frac{\zeta(3)}{16\pi^4}
\,\frac{(2\pi\ell_s^2)^3}{L^2\sqrt{1+f^2}} \int d^6 \sigma
\sqrt{-g}\, \nabla_\mu F_{\alpha \beta} \nabla^\mu F^{\alpha
\beta}\, F^2\ .
 \eeq
In fact, this term only becomes relevant for the calculation of the
vector correlators because of the flux background on the compact
$S^2$. After integrating over the internal two-sphere, this
interaction reduces to
 \beq\labell{gaugeActionmod1}
-\frac{1}{4 g_4^2}\, \frac{\zeta(3)}{2\pi}
\,\frac{\ell_s^2\,f^2}{\sqrt{1+f^2}} \int d^4 \sigma \sqrt{-g}\,
\nabla_\mu F_{\alpha \beta} \nabla^\mu F^{\alpha \beta}\ .
 \eeq
Note that the background flux \reef{magflux} remains unmodified by
this new interaction. When considering linearized fluctuations of
the vector field, we can work with a perturbative expansion in
$\ls^2/L^2=1/\sqrt{\lambda}$. We begin with the Ansatz
 \beq
 F = F^{\!(0)}
+ \frac{\ls^2}{L^2 }\, F^{\!(1)} + \cdots\ . \labell{exxpand}
 \eeq
The equations determining the leading $0^{th}$ order field are still
the same Maxwell equations \reef{maxwell},  while for $ F^{\!(1)}$,
we have
 \beq\labell{maxwellmod} \nabla_\nu
F^{\!(1) \,\nu \mu} \, = \, \frac{\zeta(3)}{2\pi}\frac{
L^2\,f^2}{\sqrt{1+f^2}} \nabla_\nu \nabla_\alpha \nabla^\alpha
F^{\! (0)\, \nu \mu} \, .
 \eeq
Writing this out explicitly for $\mu = y$ gives
 \bea
&&\partial_u \frac{h}{\sqrt{H}} \partial_u A^{\!(1)}_y  +
\frac{\sqrt{H}}{h}(\tom^2 - h \tlq^2) \partial_u A^{\!(1)}_y
\labell{gauge4mod} \\
&&\qquad = \, \frac{\zeta(3) f^2 }{2 \pi  \sqrt{1+f^2}}
\left(\partial_u\, u^2 \sqrt{\frac{h}{H}} \partial_u\, \frac{h}{u^2
\sqrt{H}}
\partial_u\, u^2 \sqrt{\frac{h}{H}} \partial_u A^{\!(0)}_y \, + \, \partial_u
\,\frac{u^2}{\sqrt{H}} (\tom^2 - \tlq^2 h)\partial_u A^{\!(0)}_y \,
\right.\nn
&&\qquad\ \ + \, \left. \frac{\tom^2 u^2}{\sqrt{h}}\partial_u\,
\frac{h}{u^2\sqrt{H}} \partial_u\, \frac{u^2}{\sqrt{h}} A_y - \tlq^2
u^2 \partial_u\, \frac{h}{u^2 \sqrt{H}}\partial_u\, u^2 A^{\!(0)}_y
+ \frac{u^2 \sqrt{H}}{h^2}\left(\tom^2 - \tlq^2 h\right)^2
A^{\!(0)}_y\right) \, . \nonumber
 \eea
Note that we adopt the convention above that the derivatives
$\partial_u$ act on all factors to their right. At the horizon, $u
\to 1$, we again wish to impose infalling boundary conditions. If we
substitute the expansion \reef{nearform} for $A_y$ as $u\to 1$, the
right hand side of \reef{gauge4mod} diverges as $h^{i\tom/4 -2}$ and
hence we expect to find singular behaviour in $A^{\!(1)}_y$ there.
Hence we begin by isolating this singular behaviour in a particular
solution to \reef{gauge4mod}: $A^{\!(1),NHG}_{ y} \sim h^{i \tom/4}
(\frac{a}{h} + b + c \log h \, + d\, h\,\log h)$ with appropriate
constants $a,b,c,d$. This particular solution is holds to order
$h^0$ near $u\to 1$ and is well-behaved in the rest of the geometry.
Implicitly, it also satisfies the desired infalling boundary
conditions. Next we add to $A^{\!(1),NHG}_{ y}$ a contribution which
is regular at the horizon and takes the form on an infalling
homogenous solution near $u\to1$ as described by \reef{nearform} and
\reef{nearform1}. That is, we use the Ansatz
 \beq
 A^{\!(1)}_{ y} \, = \, A^{\!(1),NHG}_{ y}+
 h^{i \tom/4} \left(1 + \left(\frac{i\tom}{4}
\frac{3+5f^2}{1+f^2}+\frac{\tlq^2}{4+2 i \tom} \right)(1-u)\right)
\mathscr{F}(u)
 \ .
 \labell{fullsolu}
 \eeq
where $\partial_u \mathscr{F}(u)|_{u\to 1} = 0$. This Ansatz is
constructed so that $\mathscr{F}(u)$ is well behaved everywhere and
we proceed by calculating this profile numerically. In practise, we
apply the boundary condition at some small, but finite, $(1-u)$, so
need we go one order higher in $h$ and increase the accuracy for
solving $A^{\!(0)}$.

The correlator is then found by substituting our original Ansatz
\reef{exxpand} into \reef{green},
 \beq
C_{yy} \, = \, \frac{\varepsilon_0}{\sqrt{1+f^2}}\frac{\partial_u
(A^{\!(0)}_y + \ls^2/L^2\, A^{\!(1)}_y)}{A^{\!(0)}_y + \ls^2/L^2\,
A^{\!(1)}_y} \, \simeq \,
\frac{\varepsilon_0}{\sqrt{1+f^2}}\left[\frac{\partial_u A^{\!(0)}_y
}{A^{\!(0)}_y} + \frac{\ls^2}{L^2}\, \partial_u
\left(\frac{A^{\!(1)}_y}{A^{\!(0)}_y} \right)\right]_{u \to 0}
+\cdots\, .\labell{newer1}
 \eeq
From this expression, we can see that the normalization of any
homogenous solution in $A^{\!(1)}_y$ does not effect the correlator,
as we expect for a gauge-invariant quantity. The expressions for
$C_{tt}$ and $C_{xx}$ following from \reef{green1} are similar to
that above.

The second term on the right-hand side of \reef{newer1} yields the
correction to the correlator due to the fact that the solutions for
the fluctuations $A_\mu (u)$ are modified. In addition,
\reef{gaugeActionmod1} also contributes to the overall value of the
bulk action and hence provides an additional modification of the
correlators. Proceeding as in \reef{gaugeAct3}, we now get two
boundary terms at order $\ls^2$, since $D_\mu F_{\alpha \beta} D^\mu
F^{\alpha\beta} = D_\mu (F_{\alpha \beta}D^\mu F^{\alpha\beta}) -
D_\alpha (A_\beta D_\mu D^\mu F^{\alpha\beta} ) + A_\beta
(e.o.m.)^\beta$, where the last term combines with contributions
from the leading Maxwell term in the action to vanish by the
equations of motion. These expressions lead to a number of new
contributions to the flux \reef{flux}, which in principle even
contribute to off-diagonal correlators. We might add that there is a
further ambiguity in these expressions because the effective action
\reef{gaugeActionmod1} was constructed from examining string
scattering amplitudes \cite{Kitazawa} and so it is only determined
up to total derivatives or boundary terms. Explicitly comparing
\cite{Kitazawa} and \cite{koerber}, one finds that in fact their
results differ by such a boundary term. However, this ambiguity does
not contribute in our background \reef{newind} and in fact, of the
myriad of potential boundary contributions, only a single term
survives
 \beq
\Delta C_{yy} \, = \, - \frac{\ls^2 \, f^2
\zeta(3)}{2\pi\sqrt{1+f^2}}\, \frac{\varepsilon_0}{ \sqrt{1+f^2} }
\left. \frac{\nabla^2
\partial_u A_y}{A_y}  \right|_{u\to 0} \ . \labell{newer2}
 \eeq
Similar expressions survive for $C_{tt}$ and $C_{xx}$ while the
potential contributions to the off-diagonal correlators vanish.
Combining the results in \reef{newer1} and \reef{newer2}, we can
write the total change to the correlator as
 \beq
\delta C_{yy} \, = \,
\frac{\ls^2}{L^2}\,\frac{\varepsilon_0}{\sqrt{1+f^2}}\,
\left[\partial_u \left(\frac{A^{\!(1)}_y}{A^{\!(0)}_y} \right)-
\frac{ L^2 f^2 \zeta(3)}{2\pi\sqrt{1+f^2}} \frac{\nabla^2
\partial_u A^{\!(0)}_y}{A^{\!(0)}_y}  \right]_{u\to 0} \ . \labell{newer3}
 \eeq
Again similar expressions arise for $\delta C_{tt}$ and $\delta
C_{xx}$.

The two previous formulae, \reef{newer2} and \reef{newer3}, still
require a precise definition for $\nabla^2 \partial_u A_y$. This
expression should understood as the covariant tensor expression
$\nabla^2 F_{uy}$ which when evaluated asymptotically yields a
remarkably simple result:
 \beq
\nabla^2 \partial_u A_y= -\frac{2}{L^2}\,\frac{1}{1+f^2}\,\partial_u
A_y +\cdots \labell{stepp}
 \eeq
where the implicit terms decay rapidly enough as $u\to0$ that they
will not contribute to the correlator. Hence \reef{newer2} can be
greatly simplified to
 \beq
\Delta C_{yy} \, = \,  \frac{\ls^2 }{L^2}\frac{  f^2
\zeta(3)}{\pi(1+f^2)^{3/2}} \frac{\varepsilon_0}{  \sqrt{1+f^2} }
\left. \frac{\partial_u A_y}{A_y}  \right|_{u\to 0} \ .
\labell{newer4}
 \eeq
The last factor has exactly the same form as the leading correlator
\reef{green} and so this contribution can be interpreted in terms of
a rescaling of the pre-factor $\varepsilon_0$:
 \beq\labell{permmod} \varepsilon_0 \to \varepsilon_0
\left(1 + \frac{1}{\sqrt{\lambda}}\frac{ f^2\,
\zeta(3)}{\pi(1+f^2)^{3/2}}  \right) \ ,
 \eeq
where we have replaced $\ls^2/L^2=1/\sqrt{\lambda}$. Remarkably our
numerical calculations show that the first contribution to $\delta
C_{yy}$ in \reef{newer3} also produces a shift of $\varepsilon_0$
with precisely the same $f$ dependence -- see figure \ref{epsmodf}.

At this point, several comments are in order. We already pointed out
that the square-root form of the DBI action already incorporates an
infinite set of stringy $\alpha'$ corrections. While this form was
incorporated in our leading order calculations, \eg \reef{intern},
it did not appear to introduce any $1/\lambda$ corrections. Of
course, these factors are hidden in the definition \reef{needy} of
$f$ and for finite values of $f$, we are actually introducing a
magnetic flux quantum number $q\sim O(\sqrt{\lambda})$. In this
context, it is not quite correct to say the interaction
\reef{gaugeActionmod} is the leading term to modify the correlators.
One can well imagine that there will be other interactions which,
although they appear to be of higher order in the $\alpha'$
expansion, will modify the correlator with contributions of order
$f^n/\sqrt{\lambda}$ where $n>2$. Of course, these contributions
will be suppressed in a regime where $f\ll 1$. Considering the
possible tensor structure of the relevant higher order interactions,
it seems that this class of contributions will always be arising
from an equation of motion of the form \reef{maxwellmod}.

\section{Hyperbolic tangent potential}\label{tanhpot}

To get more insight into the appearance of the finite temperature
effects in the frequency dependence of the conductivity, we can
study a qualitatively similar problem that has an analytic solution.

\subsection{Finding the spectral curves}

Let us modify the effective Schr\"odinger equation in \reef{schroeq}
to write it in terms of the complex frequency $\nu$ of section
\ref{quasinorm}, \ie $\tilde{\nu} \equiv \sqrt{\tom^2 - \tlq^2}$,
 \beqa\labell{schroeq2}
&&\qquad\qquad\ \ \ \ \ \left( - \partial_\rho ^{\, 2}  +  V\right)
A_y \, = \,
\tilde{\nu}^2 A_y \\
&&{\rm with}\qquad V=-\tlq^2\,u^4 \quad {\rm and}\quad \rho \, = \,
\int_0^u d\tilde{u} \frac{\sqrt{H}}{h} \ .\nonumber
 \eeqa
Recall in terms of the new radial coordinate $\rho$, the asymptotic
boundary is mapped to $\rho=0$ and the horizon, to $\rho\to\infty$.
In fact, it is straightforward to evaluate the integral above to
find $\rho(u)$ in terms of incomplete elliptic integrals of the
third kind or alternatively, in terms of the Appell hypergeometric
function \cite{appell} $\rho = \sqrt{1+f^2} u\, F_1
\left(\frac{1}{4};\frac{1}{2},1;\frac{5}{4};-f^2 u^4, u^4\right)$.
Note that the only difference between \reef{schroeq} and
\reef{schroeq2} is that we have subtracted $\tlq^2\,A_y$ from both
sides in the equation above. Hence in the present form, the
effective potential $V$ vanishes at $\rho=0$ and approaches
$-\tlq^2$ as $\rho\to\infty$.

The equation of motion in the form of \reef{schroeq2} can be
examined in three distinct regions which for $f \gg 1$, where the
potential looks roughly like
 \bea \nonumber
V \ \sim \ 0 & : & \rho \lesssim  f^{1/2} \\ \nonumber
V \ \sim \ - \tlq^2 ( 2 f^{1/2} - \rho)^{-4} & : & f^{1/2}
\lesssim \rho \lesssim 2 f^{1/2}  - 1 \\
V \ \sim \ - \tlq^2 + \tlq^2 e^{-4 (\rho -  2 f^{1/2}  + 1)} & : & 2
f^{1/2}  - 1 \lesssim \rho \ .
 \eea
Of course, the full potential is smooth across these three regions.
\FIGURE{
\includegraphics[width=0.48 \textwidth]{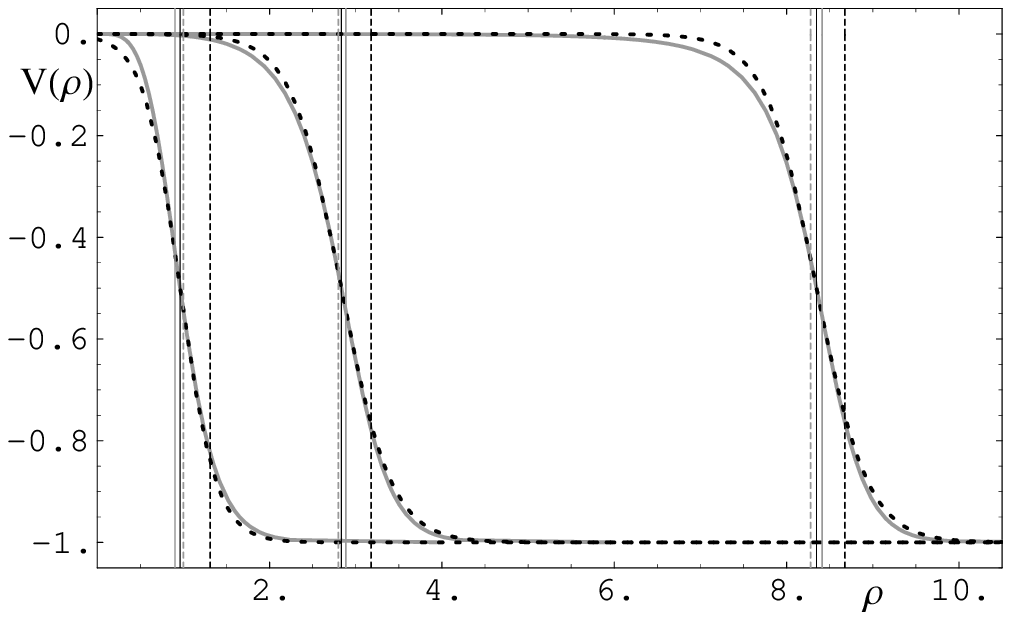}
\caption{The exact potential $V$ (solid grey) and the $\tanh$
approximation $V_{app}$ (black dotted), for $f \in \{0,4,25\}$ with
$\tlq=1$. We also indicate a number of interesting values of $\rho$
with narrow vertical lines: $\rho(u_{half})$, solid black;
$\rho(u_{min})$, solid grey; $\rho=\frac{\pi\,T}{2\,T_{eff}}$,
dashed black; and $\rho=\pi\,D\,T$, grey
dashed.}\label{tanhpotplot}}
To model the ``smooth'' step with a finite slope and a fast tailoff
in the asymptotic regions, we will use
 \beq
V_{app} \, = \, - \tlq^2 \frac{1\, +\, \tanh \, \mathfrak{n} (\rho-\rho_0)}{2}
 \ , \labell{tanhpoteq}
 \eeq
which allows us to find exact solution for \reef{schroeq2}. We fix
the ``step'' position $\rho_0$ either as location where the the
potential has half its minimum, \ie at $u_{half}^4= 1/2$ or the
point where the slope of the potential is minimum, \ie at
 \beq
u_{min}^4 = \frac{5f^2 - 7 + \sqrt{(7 - 5f^2)^2 +108 f^2}}{18 f^2}\
. \labell{noose}
 \eeq
To fix the parameter $\mathfrak{n}$ that we will use to characterize
the slope of the step of the potential, we simply take the slope of
the full potential at either of the corresponding points:
$\mathfrak{n}\equiv-2 \left.\partial_\rho V\right|_{u_{half,min}} =
8 \tlq^2 u^3 h(u) H(u)^{-1/2} \big|_{u_{half,min}}$. Working with
$u_{half}$ or $u_{min}$ will give $V_{app}$ which is a good
approximation of the full potential $V$. As shown in figure
\ref{tanhpotplot}, the potentials constructed with these two choices
typically can not be distinguished. For comparison purposes, the
figure also indicates two other interesting scales:
$\rho={\pi\,T}/{2T_{eff}}$ and $\pi\,D\,T$. There we see that both
of these scales are close to the width of the effective potential
but $\rho=\pi\,D\,T$ is particularly close to the center of the
step.

Note that $V(\rho = 0) = 0$ and $\partial_\rho V(\rho = 0) = 0$
while neither of these properties holds for the approximation
\reef{tanhpoteq}. We will take account of this fact simply by
expanding around $\rho - \rho_0 \rightarrow - \infty$ and discarding
higher order terms when computing the values at $\rho = 0$. One can
expect this potential to be a good approximation for $\tlnu \gg 1$,
where the result is less sensitive to the details of the potential
away from it's maximum slope and for large $f$, where the results
should be dominated by the large flat part of the potential.

Using infalling boundary conditions at $\rho \rightarrow \infty$,
the solution is
 \bea \nonumber
A_y & = & A_{y 0} \, {}_{2 \!} F_1{\scriptstyle \left(1+\frac{i\tlnu}{2 \mathfrak{n}}
 (1 - \sqrt{1 + \tlq^2/\tlnu^2}),\,  \frac{i\tlnu}{2 \mathfrak{n}}  (1 - \sqrt{1 +
  \tlq^2/\tlnu^2});\, 1- \frac{i\tlnu}{\mathfrak{n}} \sqrt{1 + \tlq^2/\tlnu^2};\,
   \frac{1}{1+e^{2 \mathfrak{n} (\rho-\rho_0)}}  \right)} \\
& & \ \ \ \ \ \ (1- \tanh  \, \mathfrak{n} (\rho-\rho_0) )^{-\frac{i\tlnu}{2\mathfrak{n}}
\sqrt{1 + \tlq^2/\tlnu^2}} (1 + \tanh  \, \mathfrak{n} (\rho-\rho_0) )^{\frac{i\tlnu}{2\mathfrak{n}}}
 \eea
and the asymptotic expansion gives us
 \bea \nonumber
A_y & = & A_{y 0}\, e^{i\tlnu (\rho-\rho_0)} \left(
\frac{\left(\sqrt{1 + \tlq^2/\tlnu^2}-1\right)\,\, \Gamma
\!\left(\frac{i\tlnu}{2 \mathfrak{n}}
 (1 - \sqrt{1 + \tlq^2/\tlnu^2}) \right) ^2\,\, \Gamma\! \left( - \frac{i\tlnu}{\mathfrak{n}}\right)}{ \left(
 \sqrt{1 + \tlq^2/\tlnu^2} + 1\right) \, \Gamma\! \left( - \frac{i\tlnu}{2 \mathfrak{n}}  ( \sqrt{1 + \tlq^2/\tlnu^2}
  + 1)\right)^2 \, \Gamma\! \left( \frac{i\tlnu}{\mathfrak{n}} \right)}  \, +\, e^{- 2 i\tlnu (\rho-\rho_0)} \right) \\
& \equiv&  A_{y 0}\, e^{i\tlnu (\rho-\rho_0)} \left(\aleph  \, +\,
e^{- 2 i\tlnu (\rho-\rho_0)} \right) \ , \labell{bangbang}
 \eea
for which we redefined $A_{y0}$. In the opposite limit, as $\rho
\rightarrow \infty$, the solution converges exponentially to $A_y
\propto e^{-i\tlnu \rho \sqrt{1 + \tlq^2/\tlnu^2} }$, which is the
desired infalling wave solution at the horizon.

The transverse conductivity can now be written in a compact
suggestive form in terms of the implicitly defined $\aleph$:
 \beq
\tilde{\sigma}_{yy} \, = \, \varepsilon_0\, \Re \frac{\tlnu}{\tom}
\frac{\aleph  \, -\, e^{ 2 i\tlnu \rho_0}}{\aleph \, +\, e^{ 2
i\tlnu \rho_0}} \, , \labell{tanhpotcon}
 \eeq
which is of the same form as the result \reef{squarecon}, but with
different coefficients. We again see how the oscillatory pattern
arises from interference in the asymptotic region due to the
potential ``step'', and how the effective temperature arises from
tunnelling through the potential out of the ``heat bath'' in the
near horizon region. For $|\tlnu| \gg \mathfrak{n}$, we find the
limit
  \beq
\aleph \, \stackrel{|\tlnu| \rightarrow \infty}{\longrightarrow} \,
\frac{4 \mathfrak{n} \tlnu}{\mathrm{sgn}(\Re \tlnu)\, \pi\tlq^2}
e^{\mathrm{sgn}(\Re \tlnu)\, \pi\tlnu/\mathfrak{n}} \left(1 +
\mathscr{O} \left({\scriptstyle \frac{\tlq}{\tlnu}}\, \log \,
\tlnu/\mathfrak{n}\right)\right) \ ,
 \eeq
This result shows that the subleading term in the conductivity
(beyond the low temperature limit \reef{looow}) has an exponential
frequency dependence for $\tom> \tlq$, as we found in section
\ref{efftf2}. We can also see how the results depend on
$\mathfrak{n}$, \ie on the slope of the step in the potential. We
compare the present results \reef{tanhpotcon} for the conductivity
to our numerical results and to the perturbative approximation in
figure \ref{compplot}. There we can see that the various approaches
are in very close agreement for $\tom \gtrsim \tlq$.

\subsection{Exact pole structure}\label{ex_pol_str}

We can look at the solutions \reef{bangbang} and identify the
quasinormal modes by imposing the asymptotic boundary condition:
$A_y(\rho = 0) = 0$. The quasinormal frequencies are then simply
determined by the equation
 \beq e^{2 i\tlnu \rho_0} \, = \, -
\frac{\left( \sqrt{1 + \tlq^2/\tlnu^2} -1\right)\, \Gamma\!
\left(\frac{i\tlnu}{2 \mathfrak{n}}  (1 - \sqrt{1 + \tlq^2/\tlnu^2})
\right) ^2 \, \Gamma\!\left( - \frac{i\tlnu}{\mathfrak{n}}\right)}{
\left( \sqrt{1 + \tlq^2/\tlnu^2} - 1\right)\, \Gamma\! \left( -
\frac{i\tlnu}{2 \mathfrak{n}}  (1+ \sqrt{1 +
\tlq^2/\tlnu^2})\right)^2\, \Gamma\! \left(
\frac{i\tlnu}{\mathfrak{n}} \right)}
 \eeq
which simplifies in the limit of $|\tlnu| \gg \mathfrak{n}, \tlq$ to
 \beq\label{tanhexpoleapp}
e^{(2 i \rho_0 - \mathrm{sgn}(\Re \tlnu)\, \pi/\mathfrak{n})\tlnu}
\, \simeq \, - \frac{4 \mathfrak{n} \tlnu}{\mathrm{sgn}(\Re \tlnu)\,
\pi\tlq^2}  \  .
 \eeq
These quasinormal frequencies also give the location of the poles in
the spectral function \cite{Son:2002sd,solo}, with the exception of
the asymptotically constant solution, $\tlnu = 0$. In the
terminology of section \ref{quasinorm}, the present approximation
yields:
 \beq
\tlnu_0 = \frac{4\pi \rho_0}{4\rho_0^2+\pi^2/\mathfrak{n}^2}\
,\qquad \frac{\tlgam_0}{\tlnu_0} = \frac{\pi}{2 \rho_0
\mathfrak{n}}\ . \labell{woppy}
 \eeq
Further as mentioned in section \ref{quasinorm}, the subleading
contributions are logarithmic in $n$, giving us
 \beq
\delta \tlnu_n = \frac{\tlgam_0 - i \tlnu_0}{2\pi}\big(\ln n  + \ln
\frac{4 \mathfrak{n} ( i \gamma_0 -\tlnu_0)}{\pi\tlq^2} \big) \ ,
\labell{submarine}
 \eeq
in contrast to the constant shift as the largest subleading term for
the full correlator.

The residues in the Green's function, before taking the imaginary
part, are given by
 \beq
\mathscr{R}_{yy}^{(n)} \, \equiv \,
\frac{\varepsilon_0}{\sqrt{1+f^2}}\, \Res{\tlnu = \tlnu_n} \,
\frac{\partial_u A_y}{A_y}\bigg|_{u\rightarrow 0^+} \, = \,
\varepsilon_0\, \frac{ - 2 i \tlnu_n }{2 i \rho_0 - \partial_\tlnu
\log \gamma|_{\tlnu = \tlnu_n}} \ .
 \eeq
One finds that $ \partial_\tlnu \log \gamma$ can be expressed in
terms of digamma functions, and asymptotes to
 \beq
\frac{\varepsilon_0}{\sqrt{1+f^2}}\, \Res{\tlnu = \tlnu_n} \,
\frac{\partial_u A_y}{A_y} \bigg|_{u\rightarrow +0} \, \simeq \,
\varepsilon_0\, \frac{- 2\tlnu_n }{2 \rho_0 + i\pi/\mathfrak{n}} \ ,
 \eeq
where we used implicitly $n \in \mathbb{Z}$ to label all the poles.

We can use the knowledge of the poles to verify that the spectral
function is indeed approximated extremely well by the regularized
sum of poles plus the term linear in $\tom$ as given in
\reef{omansatztrans},
 \bea\labell{expolesumtrans}
C_{y y} &=& \varepsilon_0 \sum_{n \neq 0}\mathscr{R}_{yy}^{(n)}\frac{\tlnu_n \,
\mathrm{sgn}(\Re \tlnu_n)}{\big( \tlq^2 +  \tlnu_n^2 \big)^{1/2}}\left( \frac{1}{\tom
- \mathrm{sgn}(\Re \tlnu_n)\big( \tlq^2 + \tlnu_n^2 \big)^{1/2}} +
\frac{ \mathrm{sgn}(\Re \tlnu_n)}{ \big( \tlq^2 +\tlnu_n^2 \big)^{1/2}}\right)  \nonumber \\
&+& \varepsilon_0 \lim_{n\rightarrow + \infty} \frac{\tom}{\pi} \log
\frac{\tlnu_n-\tlnu_{n+1}}{\tlnu_{-(n+1)}-\tlnu_{-n}}
 \ ,
 \eea
or from \reef{omansatzlong}
 \beq\labell{expolesumlong}
C_{xx} \, = \, \varepsilon_0 \sum_{n \neq
0}\left(\mathscr{R}_{xx}^{(n)}\frac{\tlnu_n \, \mathrm{sgn}(\Re
\tlnu_n)}{\big( \tlq^2 +  \tlnu_n^2
\big)^{1/2}}\frac{\mathscr{R}_{xx}^{(n)}}{\tlnu - \mathrm{sgn}(\Re
\tlnu_n)\big( \tlq^2 + \tlnu_n^2 \big)^{1/2}} \right) \ ,
 \eeq
which we show in figure \ref{compplot}. We also use the exact
location of the poles to check how well the algorithm from section
\ref{quasinorm} finds the location and residue of the poles, which
we show in figure \ref{poleplot} and discuss in section
\ref{quasinorm}.


\begin{thebibliography}{99}

\bibitem 
{juan} J.M.~Maldacena,
``The large N limit of superconformal field theories and supergravity,''
Adv.\ Theor.\ Math.\ Phys.\  {\bf 2} (1998) 231
[arXiv:hep-th/9711200].

\bibitem 
{adscft} S.S.~Gubser, I.R.~Klebanov and A.M.~Polyakov,
``Gauge theory correlators from non-critical string theory,''
Phys.\ Lett.\ B {\bf 428} (1998) 105  [arXiv:hep-th/9802109];\\
E.~Witten, 
Adv.\ Theor.\ Math.\ Phys.\  {\bf 2} (1998) 253
[arXiv:hep-th/9802150].

\bibitem 
{bigRev} O.~Aharony, S.S.~Gubser, J.M.~Maldacena, H.~Ooguri and
Y.~Oz,
``Large N field theories, string theory and gravity,''
Phys.\ Rept.\  {\bf 323} (2000) 183 [arXiv:hep-th/9905111].

\bibitem 
{talks} See, for example:\\
W.~Zajc, ``Quark Gluon Plasma at RHIC (and in QCD and String
Theory),'' presented at PASCOS 08 --- see
http://pirsa.org/08060040/;\\
K.~Rajagopal, ``Quark Gluon Plasma in QCD, at RHIC, and in String
Theory,'' presented at PASCOS 08 --- see
http://pirsa.org/08060041/;\\
D.~Mateos,
  ``String Theory and Quantum Chromodynamics,''
  Class.\ Quant.\ Grav.\  {\bf 24} (2007) S713
  [arXiv:0709.1523 [hep-th]];\\
S.~S.~Gubser,
  ``Heavy ion collisions and black hole dynamics,''
  Gen.\ Rel.\ Grav.\  {\bf 39} (2007) 1533
  [Int.\ J.\ Mod.\ Phys.\  D {\bf 17} (2008) 673];\\
D.~T.~Son,
  ``Gauge-gravity duality and heavy-ion collisions,''
  AIP Conf.\ Proc.\  {\bf 957} (2007) 134.



\bibitem 
{pavel} C.~P.~Herzog, P.~Kovtun, S.~Sachdev and D.~T.~Son,
  Phys.\ Rev.\  D {\bf 75} (2007) 085020
  [arXiv:hep-th/0701036].

\bibitem 
{more} See, for example:\\
S.~A.~Hartnoll, P.~K.~Kovtun, M.~Muller
and S.~Sachdev,
  ``Theory of the Nernst effect near quantum phase transitions in condensed
  matter, and in dyonic black holes,''
  Phys.\ Rev.\  B {\bf 76} (2007) 144502
  [arXiv:0706.3215 [cond-mat.str-el]];
S.~A.~Hartnoll and C.~P.~Herzog,
  ``Ohm's Law at strong coupling: S duality and the cyclotron resonance,''
  Phys.\ Rev.\  D {\bf 76} (2007) 106012
  [arXiv:0706.3228 [hep-th]];
  C.~P.~Herzog, P.~K.~Kovtun and D.~T.~Son,
  ``Holographic model of superfluidity,''
  arXiv:0809.4870 [hep-th];
  S.~S.~Lee,
  ``A Non-Fermi Liquid from a Charged Black Hole: A Critical Fermi Ball,''
  arXiv:0809.3402 [hep-th];
  A.~O'Bannon,
  ``Holographic Thermodynamics and Transport of Flavor Fields,''
  arXiv:0808.1115 [hep-th];
  N.~Evans and E.~Threlfall,
  ``R-Charge Chemical Potential in the M2-M5 System,''
  arXiv:0807.3679 [hep-th];\\
  S.~S.~Gubser and F.~D.~Rocha,
  ``The gravity dual to a quantum critical point with spontaneous symmetry
  breaking,''
  arXiv:0807.1737 [hep-th];
S.~A.~Hartnoll, C.~P.~Herzog and G.~T.~Horowitz,
  ``Building an AdS/CFT superconductor,''
  arXiv:0803.3295 [hep-th];
M.~M.~Roberts and S.~A.~Hartnoll,
  ``Pseudogap and time reversal breaking in a holographic superconductor,''
  arXiv:0805.3898 [hep-th];
  K.~Maeda and T.~Okamura,
  ``Characteristic length of an AdS/CFT superconductor,''
  arXiv:0809.3079 [hep-th];
  M.~M.~Roberts and S.~A.~Hartnoll,
  ``Pseudogap and time reversal breaking in a holographic superconductor,''
  JHEP {\bf 0808}, 035 (2008)
  [arXiv:0805.3898 [hep-th]];
  S.~S.~Gubser and S.~S.~Pufu,
  ``The gravity dual of a p-wave superconductor,''
  arXiv:0805.2960 [hep-th];
  W.~Y.~Wen,
  ``Inhomogeneous magnetic field in AdS/CFT superconductor,''
  arXiv:0805.1550 [hep-th];
  T.~Albash and C.~V.~Johnson,
  ``A Holographic Superconductor in an External Magnetic Field,''
  JHEP {\bf 0809}, 121 (2008)
  [arXiv:0804.3466 [hep-th]];
  E.~Nakano and W.~Y.~Wen,
  ``Critical Magnetic Field In A Holographic Superconductor,''
  Phys.\ Rev.\  D {\bf 78}, 046004 (2008)
  [arXiv:0804.3180 [hep-th]];
  D.~Minic and J.~J.~Heremans,
  ``High Temperature Superconductivity and Effective Gravity,''
  arXiv:0804.2880 [hep-th];
  S.~S.~Gubser,
  ``Colorful horizons with charge in anti-de Sitter space,''
  arXiv:0803.3483 [hep-th];
  M.~Ammon, J.~Erdmenger, M.~Kaminski and P.~Kerner,
``Superconductivity from gauge/gravity duality with flavor,''
  arXiv:0810.2316 [hep-th];
P.~Basu, A.~Mukherjee and H.~H.~Shieh,
  ``Supercurrent: Vector Hair for an AdS Black Hole,''
  arXiv:0809.4494 [hep-th];
S.~A.~Hartnoll, C.~P.~Herzog and G.~T.~Horowitz,
  arXiv:0810.1563 [hep-th];
E.~I.~Buchbinder, A.~Buchel and S.~E.~Vazquez,
  ``Sound Waves in (2+1) Dimensional Holographic Magnetic Fluids,''
  arXiv:0810.4094 [hep-th];
S.~A.~Hartnoll and C.~P.~Herzog,
  ``Impure AdS/CFT,''
  Phys.\ Rev.\  D {\bf 77} (2008) 106009
  [arXiv:0801.1693 [hep-th]];
M.~Fujita, Y.~Hikida, S.~Ryu and T.~Takayanagi,
  ``Disordered Systems and the Replica Method in AdS/CFT,''
  arXiv:0810.5394 [hep-th].


\bibitem 
{hallo} S.A.~Hartnoll and P.~Kovtun, ``Hall conductivity from dyonic
black holes,''  arXiv:0704.1160 [hep-th];

\bibitem 
{hall1} E.~Keski-Vakkuri and P.~Kraus,
  ``Quantum Hall Effect in AdS/CFT,''
  arXiv:0805.4643 [hep-th].

\bibitem 
{nonrel} See, for example:\\
D.~T.~Son,
  ``Toward an AdS/cold atoms correspondence: a geometric realization of the
  Schr\"odinger symmetry,''
  Phys.\ Rev.\  D {\bf 78} (2008) 046003
  [arXiv:0804.3972 [hep-th]];
    K.~Balasubramanian and J.~McGreevy,
  ``Gravity duals for non-relativistic CFTs,''
  Phys.\ Rev.\ Lett.\  {\bf 101}, 061601 (2008)
  [arXiv:0804.4053 [hep-th]];
W.~D.~Goldberger,
  ``AdS/CFT duality for non-relativistic field theory,''
  arXiv:0806.2867 [hep-th];
J.~L.~B.~Barbon and C.~A.~Fuertes,
  ``On the spectrum of nonrelativistic AdS/CFT,''
  JHEP {\bf 0809} (2008) 030
  [arXiv:0806.3244 [hep-th]];
  C.~P.~Herzog, M.~Rangamani and S.~F.~Ross,
  ``Heating up Galilean holography,''
  arXiv:0807.1099 [hep-th];
J.~Maldacena, D.~Martelli and Y.~Tachikawa,
  ``Comments on string theory backgrounds with non-relativistic conformal
  symmetry,''
  JHEP {\bf 0810} (2008) 072
  [arXiv:0807.1100 [hep-th]];
A.~Adams, K.~Balasubramanian and J.~McGreevy,
  ``Hot Spacetimes for Cold Atoms,''
  arXiv:0807.1111 [hep-th];
D.~Minic and M.~Pleimling,
  ``Non-relativistic AdS/CFT and Aging/Gravity Duality,''
  arXiv:0807.3665 [cond-mat.stat-mech];
J.~W.~Chen and W.~Y.~Wen, ``Shear Viscosity of a Non-Relativistic
Conformal Gas in Two Dimensions,''
  arXiv:0808.0399 [hep-th];
S.~Kachru, X.~Liu and M.~Mulligan,
  arXiv:0808.1725 [hep-th];
P.~Kovtun and D.~Nickel,
  ``Black holes and non-relativistic quantum systems,''
  arXiv:0809.2020 [hep-th];
C.~Duval, M.~Hassaine and P.~A.~Horvathy,
  ``The geometry of Schr\"odinger symmetry in gravity
  background/non-relativistic CFT,''
  arXiv:0809.3128 [hep-th];
D.~Yamada,
  ``Thermodynamics of Black Holes in Schr\"odinger Space,''
  arXiv:0809.4928 [hep-th];
S.~A.~Hartnoll and K.~Yoshida,
  ``Families of IIB duals for nonrelativistic CFTs,''
  arXiv:0810.0298 [hep-th].
S.~S.~Pal,
  ``Null Melvin Twist to Sakai-Sugimoto model,''
  arXiv:0808.3042 [hep-th].
S.~Pal,
  ``More gravity solutions of AdS/CMT,''
  arXiv:0809.1756 [hep-th].
M.~Schvellinger,
  ``Kerr-AdS black holes and non-relativistic conformal QM theories in diverse dimensions,''
  arXiv:0810.3011 [hep-th].
F.~L.~Lin and S.~Y.~Wu,
  ``Non-relativistic Holography and Singular Black Hole,''
  arXiv:0810.0227 [hep-th].
C.~Leiva and M.~S.~Plyushchay,
  ``Conformal symmetry of relativistic and nonrelativistic systems and  AdS/CFT correspondence,''
  Annals Phys.\  {\bf 307} (2003) 372
  [arXiv:hep-th/0301244].

\bibitem 
{subir} See, for example: S.~Sachdev, {\it Quantum Phase
Transitions}, Cambridge University Press (1999).

\bibitem 
{land} See, for example: M.~R.~Douglas and S.~Kachru,
  ``Flux compactification,''
  Rev.\ Mod.\ Phys.\  {\bf 79} (2007) 733
  [arXiv:hep-th/0610102].

\bibitem 
{lisa} A.~Karch and L.~Randall,
  ``Localized gravity in string theory,''
  Phys.\ Rev.\ Lett.\  {\bf 87} (2001) 061601
  [arXiv:hep-th/0105108];
JHEP {\bf 0106} (2001) 063 [arXiv:hep-th/0105132].

\bibitem 
{hirosi} O.~DeWolfe, D.Z.~Freedman and H.~Ooguri, ``Holography and
defect conformal field theories,'' Phys.\ Rev.\  D {\bf 66} (2002)
025009 [arXiv:hep-th/0111135].

\bibitem 
{jaume} J.~Gomis and C.~Romelsberger, ``Bubbling defect CFT's,''
JHEP {\bf 0608} (2006) 050 [arXiv:hep-th/0604155];\\
E.~D'Hoker, J.~Estes and M.~Gutperle, ``Exact half-BPS type IIB
interface solutions. II: Flux solutions and multi-janus,''
  JHEP {\bf 0706} (2007) 022
  [arXiv:0705.0024 [hep-th]].

\bibitem 
{rey} S.Y.~Rey, ``Quantum Phase Transitions from String Theory,''
talk at Strings 2007 --- see:
http://www.ift.uam.es/strings07/010\_marco.htm

\bibitem 
{quantumhall}
  J.~L.~Davis, P.~Kraus and A.~Shah,
  ``Gravity Dual of a Quantum Hall Plateau Transition,''
  arXiv:0809.1876 [hep-th].

\bibitem 
{karchkatz} A.~Karch and E.~Katz, ``Adding flavor to AdS/CFT,'' JHEP
{\bf 0206} (2002)  043 [arXiv:hep-th/0205236];\\
O.~Aharony, A.~Fayyazuddin and J.~M.~Maldacena,
  JHEP {\bf 9807} (1998) 013
  [arXiv:hep-th/9806159].

\bibitem 
{johanna} J.~Babington, J.~Erdmenger, N.J.~Evans, Z.~Guralnik and
I.~Kirsch, ``Chiral symmetry breaking and pions in
non-supersymmetric gauge/gravity duals,'' Phys.\ Rev.\ D {\bf 69}
(2004) 066007
[arXiv:hep-th/0306018];\\
I.~Kirsch, ``Generalizations of the AdS/CFT correspondence,''
Fortsch.\ Phys.\  {\bf 52} (2004) 727 [arXiv:hep-th/0406274].\\

\bibitem 
{recent} T.~Albash, V.~Filev, C.V.~Johnson and A.~Kundu, ``A
topology-changing phase transition and the dynamics of flavour,''
arXiv:hep-th/0605088;\\
T.~Albash, V.~Filev, C.V.~Johnson and A.~Kundu, ``Global currents,
phase transitions, and chiral symmetry breaking in large $\nc$ gauge
theory,'' arXiv:hep-th/0605175;\\
V.G.~Filev, C.V.~Johnson, R.C.~Rashkov and K.S.~Viswanathan,
``Flavoured large N gauge theory in an external magnetic field,''
arXiv:hep-th/0701001;\\
A.~Karch and A.~O'Bannon, ``Chiral transition of N=4 super
Yang-Mills with flavor on a 3-sphere,'' Phys.\ Rev.\ D {\bf 74}
(2006) 085033 [arXiv:hep-th/0605120].

\bibitem 
{long} D.~Mateos, R.C.~Myers and R.M.~Thomson, ``Holographic phase
transitions with fundamental matter,'' Phys.\ Rev.\ Lett.\  {\bf 97}
(2006) 091601 [arXiv:hep-th/0605046];\\
D.~Mateos, R.~C.~Myers and R.~M.~Thomson,
  JHEP {\bf 0705} (2007) 067
  [arXiv:hep-th/0701132].

\bibitem 
{meson} M.~Kruczenski, D.~Mateos, R.C.~Myers and D.J.~Winters,
``Meson spectroscopy in AdS/CFT with flavour,'' JHEP {\bf 0307}
(2003) 049 [arXiv:hep-th/0304032].

\bibitem 
{ramallo} D.~Arean and A.V.~Ramallo, ``Open string modes at brane
intersections,'' JHEP {\bf 0604} (2006) 037
[arXiv:hep-th/0602174];\\
D.~Arean, A.~V.~Ramallo and D.~Rodriguez-Gomez, ``Mesons and Higgs
branch in defect theories,''
  Phys.\ Lett.\  B {\bf 641} (2006) 393
  [arXiv:hep-th/0609010].


\bibitem 
{holomeson} R.C.~Myers and R.M.~Thomson,
  ``Holographic mesons in various dimensions,''
  [arXiv:hep-th/0605017].

\bibitem 
{matt} M.~C.~Wapler, ``Holographic studies of the physics of strongly coupled defect field theories'', PhD thesis, University of Waterloo, in preparation.

\bibitem 
{wittt} E.~Witten,
  ``Anti-de Sitter space, thermal phase transition, and confinement in  gauge
  theories,''
  Adv.\ Theor.\ Math.\ Phys.\  {\bf 2} (1998) 505
  [arXiv:hep-th/9803131].

\bibitem 
{joh} J.~Erdmenger, Z.~Guralnik and I.~Kirsch,
  ``Four-dimensional superconformal theories with interacting boundaries or
  defects,''
  Phys.\ Rev.\  D {\bf 66} (2002) 025020
  [arXiv:hep-th/0203020].

\bibitem 
{source} A.~Kapustin and S.~Sethi,
  ``The Higgs branch of impurity theories,''
  Adv.\ Theor.\ Math.\ Phys.\  {\bf 2} (1998) 571
  [arXiv:hep-th/9804027].

\bibitem 
{neil2} N.~R.~Constable, R.~C.~Myers and O.~Tafjord,
  ``The noncommutative bion core,''
  Phys.\ Rev.\  D {\bf 61} (2000) 106009
  [arXiv:hep-th/9911136].

\bibitem 
{BF} P.~Breitenlohner and D.Z.~Freedman, ``Positive Energy In
Anti-De Sitter Backgrounds And Gauged Extended Supergravity,''
Phys.\ Lett.\  B {\bf 115} (1982) 197;\\
L.~Mezincescu and P.K.~Townsend, ``Stability At A Local Maximum In
Higher Dimensional Anti-De Sitter Space And Applications To
Supergravity,'' Annals Phys.\  {\bf 160} (1985) 406.

\bibitem 
{us} R.~Emparan, G.T.~Horowitz and R.C.~Myers, ``Black holes radiate
mainly on the brane,'' Phys.\ Rev.\ Lett.\  {\bf 85} (2000) 499
[arXiv:hep-th/0003118].

\bibitem 
{nonabDBI} R.~C.~Myers,
  ``Dielectric-branes,''
  JHEP {\bf 9912} (1999) 022
  [arXiv:hep-th/9910053];\\
W.~Taylor and M.~Van Raamsdonk,
  ``Multiple Dp-branes in weak background fields,''
  Nucl.\ Phys.\  B {\bf 573} (2000) 703
  [arXiv:hep-th/9910052].

\bibitem 
{strace} A.~A.~Tseytlin,
  ``On non-abelian generalisation of the Born-Infeld action in string
  theory,''
  Nucl.\ Phys.\  B {\bf 501} (1997) 41
  [arXiv:hep-th/9701125].

\bibitem 
{nreview} R.~C.~Myers,
  ``Nonabelian phenomena on D-branes,''
  Class.\ Quant.\ Grav.\  {\bf 20} (2003) S347
  [arXiv:hep-th/0303072].

\bibitem 
{tilt} A.~Hashimoto and W.~Taylor,
  ``Fluctuation spectra of tilted and intersecting D-branes from the
  Born-Infeld action,''
  Nucl.\ Phys.\  B {\bf 503} (1997) 193
  [arXiv:hep-th/9703217];\\
  P.~Bain, ``On the non-Abelian Born-Infeld action,''
  arXiv:hep-th/9909154.

\bibitem 
{ConstableRobEtc} N.~R.~Constable, R.~C.~Myers and O.~Tafjord,
  ``Non-Abelian brane intersections,''
  JHEP {\bf 0106} (2001) 023
  [arXiv:hep-th/0102080].

\bibitem 
{Son:2002sd} D.T.~Son and A.O.~Starinets, ``Minkowski-space
correlators in AdS/CFT correspondence: Recipe and applications,''
JHEP {\bf 0209}, 042 (2002) [arXiv:hep-th/0205051].

\bibitem 
{Kovtun:2005ev} P.K.~Kovtun and A.O.~Starinets, ``Quasinormal modes
and holography,'' Phys.\ Rev.\ D {\bf 72}, 086009 (2005)
[arXiv:hep-th/0506184].

\bibitem 
{hydropaper} L.P.~Kadanoff and P.C.~Martin, ``Hydrodynamic Equations
and Correlation Functions,'' Ann. Phys. {\bf 24} (1963) 419.

\bibitem 
{Chaikin}
See, for example:\\
P.~M. Chaikin and T.~C.~Lubensky {\it Principles of condensed matter
physics}, (Cambridge University Press, 1995)

\bibitem 
{fluepi} For example, see:
http://mathworld.wolfram.com/ConfluentHypergeometricLimitFunction.html

\bibitem 
{RobSpec}
  R.~C.~Myers, A.~O.~Starinets and R.~M.~Thomson,
  ``Holographic spectral functions and diffusion constants for fundamental
  matter,''
  JHEP {\bf 0711} (2007) 091
  [arXiv:0706.0162 [hep-th]].

\bibitem 
{sugra} A.~Nunez and A.~O.~Starinets,
  ``AdS/CFT correspondence, quasinormal modes, and thermal correlators in N  =
  4 SYM,''
  Phys.\ Rev.\  D {\bf 67} (2003) 124013
  [arXiv:hep-th/0302026];\\
A.~O.~Starinets,
  ``Quasinormal modes of near extremal black branes,''
  Phys.\ Rev.\  D {\bf 66} (2002) 124013
  [arXiv:hep-th/0207133].

\bibitem 
{fast} R.~C.~Myers and A.~Sinha,
  ``The fast life of holographic mesons,''
  JHEP {\bf 0806} (2008) 052
  [arXiv:0804.2168 [hep-th]].

\bibitem 
{ads4} A.~S.~Miranda, J.~Morgan and V.~T.~Zanchin,
  ``Quasinormal modes of plane-symmetric black holes according to the AdS/CFT
  correspondence,''
  arXiv:0809.0297 [hep-th].

\bibitem 
{Kitazawa}
  Y.~Kitazawa,
  ``Effective lagrangian for open superstring from five point function,''
  Nucl.\ Phys.\  B {\bf 289} (1987) 599.

\bibitem 
{koerber}
  P.~Koerber and A.~Sevrin,
  ``The non-Abelian Born-Infeld action through order $\alpha'^3$,''
  JHEP {\bf 0110} (2001) 003
  [arXiv:hep-th/0108169];
  ``Testing the $\alpha'^3$ term in the non-abelian open superstring  effective
  action,''
  JHEP {\bf 0109} (2001) 009
  [arXiv:hep-th/0109030].

\bibitem 
{rsquared}
  C.~P.~Bachas, P.~Bain and M.~B.~Green,
  ``Curvature terms in D-brane actions and their M-theory origin,''
  JHEP {\bf 9905} (1999) 011
  [arXiv:hep-th/9903210].

\bibitem 
{bad} M.~Wijnholt,
  ``On curvature-squared corrections for D-brane actions,''
  arXiv:hep-th/0301029.

\bibitem 
{findens} S.~Kobayashi, D.~Mateos, S.~Matsuura, R.C.~Myers and
R.M.~Thomson, ``Holographic phase transitions at finite baryon
density,'' JHEP {\bf 0702} (2007) 016  [arXiv:hep-th/0611099];\\
D.~Mateos, S.~Matsuura, R.~C.~Myers and R.~M.~Thomson, ``Holographic
phase transitions at finite chemical potential,''
  JHEP {\bf 0711} (2007) 085
  [arXiv:0709.1225 [hep-th]].

\bibitem 
{sl2z} E.~Witten, ``SL(2,Z) action on three-dimensional conformal
field theories with Abelian symmetry,''
  arXiv:hep-th/0307041.

\bibitem 
{matt2} M.~Wapler, ``Condensed matter properties of strongly coupled
defects,'' in preparation.

\bibitem 
{Kovtun:2003wp} P.~Kovtun, D.T.~Son and A.O.~Starinets, ``Holography
and hydrodynamics: Diffusion on stretched horizons,'' JHEP {\bf
0310}, 064 (2003) [arXiv:hep-th/0309213].

\bibitem 
{Policastro:2002se}  G.~Policastro, D.T.~Son and A.O.~Starinets,
``From AdS/CFT correspondence to hydrodynamics,'' JHEP {\bf 0209},
043 (2002) [arXiv:hep-th/0205052].

\bibitem 
{andrei2} A.~O.~Starinets,
  ``Quasinormal spectrum and the black hole membrane paradigm,''
  arXiv:0806.3797 [hep-th];
see, also: http://pirsa.org/06060018

\bibitem 
{M2hydro} C.P.~Herzog, ``The hydrodynamics of M-theory,'' JHEP {\bf
0212} (2002) 026 [arXiv:hep-th/0210126].

\bibitem 
{appell} For example, see:
http://mathworld.wolfram.com/AppellHypergeometricFunction.html

\bibitem 
{solo} D.~Birmingham, I.~Sachs and S.~N.~Solodukhin,
  ``Conformal field theory interpretation of black hole quasi-normal modes,''
  Phys.\ Rev.\ Lett.\  {\bf 88} (2002) 151301
  [arXiv:hep-th/0112055].


\end{thebibliography}
\end{document}